\documentclass[art,nonblindrev]{informs3} 
\graphicspath{ {./figures/} }
\usepackage{amsmath}
\usepackage{amssymb}
\usepackage{graphicx}
\usepackage{array}
\usepackage{booktabs}
\usepackage{multirow}
\usepackage{caption}
\usepackage{subcaption}
\usepackage{comment}
\usepackage{bbm}
\usepackage[toc,page]{appendix}
\usepackage{hyperref}

\OneAndAHalfSpacedXI

\graphicspath{ {./figures/} }

\usepackage{natbib}
 \bibpunct[, ]{(}{)}{,}{a}{}{,}%
 \def\bibfont{\small}%

\renewcommand{\footnoterule}{%
  \kern -3pt
  \hrule width \textwidth height 1pt
  \kern 2pt
}

\TheoremsNumberedThrough     
\ECRepeatTheorems

\EquationsNumberedThrough    

\MANUSCRIPTNO{MS-xxxx-xxxx.xx}

\begin{document}


\RUNAUTHOR{Su, Tretyakov and Newton}

\RUNTITLE{Deep Learning of Transition Probability Densities}

\TITLE{Deep Learning of Transition Probability Densities for Stochastic Asset Models with Applications in Option Pricing}

\ARTICLEAUTHORS{%
\AUTHOR{Haozhe Su}
\AFF{Nottingham Business School, Nottingham Trent University, Nottingham NG1 4FQ, UK, \EMAIL{Haozhe.Su@ntu.ac.uk}} 
\AUTHOR{M.V. Tretyakov}
\AFF{School of Mathematical Sciences, University of Nottingham, Nottingham NG7 2RD, UK, \EMAIL{Michael.Tretyakov@nottingham.ac.uk}}
\AUTHOR{David P. Newton}
\AFF{School of Management, University of Bath, Bath BA2 7AY, UK, \EMAIL{dpn25@bath.ac.uk}}
} 

\ABSTRACT{%
\vspace{3cm}

Transition probability density functions (TPDFs) are fundamental to computational finance, including option pricing and hedging.
Advancing recent work in deep learning, we develop novel neural TPDF generators through solving backward Kolmogorov equations in parametric space for cumulative probability functions. The generators are ultra-fast, very accurate and can be trained for any asset model described by stochastic differential equations. These are \textquotedblleft single
solve\textquotedblright , so they do not require retraining when
parameters of the stochastic model are changed (e.g. recalibration of volatility).
Once trained, the neural TDPF generators can be transferred to less powerful computers where they can be used for e.g. option pricing at speeds as fast as if the TPDF were known in a closed form.

We illustrate the computational efficiency of the proposed neural approximations of TPDFs by inserting them into numerical option pricing methods. We demonstrate a wide range of applications including the Black-Scholes-Merton model, the standard Heston model, the SABR model, and  jump-diffusion models. These numerical experiments confirm the ultra-fast speed and high accuracy of the developed neural TPDF generators.

} 


\maketitle

%


\section{Introduction}

In computational finance, such common tasks as option pricing and estimating sensitivities (see, e.g., \citet{GLA03,andricopoulos2003universal, andricopoulos2007extending,chen2014advancing,su2020option} and references therein) and simulating likelihood estimators \citep{ait2002maximum,ait2007maximum,yu2007closed,giesecke2019simulated} greatly benefit from access to fast and accurate evaluation of transition probability density functions (TPDFs) for stochastic differential equations (SDEs) modelling asset prices.

It is only in a few cases that we are able to find a TPDF in closed form; thus, we rarely have an exact formula which does not require numerical methods to calculate the TPDF. The well-known case of geometric Brownian motion, which is behind the Black-Scholes equation for European options,  is one such with a closed form TPDF. In the absence of a closed form solution, sometimes we are able to obtain a semi-closed form for the TPDF. These cases require additional calculations such as use of Bessel functions or calculations of integrals in the complex plane. For example, the CEV process suggested by \citet{cox1996constant} has a semi-closed form TPDF (see, e.g., \citealp{chen2014advancing}). \citet{lewis2016option} presented the semi-closed form of the joint transition density for \citeauthor{heston1993closed}'s \citeyearpar{heston1993closed} stochastic volatility process and a 3/2-stochastic volatility process, a variation of the Heston process
(\citealp{heston1997simple}; \citealp{lewis2016option}). In certain cases, TPDFs are not known in closed form but their characteristic functions are. Densities can subsequently be calculated using an inverse Fourier transform (\citealp{o2005path};
\citealp{lord2008fast}; \citealp{su2017option}). For the majority of processes used for modelling asset prices, where we have no other ways to calculate density, we can turn to density approximations, as introduced by \citet{chen2014advancing}.

There are several directions for obtaining density approximations described in the literature. First is the class of expansion methods: \citet{ait2002maximum,ait2008closed}, \citet{ait2007maximum} rely on small-time asymptotic expansions to solve multivariate diffusion processes. This direction is subsequently extended in \citet{yu2007closed} to solve multivariate jump-diffusions; \citet{henry2008analysis} and \citet{henry2017unbiased} use different expansion approaches for multivariate diffusion processes which do not rely on small-time expansion; the polynomial expansion approach suggested by \citet{filipovic2013density} is able to evaluate the density for affine jump-diffusion. Another class of density approximation explores exact sampling based on Monte Carlo simulation, first developed by \citet{beskos2005exact} and \citet{chen2013localization} then extended in a number of papers including \citet{giesecke2013exact} and \citet{giesecke2019simulated}. The latest development in this direction, from \citet{guay2021efficient}, gives an unbiased density estimator that can be used to tackle non-affine multivariate jump diffusion processes. In addition to these two major classes of approximation techniques, there are other techniques available; for example, the saddlepoint approximation approach suggested by \citet{ai2006saddlepoint} which is based on the characteristic function.

Nevertheless, density approximations have limitations, an important one being that they cannot universally provide approximations for all underlying processes. For example, the approximation techniques used in \citet{chen2014advancing} cannot solve for jump-diffusion processes and the density approximation is only accurate when the time step is very small. On top of this, density approximations can be cumbersome and may require complex computations. Thus, universality across any of the models of asset prices is available for those willing to put in the necessary effort but its practicality is more limited than we would wish. It is true that conventional numerical methods, particularly finite differences, can also be used to solve for the TPDF (see, e.g., \citealp{hagan_arbitragefree_2014}; \citealp{floc2014finite}; \citealp{su2020option}) but they are computationally expensive. More importantly, these only allow calculation for a single set of model parameters at a time, so they are far from ideal. A faster and more generic way of calculating density is needed for practical universal application. This leads us to {\it deep learning}.

The use of deep learning in option pricing has a long history, dating back at least as far as the early 1990s (see, e.g., \citealp{malliaris1993beating}; \citealp*{hutchinson1994nonparametric}) and has built a large technical literature of applications and improvements aimed at creating more
reliable setups linking underlying data directly with option prices in the markets and in calibrating implied volatilities. In contrast, we apply deep learning to very accurately approximate the TPDF, which can then be used (among other applications) in option pricing. The difficulties in extracting densities from market data are considerable, particularly in the tails (\citealp{figlewski2009estimating}, \citeyear{figlewski2018risk}) but that is not our aim. Rather, we use deep learning in place of other mathematical techniques in the calculation of TPDFs, training {\it artificial neural networks} (NNs) on simulated data according to various models of assets underlying the options. This approach offers
the potential to obtain TPDFs to a high accuracy more quickly and easily than previously, via pre-trained networks. A novel advance in our approach is in the parametric solution of the related partial differential equations (PDEs). It is usual in option pricing to solve a PDE for fixed parameters but then to have to recalculate when changes are made (e.g. within a calibration procedure). Here we create an engine that does not require us to solve the PDE again.

Deep learning algorithms offer an effective way of solving ordinary differential equations and PDEs. Prototypes of algorithms as differential equation solvers may be found in the work of \citet{lee1990neural} and \citet{dissanayake1994neural}. Subsequently, similar feedforward approaches have been introduced by \citet{van1995neural}, \citet{lagaris1998artificial} and \citet{lagaris2000neural}. \citet{schmidhuber2015deep} and \citet{yadav2015introduction} review the area. Calculations are very fast once the learning networks are trained but they suffer the drawback of needing to sample the data using a mesh. Mesh-free approaches to solving PDE problems based on deep learning are presented by \citet{sirignano2018dgm} - see also \citet{rackauckas2020universal} and references therein. Here, we adapt the \textquotedblleft Deep Galerkin Method\textquotedblright{}
(DGM) deep learning algorithm developed by Sirignano and Spiliopoulos for computing densities of underlying processes. Through this algorithm we arrive at an approximate PDE solution represented by a deep neural network, which is trained to satisfy the differential operator, the initial
condition, and the boundary conditions. The key for this work is that deep learning algorithms for PDEs can be extended to parametric PDE problems (see, e.g., \citealp{Khoo17,Kutyniok2019,Kutyniok2020}).
By parametric problems, we understand that we are interested not just in solving a PDE in a domain of its independent variables (time and space), as standard PDE solvers do, but also in having a representation of a PDE's solution for a range of its parameters' values. There are several types of methods under the umbrella of model order reduction capable of dealing with parametric PDE problems (see, e.g., \citealp{Antoulas2015} and references therein). The type used in this paper, based on deep
learning networks, has been shown over recent years to be very robust and universal (see, e.g., \citealp{Kutyniok2019,Kutyniok2020,Khoo17} and references therein, and also its use in this paper). Parametric PDE problems arise naturally in finance. Indeed, in finance applications, we need to be able to find option prices, densities for the underlying price process, etc. not only as functions of spot price and maturity but also as functions of volatility and other parameters of the underlying process to satisfy the practical need for frequent recalibration of models. Frequent recalibration imposes the requirement for valuations to be very fast for a range of parameter values. Deep learning applied to parametric PDE problems can achieve this, as we shall demonstrate. We emphasise that standard PDE solvers (e.g. finite differences) give approximations of PDE solutions only for given, fixed values of parameters and need to be run online each time a parameter changes, making them uncompetitive for the parametric problem described above. An important feature of using deep learning to solve parametric problems, as applied to the evaluation of TPDFs, is that it is ``single solve'' such that, once trained, the algorithm delivers results for any set of parameters without the need to re-run calculations.

These advantages are the prize. Nevertheless, formulating the use of deep learning techniques to calculate transition probability densities is a challenge. We know that the solution of the Fokker-Planck (Kolmogorov forward) equation is the transition probability density. Its initial
condition is a Dirac delta function, which has zero value everywhere except at one point where it is infinite. The presence of the delta function in the initial condition makes it impractical to use deep learning methods for the Fokker-Plank equation. \citet{al2018solving} make an attempt to solve the Fokker-Planck equation with DGM for the Ornstein-Uhlenbeck process by assuming a Gaussian initial condition. Pleasing as this is, it cannot be generalised to other underlying processes simply because other processes do not possess Gaussian shapes. Here, we circumvent this difficulty by considering instead the backward Kolmogorov equation with the terminal condition being a step function, whose solution is the cumulative distribution function.
Consequently, we propose a different direction by employing deep learning to solve the governing backward Kolmogorov equation for the cumulative probability function of an underlying stochastic process, thereby retrieving the transition density through differentiating the approximate cumulative probability. The governing backward Kolmogorov equation exists for any Markovian stochastic process meaning that, in theory, the deep learning solver is able to tackle any asset price model described by SDEs.

To illustrate the computational efficiency and accuracy of the proposed neural TPDF generators, we use them as a new ``engine'' for QUAD methods for option pricing. QUAD calculates option prices through integrating the product of the payoff function and the TPDF, and does not require massive calculation between the key points in time that specify the option type \citep{andricopoulos2003universal, andricopoulos2007extending}. Lack of available closed form or accurate approximations of TDPFs limited the range of underlying processes that could be tackled by QUAD methods for derivatives pricing. \citet{chen2014advancing} finessed this through insertion of interchangeable approximation techniques. This works well but a limitation is the accuracy of the TPDF approximation engines, leading to a computational bottleneck for the more practically interesting models of the underlying. The neural TDPF generators developed in our paper not only remove the bottleneck but, more significantly, innovate the deep learning route by being \textquotedblleft single solve\textquotedblright , across all option pricing techniques using the TPDF - there is no need to recalculate densities and the network rapidly returns results for any fresh inputs. We demonstrate the new approach starting with geometric Brownian motion (the classical Black-Scholes model), then the Heston model, the SABR model (a non-affine case), Kou's double exponential jump-diffusion model, and stochastic volatility jump diffusion. In the case of jump-diffusion models, we show how to use the deep learning approach to parametrically solve partial integral differential equations (PIDEs) in order to evaluate the corresponding TPDF. By doing so, we simultaneously contribute to the literature on deep PDE solvers by extending it to solving PIDEs. We also advance the use of deep learning for solving parametric PDE/PIDE problems by extensive computational investigation of the choices of loss functions, NN hyperparameters and NN architecture, as well as testing the performance of various GPU cards.


Our paper introduces a novel numerical approach for solving TPDFs, offering a fast and universal approximation method. We pioneer the utilisation of deep learning in successfully addressing this issue. Further, we  comprehensively assess the accuracy of neural TPDFs through option pricing results with QUAD, thereby enhancing our understanding of the implementation of deep learning techniques for solving PDEs.

\section{Methods}

We set the scene by expressing option prices via TPDFs of the underliers, since we will test the performances of NN approximated densities in the application setting of option pricing. For clarity of exposition, we begin with a simple one-dimensional model. In Section \ref{subsec:Preliminaries-on-transition}, we show how to formulate PDE problems to find TPDFs. In Section \ref{subsec:The-backward-Kolmogorov}, we apply deep learning techniques to approximating parametric PDEs and using this to solve backward Kolmogorov equations in order to find parametric TPDFs in a computationally effective manner.

\subsection{Option pricing via the transition density of an underlier}

Before moving on to financially more interesting cases, we consider the simplest model where a single underlying, $S(s),$ follows a stochastic differential equation (SDE) written under the forward measure:
\begin{equation}
dS=\sigma(s,S)dW,\,\,S(t)=S_{0}.\label{eq:general stock price}
\end{equation}
Here, $W(s)$ is a one-dimensional standard Wiener process and the
volatility $\sigma(s,x)$ is a deterministic function. For example,
the famous Black-Scholes option pricing formula for European style options rests on the
model in which the asset's price follows geometric Brownian motion:
\begin{equation}
dS=\sigma SdW,\,\,S(t)=S_{0}.\label{eq:Stock price SDE forward}
\end{equation}
Neglecting the discounting factor, the European option price $V(0,S_{0})$ at time $t=0$ on an underlying
$S_{0,S_{0}}(s)$ with the spot price $S_{0,S_{0}}(0)=S_{0}$ and
with a payoff function $f(x)$ and maturity $T$ is expressed by
\begin{equation}
V(0,S_{0})=E[f(S_{0,S_{0}}(T))].
\end{equation}
Since the solution of SDE \eqref{eq:general stock price} is a Markov
process, we can write
\begin{equation}
V(0, S_{0})=\int_{0}^{+\infty}f(y)p(0,S_{0};T,y)dy,\label{eq:QUAD_S0}
\end{equation}
where $p(t,S_{0};T,y)$ is the TPDF for the random variable $S_{t,S_{0}}(T)$
which is a solution of \eqref{eq:general stock price} at time $T$ starting from the point $S_{0}$ at time $t$.

Computationally, although working in the asset price space is viable, it is preferable to use the log-asset price space since this change of variables allows us to stretch the phase space, i.e., the left bound of the integration becomes $-\infty$. Consequently, this change of variables is beneficial
for training a NN for the backward Kolmogorov equation, as we will discuss later (see Section \ref{subsec:The-backward-Kolmogorov}). By Ito's formula, we obtain the SDE in the log-space, i.e.
for $X(s)=\ln S(s)$:
\begin{equation}
dX=-\frac{1}{2}\sigma^{2}(s,e^{X(s)})ds+\sigma(t,e^{X(s)})dW,\,X(t)=x_{0}=\ln S_{0}.\label{eq:SDE log price}
\end{equation}
 After this change of variable, the pricing formula \eqref{eq:QUAD_S0} becomes
\begin{equation}
V(0,S_{0})=\int_{-\infty}^{+\infty}f(e^{y})p(0,x_{0};T,y)dy,\label{eq:QUAD_simplest}
\end{equation}
where $p(t,x;T,y)$ is the TPDF for the solution $X_{t,x}(T)$ of
the SDE \eqref{eq:SDE log price} at time $T$ starting from $x$ at time $t$. When the volatility $\sigma$ is time independent, without loss of generality, we can start the solution $X(s)$  at time $t=0$ and use the simpler notation $p(x;s,y)$ or in short $p(s,y)$ instead of $p(0,x;s,y)$ which can also be used when we aim to find an option price at time $t=0$.

Thus, in Eq. \eqref{eq:QUAD_simplest}, if we know the TPDF $p(t,x;T,y)$,
we can readily calculate the price of the option, $V(0,S_{0})$. To
this end, the integral in Eq. \eqref{eq:QUAD_simplest} can be approximated
via quadrature after appropriate truncation of the integration range
$[-\infty,+\infty]$. Details of implementation for a given density
can be found in the early papers in the QUAD series (\citealp{andricopoulos2003universal,andricopoulos2007extending}).

\subsection{Preliminaries on transition densities\label{subsec:Preliminaries-on-transition}}

It is straightforward to write the Fokker-Planck (forward Kolmogorov) PDE for a TPDF. However, its initial condition contains the Dirac delta function which hinders the effective application of numerical techniques including the deep learning approach we present in this paper. We circumvent this problem by proposing a novel approach through solving the backward Kolmogorov PDE for the cumulative distribution function (CDF) and then differentiating it to obtain
the TPDF.

\subsubsection{The forward equation for transition density.}

The Fokker-Planck (forward Kolmogorov) equation governs the time evolution
of the transition probability for SDEs' solutions \citep{GIS68,FRE85,gardiner2004}.
In the case of the SDE \eqref{eq:SDE log price}, it takes the form
\begin{equation}
\frac{\partial}{\partial s}p(s,y)=\frac{1}{2}\frac{\partial^{2}}{\partial y^{2}}\left[\sigma^{2}(s,e^{y})p(s,y)\right]+\frac{1}{2}\frac{\partial}{\partial y}\left[\sigma^{2}(s,e^{y})p(s,y)\right],\:s>0,\:y\in\mathbb{R},\label{eq:forward equation}
\end{equation}
with initial condition for a fixed $x$
\begin{equation}
p(0,y)=\delta(y-x),\:y\in\mathbb{R},\label{eq:delta}
\end{equation}
where $\delta(\cdot)$ is the Dirac delta function. Recall that the full notation for the TPDF
is $p(s,y)=p(0,x;s,y)$ and note that the PDE problem Eqs. \eqref{eq:forward equation}-\eqref{eq:delta} is with respect to the variables $(s,y)$ while $x$ is a parameter. Also,  the Dirac delta function $\delta(y-x)$ is a function with zero value everywhere except at $x$:
\begin{equation}
\delta(y-x)=\begin{cases}
+\infty, & y=x,\\
0, & y\neq x,
\end{cases}
\end{equation}
and $\int_{-\infty}^{+\infty}\delta(y-x)dy=1$. Because the function is infinite at the point $y=x$,
it is difficult to solve the problem \eqref{eq:forward equation}-\eqref{eq:delta} numerically. This limitation extends to approximation via deep learning. Hence, we need to find a way to deal with it. To this end, instead of solving for TPDF directly, we first solve for the CDF and then
use the CDF to calculate the TPDF. This method, as we will show, is efficient in dealing with the evaluation of TPDFs.

\subsubsection{The backward Kolmogorov equation for cumulative distribution function.
\label{subsec:The-backward-Kolmogorov}}

We define the CDF of the log-stock price process $X_{t,x}(T)$ at
terminal time $T$ and the starting point $(t,x)$ as
\begin{align}
C(t,x;T,y): & =\text{Prob}(X_{t,x}(T)\leq y)=\int_{-\infty}^{y}p(t,x;T,z)dz,
\end{align}
where $\text{Prob}(X_{t,x}(T)\leq y)$ is the probability of the random
variable $X_{t,x}(T)$ not being greater than $y$.

In the case of the SDE \eqref{eq:SDE log price}, the function $C(t,x)=C(t,x;T,y)$
satisfies the backward Kolmogorov equation (see, e.g. \citealp{GIS68,FRE85,gardiner2004}):
\begin{equation}
\frac{\partial}{\partial t}C(t,x)+\frac{1}{2}\sigma^{2}(t,e^{x})\frac{\partial^{2}}{\partial x^{2}}C(t,x)-\frac{1}{2}\sigma^{2}(t,e^{x})\frac{\text{\ensuremath{\partial}}}{\partial x}C(t,x)=0,\:t\in[0,T),\:x\in\mathbb{R},\label{eq: backward equation}
\end{equation}
with the terminal condition for a fixed $y$
\begin{equation}
C(T,x)=\mathbbm{1}(x\leq y)=\begin{cases}
1, & x\leq y,\\
0, & x>y,
\end{cases}\:\:\:x\in\mathbb{R},\label{eq:terminal condition for backward}
\end{equation}
where $C(T,x)=C(T,x;T,y)$.
We note that the problem \eqref{eq: backward equation}-\eqref{eq:terminal condition for backward}
is written with negative direction in time, while the Fokker-Plank
equation \eqref{eq:forward equation} has positive direction in time.
The PDE problem \eqref{eq: backward equation}-\eqref{eq:terminal condition for backward}
is with respect to the variables $(t,x)$, and $T$ and $y$ are parameters.
It is easier to solve numerically Eqs. \eqref{eq: backward equation}-\eqref{eq:terminal condition for backward} than Eqs. \eqref{eq:forward equation}-\eqref{eq:delta} since the terminal condition \eqref{eq:terminal condition for backward} contains the bounded function.

Using a deep NN, we approximate $C(t,x;T,y)$
and then, using an automatic differentiation built-in function from
a deep learning library, such as TensorFlow, or finite differences,
we evaluate the transition probability density $p(t,x;T,y)$ by differentiating
$C(t,x;T,y)$ with respect to $y$; that is
\begin{equation}
p(t,x;T,y)=\frac{\partial}{\partial y}C(t,x;T,y).\label{eq: C to p}
\end{equation}
For asset price models of practical interest, the CDF $C(t,x;T,y)$ is a
smooth function for $t<T$, hence the differentiation in Eq. \eqref{eq: C to p}
is well defined. Consequently, numerical differentiation of an approximate
$C(t,x;T,y)$ can be done accurately. We discuss this in more detail
in Section \ref{sec:geometric-Brownian-motion}.

The solution to the problem \eqref{eq: backward equation}-\eqref{eq:terminal condition for backward} depends not only on the independent variables $(t,x)$ but also on the set of parameters $T,y$ and the volatility $\sigma(t,x)$, which can be parametrically encoded. A universal option pricing engine (as well as evaluation of sensitivities) requires an engine which can quickly and sufficiently accurately produce the TPDF for a range of values of the parameters so that we do not need to solve a PDE again, e.g. after recalibration of the volatility. Thus, we need to solve a parametric PDE problem, not just for fixed values of parameters (and the variables $(t,x)$) but for a set of them. Traditional numerical methods (finite difference, finite element, Monte Carlo, etc.) are designed to solve PDE problems with fixed parameters. In contrast, model order reduction methods for PDEs (\citealp{Antoulas2015}), and the most modern and universal of them, based on deep learning (\citealp{Khoo17,Kutyniok2019,Kutyniok2020}), are aimed at parametric PDE problems.

\subsection{Deep learning as a PDE solver\label{subsec:Deep-Learning-as}}

Recent developments in deep learning allow us to solve parametric PDE problems effectively. In Section~\ref{sec:constructing_loss} we present the key ingredients of the Deep Galerkin Method (DGM) proposed in \citet{sirignano2018dgm} for solving PDEs with fixed parameters. In Section~\ref{sec:NN} we recall the fundamentals of deep learning and also explain how to extend DGM to solving parametric problems. Implementation of DGM methods requires us to have a NN of suitable architecture. We find that even a feedforward network, the simplest NN architecture, can be used successfully for neural TPDF generators, though here, for consistency of the presentation, we prefer to use the NN architecture introduced together with DGM in \citet{sirignano2018dgm} (henceforth we refer to it as the DGM NN), which we find effective for our purposes (see Section~\ref{sec:DGMNN}). In Appendix~\ref{appdx: MLP}, we compare the performance of the feedback forward NN (its simplest version - the multilayer perceptron) with the DGM NN.

\subsubsection{Constructing the loss function.}
\label{sec:constructing_loss}

We start with a formal description of DGM used to solve a PDE problem under fixed parameters, which
we later extend into solving a parametric PDE problem.
Let  $G\subset\mathbb{R}^{d}$ be a domain with the boundary $\partial G$ and $\mathcal{L}$ be an elliptic operator. Consider the parabolic PDE
\begin{align}
\frac{\partial}{\partial t}u(t,\boldsymbol{x})+\mathcal{L}u(t,\boldsymbol{x})=0,\thinspace\thinspace & (t,\boldsymbol{x})\in[0,T)\times G,\label{eq:pde1}
\end{align}
with terminal condition
\begin{equation}
u(T,\boldsymbol{x})=u_{T}(\boldsymbol{x}),\thinspace\thinspace\boldsymbol{x}\in G,\label{eq:pde2}
\end{equation}
 and boundary condition
\begin{equation}
u(t,\boldsymbol{x})=g(t,\boldsymbol{x}),\:\:(t,\boldsymbol{x})\in[0,T]\times\partial G.\label{eq:pde3}
\end{equation}
We assume that this problem has a classical solution $u(t,\boldsymbol{x})$
(see, e.g., \citealp{Ladyzh}).

We note that Eqs. \eqref{eq: backward equation}-\eqref{eq:terminal condition for backward}
are of the form of Eqs. \eqref{eq:pde1}-\eqref{eq:pde2} and that the
backward Kolmogorov equation for any SDE driven by Wiener process has this form (see further
examples in Sections~\ref{sec:geometric-Brownian-motion}-\ref{sec:SABR}).

Next, we would like to use deep learning algorithms to ``learn''
the function $u(t,\boldsymbol{x})$ and, hence, approximate
the solution to the PDE problem \eqref{eq:pde1}-\eqref{eq:pde2}.
In other words, we wish to train a deep NN $f(t,\boldsymbol{x};\boldsymbol{\mathbf{\theta}})$
to approximate $u(t,\boldsymbol{x})$, where $\boldsymbol{\mathbf{\theta}}$
are trainable parameters of the NN.

In order to train the NN, we need to construct a loss (objective) function. In machine learning and deep learning, the loss function quantifies the performance of the approximation. In the PDE context, the loss function measures the fit (in other words, the residual) of the differential operator (i.e., the left-hand side of the equation \eqref{eq:pde1}), the terminal condition \eqref{eq:pde2} and the boundary condition  \eqref{eq:pde3}. The loss function $L(f)$ used in this paper has the form
\begin{align}
L_{1}(f) & =\left\Vert \frac{\partial}{\partial t}f(t,\boldsymbol{x};\boldsymbol{\mathbf{\theta}})+\mathcal{L}f(t,\boldsymbol{x};\boldsymbol{\mathbf{\theta}})\right\Vert _{[0,T]\times G,\nu_{1}}^{2},\\
L_{2}(f) & =\left\Vert f(T,\boldsymbol{x};\boldsymbol{\mathbf{\theta}})-u_{T}(\boldsymbol{x})\right\Vert _{G,\nu_{2}}^{2},\\
L(f) & =\lambda_{L}L_{1}(f)+L_{2}(f),\label{eq:loss}
\end{align}
where $L_{1}(f)$ is a loss in the differential operator term (the
residual) and $L_{2}(f)$ is a loss in the terminal condition term.
Here $\left\Vert f(y)\right\Vert _{\mathcal{Y},\nu}^{2}$ is the weighted
$L^{2}$-norm defined as $\left\Vert f(y)\right\Vert _{\mathcal{Y},\nu}^{2}=\int_{\mathcal{Y}}\left|f(y)\right|^{2}\nu(y)dy,$
where $\nu(y)$ is a positive probability density on a domain $\mathcal{Y}$.
In the context of this paper's application, we choose $\nu_{1}(y)$
as the probability density for uniform distribution on the time-price
domain $\mathcal{Y}=[0,T]\times G$ and $\nu_{2}(y)$ as the probability
density for uniform distribution on the price domain $\mathcal{Y}=G$.

Our loss function \eqref{eq:loss} has two differences compared with
the loss function of the original DGM of \citet{sirignano2018dgm}.
First, we have a new hyperparameter $\lambda_{L}$ in the loss function,
which controls the relative significance  of the differential operator term in the NN training.
We take $\lambda_{L}\geq1$ because the term $L_{1}(f)$  is more related to the accuracy of the approximation
(see further discussion in Section~\ref{sec:geometric-Brownian-motion}).
 Second, we do
not include a term in the loss function corresponding to the boundary
condition \eqref{eq:pde3}. The reason is that, for most of the underlying processes that we encounter in practice, we solve
the Cauchy problem \eqref{eq: backward equation}-\eqref{eq:terminal condition for backward}
which does not have a boundary condition. As is typical for all solvers of the Cauchy problem, we solve the PDE problem in a sufficiently large computational space domain $G$, so that the domain of interest
(e.g. for pricing) is smaller than $G$ to avoid imposing boundary conditions. However, having said that, in this paper we also consider a special case of the underlying process, a constant elasticity of variance (CEV)-like process, which has a boundary condition: we illustrate this using the example of the SABR model in Section~\ref{sec:SABR}.

\subsubsection{An overview of neural network training.}\label{sec:NN}

Next, we briefly recall the deep learning terminology. As discussed in Section \ref{sec:constructing_loss},
our goal is to train the NN $f(t,\boldsymbol{x};\boldsymbol{\mathbf{\theta}})$ (with $\boldsymbol{\theta}$ being the trainable NN parameters) to accurately approximate $u(t,\boldsymbol{x})$ so that the loss function  $L(f)$ is minimised. Hence,
we have an optimisation problem: we need to find the NN parameters $\boldsymbol{\mathbf{\theta}}$ that make the loss the smallest. To capture the complexity of the PDE solution, a complex deep NN has to be used to represent the solution, and finding optimal NN parameters could become a laborious task. This task is dealt with much more efficiently using a GPU rather than a CPU, predominately thanks to GPUs having a large number of  so-called tensor cores working in parallel and aimed to accelerate matrix operations.

The NN training starts with initialisation of the parameters $\boldsymbol{\mathbf{\theta}}$ and with generating randomised points $(t,\boldsymbol{x})$ in $[0,T]\times G$. These points are fed into the input layer of the NN and propagated to evaluate $f(t,\boldsymbol{x};\boldsymbol{\mathbf{\theta}})$ (the forward propagation step) and to calculate the loss $L(f)$.
We then use an optimisation algorithm (or optimiser) to update the trainable parameter $\boldsymbol{\mathbf{\theta}}$ in order to minimise the loss function. One of the simplest examples of optimisation algorithms is gradient descent. The optimiser requires efficient calculation of gradients of the loss function with respect to the NN parameters which is a back-propagation algorithm  (see, e.g., \citealp{goodfellow2016deep}).

Let us now summarise. The idea of solving PDE problems via a NN is that, by applying the optimisation algorithm of choice and back propagation from the deep learning
\citep{goodfellow2016deep}, one can efficiently find the set of NN parameters
$\boldsymbol{\mathbf{\theta}}$ which minimises
the loss function $L(f)$. Note that the NN solves PDE problems without
building a mesh. We also remark that the deep learning approach allows
us to break away from the ``curse of dimensionality'' and efficiently
solve high-dimensional PDEs, including high-dimensional parametric
PDEs as shown in this paper.


The algorithm adapted for our setup begins by initialising its
parameters $\boldsymbol{\mathbf{\theta}}$ with some $\boldsymbol{\theta}_{0}$,
followed by the loss optimisation loops. The $n$th loop is as follows:
\begin{enumerate}
\item For the differential operator term, generate random points $(t_{n},\boldsymbol{x}_{n})$
from $[0,T]\times G$ according to the uniform distribution
density $\nu_{1}$ on $[0,T]\times G$ ; for the terminal condition
term, generate the random points $\boldsymbol{w}_{n}$ from $G$ according
to the uniform distribution density $\nu_{2}$ on $G$.

\item Calculate the mean-square error $G(\boldsymbol{\mathbf{\theta}}_{n},\boldsymbol{s}_{n})$
at the $M$ randomly sampled points $\boldsymbol{s}_{n}=\{((t_{ni},\boldsymbol{x}_{ni}),\boldsymbol{w}_{ni})_{i=1}^{M}\}$:
\begin{equation}
G(\boldsymbol{\theta}_{n},\boldsymbol{s}_{n})=\frac{\lambda_{L}}{M}\sum_{i=1}^{M}\left(\frac{\partial}{\partial t}f(t_{ni},\boldsymbol{x}_{ni};\boldsymbol{\mathbf{\theta}}_{n})+\mathcal{L}f(t_{ni},\boldsymbol{x}_{ni};\boldsymbol{\mathbf{\theta}}_{n})\right)^{2}+\frac{1}{M}\sum_{i=1}^{M}\left(f(T,\boldsymbol{w}_{ni};\boldsymbol{\mathbf{\theta}}_{n})-u_{T}(\boldsymbol{w}_{ni})\right)^{2}.\label{eq:squared error}
\end{equation}
\item Perform the optimisation algorithm to update the parameters
$\boldsymbol{\mathbf{\theta}}$ at the
random point $\boldsymbol{s}_{n}$. 
In this paper,  we use the Adaptive Movement Estimation algorithm (ADAM) introduced by \citealp{kingma2014adam}), which is one of the most popular and efficient stochastic optimisation algorithms, to perform optimisation. 
In ADAM we choose the learning rate $\alpha_{n}$ ourselves (see further details in Section~\ref{sec:DGMNN}) while we use the default values for the rest of its parameters as in \citet{kingma2014adam}.

\item Stop iterations if a convergence criterion is satisfied. The convergence
criterion could be $G(\boldsymbol{\theta}_{n},\boldsymbol{s}_{n})$
smaller than some predetermined threshold or $\left\Vert \boldsymbol{\theta}_{n+1}-\text{\ensuremath{\boldsymbol{\theta}}}_{n}\right\Vert _{2}$ smaller than some predetermined threshold. Our choice of convergence criteria in this paper is to train for a fixed number of epochs (as described in later sections) and choose the outcome (i.e., the parameters $\boldsymbol{\theta}$) of the one with the least loss.
\end{enumerate}

We recall that an epoch is a single pass of the entire data set through the NN during training whereby updating its parameters.

As it was pointed out in \citet{sirignano2018dgm}, $\boldsymbol{\mathbf{\theta}}_{n}$
will converge to a critical point of the loss function $L(f(\cdot;\boldsymbol{\mathbf{\theta}}))$
as $n\to\infty$; that is
\begin{equation}
\lim_{n\to\infty}\left\Vert \nabla_{\theta}L(f(\cdot;\boldsymbol{\theta}_{n}))\right\Vert _{2}=0.
\end{equation}
However, it is likely that $\boldsymbol{\theta}_{n}$ only converges to a local minimum rather than a global minimum since the NN $f(t,\boldsymbol{x};\boldsymbol{\theta})$ is non-convex.
It is for this reason that the hyperparameter $\lambda_{L}\ge1$ in Eq. \eqref{eq:squared error} is needed to improve the accuracy of the approximation.

The algorithm can solve the PDE problem \eqref{eq: backward equation}-\eqref{eq:terminal condition for backward}
with varying time, $t$, and spot price, $x$, with other input parameters being fixed. Its extension
to solving parametric PDE problems is explained and illustrated in later sections.

We can use different NN architectures to represent PDE solutions, even the simplest feedforward neural network architecture known as Multilayer Perceptrons (MLP).   In this paper, we mostly use the DGM NN architecture proposed by \citet{sirignano2018dgm} which we found   effective for the PDE and PIDE problems under consideration, and we present its architecture details in the following section. We briefly discuss the comparison results of MLP vs DGM in Appendix~\ref{appdx: MLP}.

\subsubsection{The DGM neural network.}\label{sec:DGMNN}

It is known (see, e.g., \citealp{sirignano2018dgm}) that the speed of NN training is
problem-dependent and that it is important to choose an appropriate, problem-specific NN structure to achieve reasonable training speed for a problem under consideration. Thanks to the activities of the machine learning community in recent years, we can gain access to numerous NN structures for PDE applications. The Deep Galerkin Method (DGM) network, used in this paper, can be considered as a variant of the widely used Long Short-Term Memory networks (LSTMs) introduced
by \citet{hochreiter1997long}. LSTMs are a special kind of recurrent NN which are useful for modelling sequential data such as time series prediction or speech recognition problems. It is shown by \citet{sirignano2018dgm} that an LSTM-type  architecture works well for PDE problems with ``sharp turns'' in the initial or terminal conditions, which is exactly the situation we have here, since we have a step function in the terminal condition of the backward Kolmogorov
equation that we approximate.

Next, we introduce the DGM NN architecture:
\begin{align*}
\mathbf{S}^{1} & =\vartheta(\mathbf{W}^{1}\mathbf{x}+\mathbf{b})\\
\mathbf{\mathbf{Z}}^{\ell} & =\vartheta(\mathbf{U}^{z,\ell}\mathbf{x}+\mathbf{W}^{z,\ell}\mathbf{S}^{\ell}+\mathbf{b}^{z,\ell}),\thinspace\thinspace\ell=1,\dots,L,\\
\mathbf{G}^{\ell} & =\vartheta(\mathbf{U}^{g,\ell}\mathbf{x}+\mathbf{W}^{g,\ell}\mathbf{S}^{\ell}+\mathbf{b}^{g,\ell}),\thinspace\thinspace\ell=1,\dots,L,\\
\mathbf{R}^{\ell} & =\vartheta(\mathbf{U}^{r,\ell}\mathbf{x}+\mathbf{W}^{r,\ell}\mathbf{S}^{\ell}+\mathbf{b}^{r,\ell}),\thinspace\thinspace\ell=1,\dots,L,\\
\mathbf{H}^{\ell} & =\vartheta(\mathbf{U}^{h,\ell}\mathbf{x}+\mathbf{W}^{h,\ell}(\mathbf{S}^{\ell}\odot\mathbf{R}^{\ell})+\mathbf{b}^{r,\ell}),\thinspace\thinspace\ell=1,\dots,L,\\
\mathbf{S}^{\ell+1} & =(1-\mathbf{G}^{\ell})\odot\mathbf{H}^{\ell}+\mathbf{Z}^{\ell}\odot\mathbf{S}^{\ell},\thinspace\thinspace\ell=1,\dots,L,\\
f(\boldsymbol{\mathbf{x}};\boldsymbol{\theta}) & =\mathbf{W}\mathbf{S}^{L+1}+\mathbf{b},
\end{align*}
where $\mathbf{x}$ denotes the input layer, $\vartheta(\cdot)$ is the activation function, $\odot$ denotes element-wise multiplication, the number of hidden layers is $L+1$ (there are 1 initial hidden layer and $L$ LSTM like layers), and $\mathbf{U},\mathbf{W},\mathbf{b}$ are the parameters of the network which together form $\boldsymbol{\theta}$. Particularly, $\mathbf{x}$ is a $D\times 1$ vector, where $D$ is the number of trainable features; $\mathbf{U}$ and $\mathbf{W}$ are $M\times D$ matrices, where $M$ is the number of nodes in the neural network; $\mathbf{b}$ is a $M\times 1$ vector. The input $\mathbf{x}$ can be just $(t,\boldsymbol{x})$ from $[0,T]\times G$ for solving a PDE problem with fixed parameters as in the previous subsection or can also include parameters of the PDE when we solve parametric PDE problems - as in the following sections. We note that we use the same letter $f$ in slightly different contexts but this should not lead to any confusion.

\begin{figure}
\begin{centering}
\subfloat[Components: input, hidden layers and output of DGM NN. We show four hidden layers
here, including one initial hidden layer and three LSTM-like layers
($L=3)$.]{\includegraphics[scale=0.3]{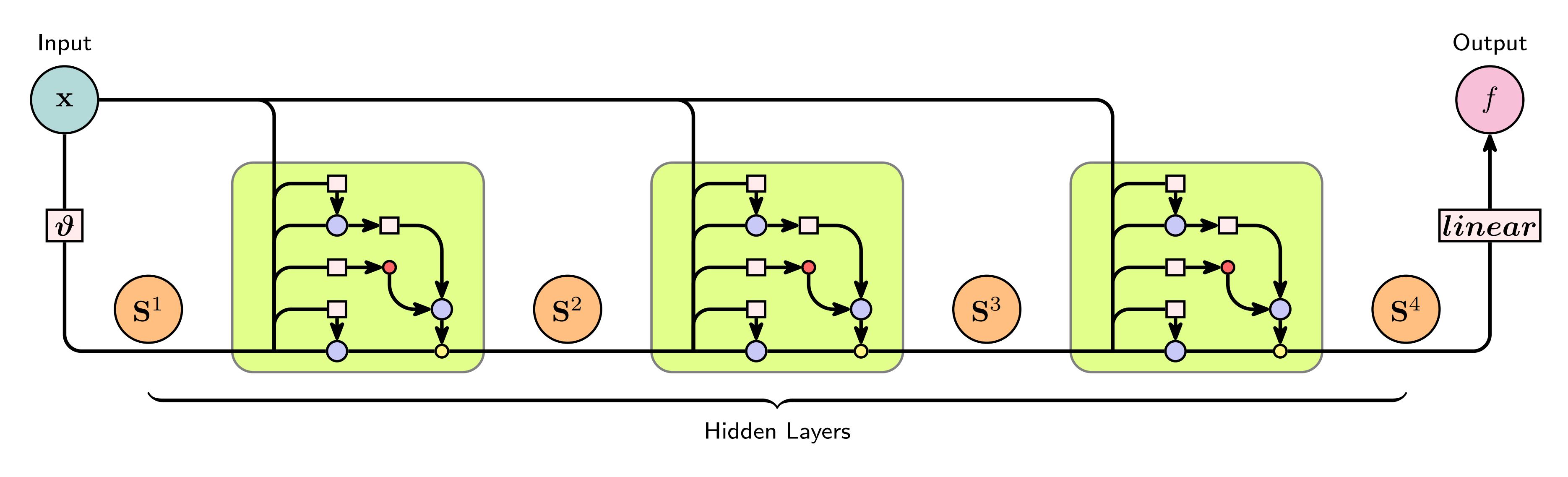}
\label{Fig: DGM nets}
}
\par\end{centering}
\begin{centering}
\subfloat[A closer look into the hidden layer.]{\includegraphics[scale=0.3]{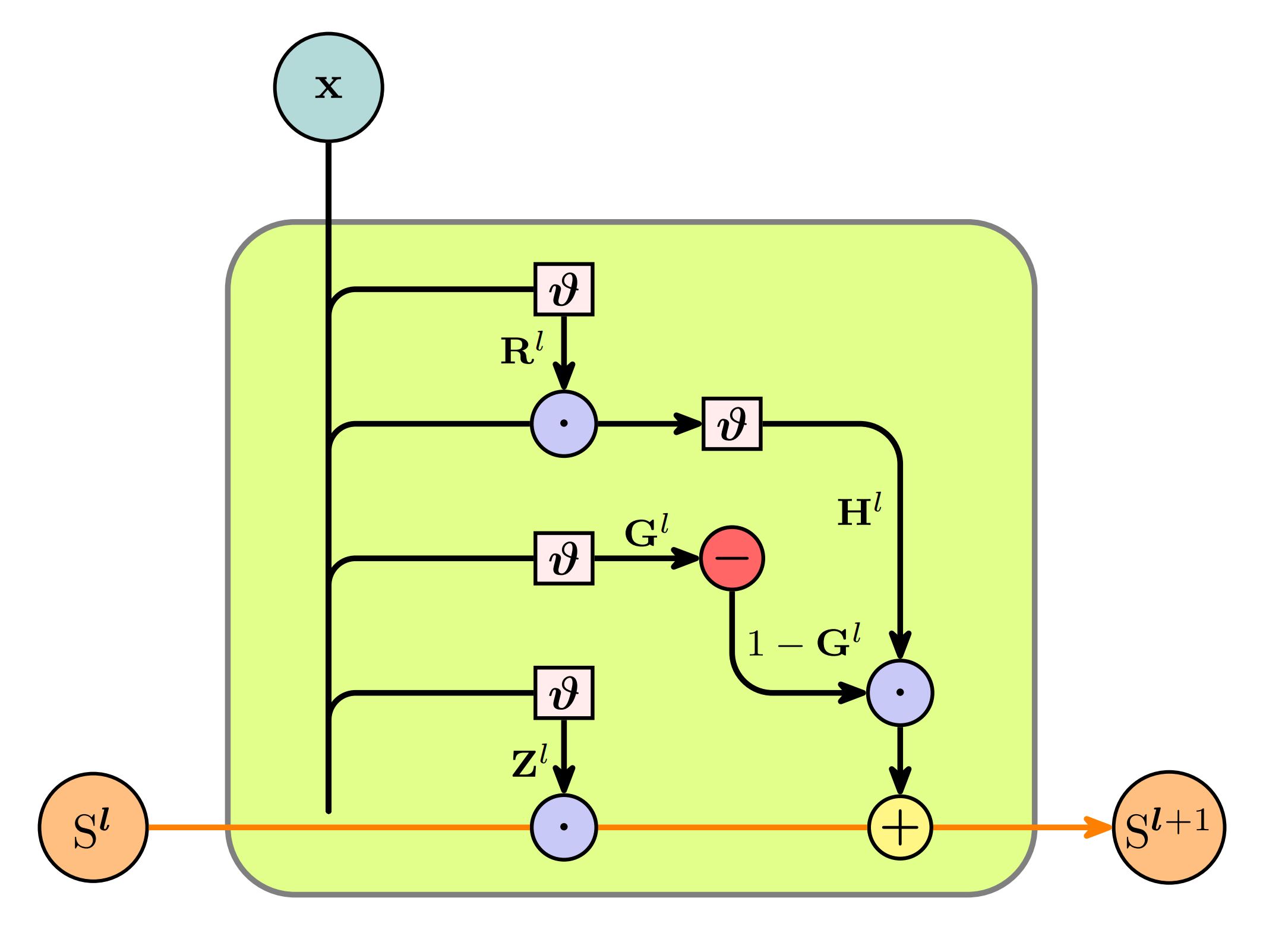}

\label{Fig: A closer look into hidden layer}

}
\par\end{centering}
\caption{Illustration of the deep Galerkin method NN architecture
used in this paper.}
\label{Fig: Neural network illustration}
\end{figure}

For completeness of the presentation, we visualise the NN used for this paper in Figure~\ref{Fig: Neural network illustration}. The input layer $\mathbf{x}$ represents the input data and in the case of parametric PDEs, that includes randomly sampled time points $t$ and all the space and parameters points $\boldsymbol{x}$. It is important to note that the input layer $\mathbf{x}$ only contains the simulated data and we use different simulated data for every five epochs. In the NN considered, $\mathbf{x}$ is not just the input of the initial layer but also the input of all the hidden layers. If we zoom in on the LSTM-like hidden layer (see Figure \ref{Fig: A closer look into hidden layer}) then we can see that apart from the initial data $\mathbf{x}$, the input of the current hidden layer also includes the output of the previous hidden layer $\mathbf{S}^{\ell}$. A hidden layer with this setting could strengthen the long-term memory ($\mathbf{S}^{\ell}$, $\ell=1,\dots,L+1$) of the network (see,
e.g., the architecture for LSTM networks \citep{hochreiter1997long} and highway networks \citep{srivastava2015training}), which, as suggested in \citet{sirignano2018dgm}, could then subsequently translate to the improved performance of capturing the nonlinearity and ``sharp
turn'' in solving PDEs as presented in this paper. Once we have calculated the output $f$, we then use it to construct the loss function before applying an efficient optimisation algorithm (like ADAM we are using here) to update the NN parameters with the aim of  minimising the loss function.

A NN with this LSTM-like architecture is quite complex, consisting of a chain of hidden layers with each hidden layer containing sub-layers. \citet{goodfellow2016deep} suggest that empirically a ``deep" NN consisting of more layers and more nodes per layer performs better than a ``shallow" NN. However, a deeper NN requires more training time with potentially only a marginal increase in accuracy. Therefore, one needs to find an optimal set of hyperparameters (the set of parameters that controls the NN architecture and the learning process) for training. \citet{sirignano2018dgm} suggest using the following hyperparameters: four hidden layers in the network ($L=3$), 50 nodes per layer ($M=50)$, using the hyperbolic tangent function as the activation function $\vartheta(\cdot)$, which guarantees that
$f(\mathbf{x};\boldsymbol{\theta})$ is smooth and, hence, is suitable for approximating classical solutions of PDEs. Xavier initialisation is used to initialise parameters and, as noted earlier, the ADAM optimisation algorithm is used to update the parameters. We utilise the default settings for ADAM in TensorFlow, but adjust the learning rate. We find that this set of hyperparameters also works well for NN training in this paper. Here, we use a fixed learning rate $\alpha_{n}=0.0001$, and we find that this $\alpha_{n}$ value gives an optimal convergence result for our problems. After training, we choose the NN with parameters such that it has the smallest total loss. We implement this algorithm in TensorFlow, which is an open-source software library developed by Google and designed specifically for deep learning. TensorFlow has a built-in function for parameter initialisation and optimisation algorithms. This allows great simplification of the NN development process\footnote{We implemented our Python code in Jupyter Notebook, which is suitable for any Jupyter Notebook environment including Google Colab. The code is available for download. The code only requires users to provide the backward Kolmogorov equation corresponding  to the SDEs of interest (see further instructions and comments in the Notebook). The code does not require a powerful machine to run (but an NVIDIA GPU is needed) and users can run the code using  Google Colab if they wish.}.

For the numerical illustrations in this paper, we use 5,000 sample points per epoch. An epoch means passing the entire 5,000 randomly generated sample points backward and forward through the NN. We note that the numbers of sample points used here are greater than the 1,000 sampling points used in \citet{sirignano2018dgm} because we focus on solving parametric PDEs and greater numbers of sample points are needed to get to the required accuracy. To increase the training efficiency, each newly generated dataset is trained on five epochs. Depending on their complexity, different models require different numbers of epochs to train in order to achieve the required accuracy. In theory, a more accurate NN approximation of the PDE/PIDE solution can be obtained with more training. However, at the same time, the cost of improvement becomes higher and users face a situation of diminishing returns when more training is used. In \citet{sirignano2018dgm}, six GPU nodes are used along with the asynchronous stochastic gradient descent method for parallel training. The training setup presented here used a single GPU. For illustration purposes, we performed deep learning training via the Google Colab Pro server, where we used a single NVIDIA Tesla P100 GPU or, alternatively, a single NVIDIA RTX 2080 GPU in a gaming laptop. In other words, even without access to high-powered computers, we are able to use the techniques developed in this paper, in a reasonable amount of time, to train the networks. We compare training time using various GPUs in Appendix~\ref{appdx:GPU}.  Moreover, once trained, these networks are used in option pricing calculations that require only minor computational power (see Appendix~\ref{appdx: calculation_time}).

\section{Tuning and performance analysis under Geometric Brownian motion
\label{sec:geometric-Brownian-motion}}

We first illustrate our implementation using geometric Brownian motion (GBM) as in the derivation of the Black-Scholes-Merton formula (\citealp{black1973pricing}; \citealp{merton1973theory}). Being
one of the simplest models for the underlying asset, the GBM model is also one of the easiest to train. It is a lot less costly than others when it comes to tuning the hyperparameter $\lambda_{L}$ and hence useful in understanding the performance of the proposed approach.

Under the forward measure of the log asset price space, the backward Kolmogorov equation for the CDF $C(t,x)=C(t,x;T,y)$ in the GBM case is
\begin{equation}
\frac{\partial C}{\partial t}-\frac{\sigma^{2}}{2}\frac{\partial C}{\partial x}+\frac{\sigma^{2}}{2}\frac{\partial^{2}C}{\partial x^{2}}=0,\:t\in[0,T),\:x\in\mathbb{R},\label{eq:BS}
\end{equation}
with terminal condition
\begin{equation}
C(T,x)=\mathbbm{1}(x\leq y)=\begin{cases}
1, & x\leq y,\\
0, & x>y.
\end{cases}\label{eq:BS2}
\end{equation}

To find the TPDF from the cumulative distribution, we must first solve the backward Kolmogorov equation in a parametric manner for a range of $y\in[y_{\min},y_{\max}]$, given a minimum value of the range $y_{\min}$ and a maximum value $y_{\max}$, so that after taking the derivative of $C(t,x;T,y)$ with respect to $y,$ we obtain the TPDF $p(t,x;T,y)$. Also, from the asset price process, we have
that the CDF is a function of the volatility, $\sigma$. Thus, the problem needs to be solved in the $(t,x,y,\sigma)$-domain $Q\subset\mathbb{R}^{4}$, which contains not only the independent variables $(t,x)\in[0,T]\times G\subset\mathbb{R}^{2}$ with respect to which the PDE is formulated, but also the parameters $(y,\sigma)\in\mathbb{R}^{2}$. Using the NN
$f(\boldsymbol{\mathbf{x}};\boldsymbol{\theta})=f(t,x,y,\sigma;\boldsymbol{\theta})$
described in the previous section, we can find a solution to this parametric problem for the backward Kolmogorov equation.

The optimisation process starts with initialising the network parameters $\boldsymbol{\theta}$. Then, for each epoch, we uniformly sample points from $Q$ before feeding them into the DGM network and using an optimisation algorithm to minimise the loss function. The process is repeated until the required accuracy is reached. The loss function in this example is defined as
\begin{align}
L_{1}(f) & =\left\Vert \left(\frac{\partial}{\partial t}-\frac{1}{2}\sigma^{2}\frac{\partial}{\partial x}+\frac{1}{2}\sigma^{2}\frac{\partial^{2}}{\partial x^{2}}\right)f(t,x,y,\sigma;\boldsymbol{\theta})\right\Vert _{[0,T]\times G,\nu_{1}}^{2},\label{eq:BSM L1}\\
L_{2}(f) & =\left\Vert f(T,x,y,\sigma;\boldsymbol{\theta})-\mathbbm{1}(x\leq y)\right\Vert _{G,\nu_{2}}^{2},\\
L(f) & =\lambda_{L}L_{1}(f)+L_{2}(f).\label{eq:BSM L}
\end{align}

The first and second partial derivatives terms $\partial f/\partial t$, $\partial f/\partial x$, and $\partial^{2}f/\partial x^{2}$ can either be computed directly using the gradient function in TensorFlow (or similar derivatives calculation functions in other deep learning libraries) or approximated via numerical differentiation with finite differences.

The domain $Q$ in which we use to train the network is $x,y\in[-2.3,2.3]$, $\sigma\in[0,0.6]$, $t\in[0,1.2].$ We note that the choice of $\pm2.3$ for the range of $x$ and $y$  corresponds approximately to $\pm3.5\sigma_{\max}t_{\max}$, where
$\sigma_{\max}$ and $t_{\max}$ are the upper bounds of $\sigma$ and $t$ parameters, respectively. The time horizon considered for the network is $T=1.2$. We also note that $T-t$ is the time to maturity, which in future we will often denote as $T$ again, but this should not cause any confusion. Although we fix the log-spot price $x_{0}=0$, we still need to train the network in $x$
as the PDE is written in $x$. We train 2 million epochs. For every five epochs, we use 5000 random points. Thus, in total, we use 2 billion random space time points to train this network. A single NVIDIA P100 GPU takes 0.022 seconds to complete 1 epoch, which means it takes about 12.5 hours to complete NN training using 2 million epochs. Once the offline training of the NN has been completed, it can be used online for the pricing of any option with this underlying model (within the range of volatility $\sigma$ and maturity $T$ for which the network was trained) and for other computational finance tasks. This is unlike conventional numerical methods, which have to be run over again (i.e. ``online") for each set of the parameters' values separately - considerably slower than taking the required values from a trained NN. In Appendices~\ref{appdx:GPU} and~\ref{appdx: calculation_time}), we report the speed of both offline training (across different GPUs) and online access time.

Once the cumulative distribution function $C(t,x;T,y)=C(t,x;T,y,\sigma)$
has been approximated (via the NN training) by $f(\boldsymbol{\mathbf{x}};\boldsymbol{\theta})=f(t,x,y,\sigma;\boldsymbol{\theta})$,
we use numerical differentiation to approximate the TPDF
$p(t,x;T,y)=p(0,x;T-t,y,\sigma)$ by $\tilde{p}(0,x;T-t,y)=\tilde{p}(0,x;T-t,y;\boldsymbol{\theta})$:
\begin{equation}
\tilde{p}(0,x;T-t,y)=\frac{f(t,x,y+\Delta,\sigma;\boldsymbol{\theta})-f(t,x,y-\Delta,\sigma;\boldsymbol{\theta})}{2\Delta}
\end{equation}
for some $\Delta>0$. In our experiments we chose $\Delta=0.005$. The choice of $\Delta$ should be neither too small nor too large: if it is too small, that could introduce oscillations; if it is too large, the approximation accuracy around the peak area of the density is insufficient. We find that a good choice for $\Delta$ is in the range of {[}0.001, 0.01{]}.

\subsection{$\lambda_{L}$ Tuning} \label{sec:tuning}

\begin{figure}
\begin{centering}
\includegraphics[scale=0.20]{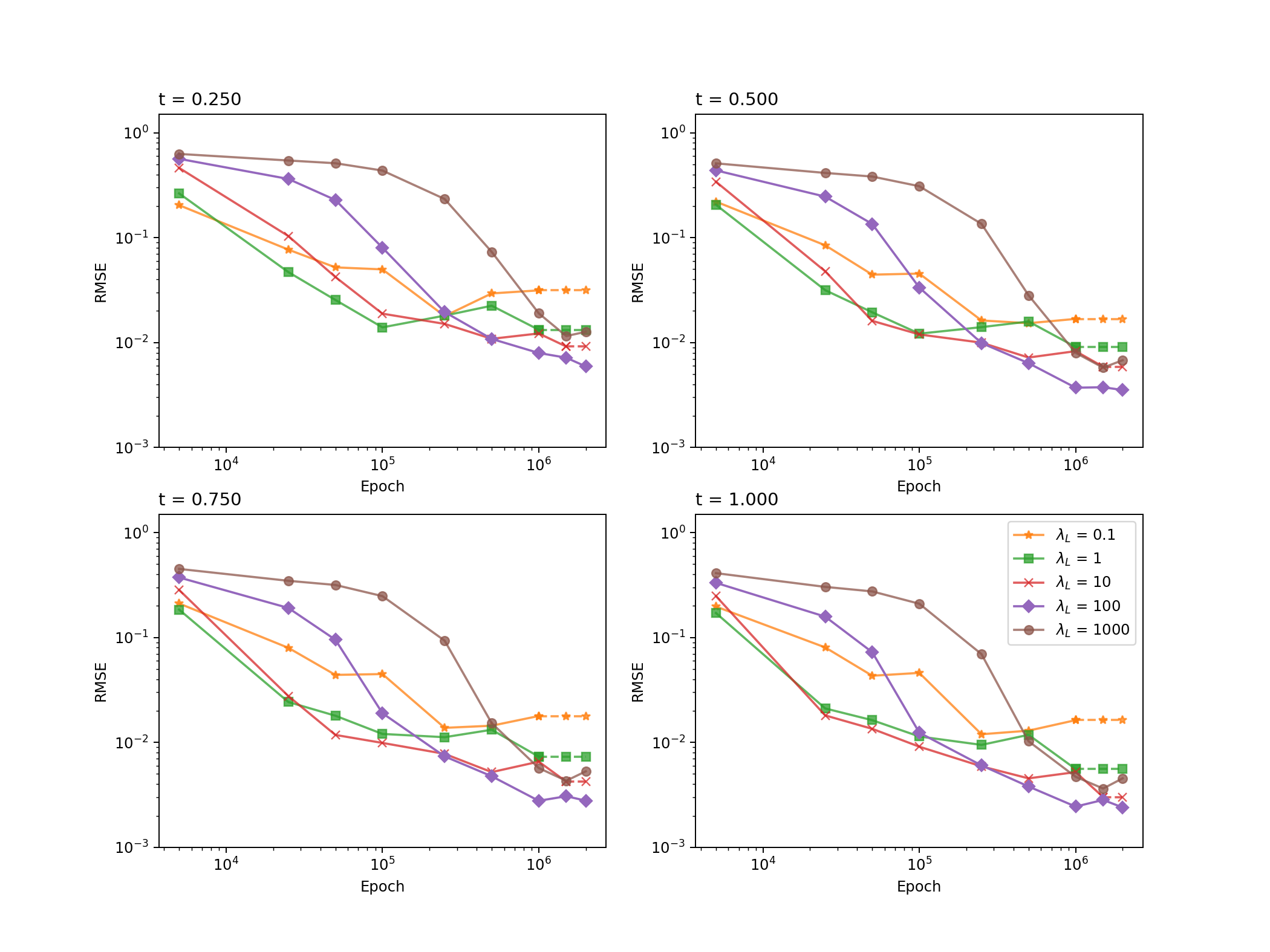}
\par\end{centering}
\caption{The root mean squared errors (RMSE) are calculated using the NN approximate TPDF benchmarked against the exact TPDF Eq.~\eqref{eq: BSM closed form density}. This figure plots
epoch against RMSE for different $\lambda_{L}=0.1,1,10,100,1000$
across various times $t=0.25$, $0.5$, $0.75$, $1.0$ set up under the Black-Scholes-Merton model. The epoch checkpoints are $5000$, $25000$, $50000$, $100000$, $250000$, $500000$,
$1000000$, $1500000$, $2000000$. Each checkpoint represents the NN with
the smallest training loss up to that epoch checkpoint. The dashed line in the graph means no better NN was found and we use the best
NN in the previous checkpoint. The domain $Q$ used to train the
NN is $x,y\in[-2.3,2.3]$, $\sigma\in[0,0.6]$, $t\in[0,1.2].$
In terms of validation, for each $t$, $\sigma=0.1$, $0.15$, $0.2$, $0.25$,
$0.3$, $0.35$, $0.4$, $0.45$,$0.5$, $0.55$, $0.6$.
The initial log price $x_{0}=0$, and the range of $y$ for each $(t,\sigma)$
set is $[-2.3,2.3]$.}
\label{Fig: lambda_L Epoch against RMSE}
\end{figure}

We use most of the NN hyperparameters as set in \citet{sirignano2018dgm} and we find their modelling set-up effective (apart from a few alterations we have adapted for this application; see Section \ref{sec:DGMNN}). We have introduced a new hyperparameter $\lambda_{L}$ in the loss function and in this subsection we use numerical experiments to understand the impact of this parameter.  This hyperparameter gives relative weighting to the two loss terms. Without the hyperparameter $\lambda_{L}$, the loss terms $L_{1}$ and $L_{2}$ bear the same weight. When $\lambda_{L}>1$, this means that the loss term with the differential operator weights more than the terminal condition term; likewise, when $\lambda_{L}<1$, it indicates that the terminal condition needs to be satisfied more compared with the differential operator. Intuitively, the loss term $L_{1}$ needs to have a higher weight than $L_{2}$ since the error $L_1$ is expected to have a higher influence on the accuracy of the NN. We tune the hyperparameter $\lambda_{L}$ and validate our assertion.
To this end, we train the NN for different values of $\lambda_{L}=0.1,1,10,100,1000$.
We compare the obtained approximation $\tilde{p}$ with the exact density $p$, which is Gaussian in the example considered:
\begin{equation}
p(0,x;T,y)=\frac{1}{\sigma\sqrt{2\pi T}}\exp\left[-\frac{(x-y+0.5\sigma^{2}T)^{2}}{2\sigma^{2}T}\right].\label{eq: BSM closed form density}
\end{equation}
For each $\lambda_{L}$, we train up to 2000000 epochs and record the best training results (i.e., with the smallest loss) within checkpoints set at 5000, 25000, 50000, 100000, 250000, 500000, 1000000, 1500000, 2000000. Figure~\ref{Fig: lambda_L Epoch against RMSE} demonstrates the tuning. Across all the considered $\lambda_{L}$, the training accuracy generally increases as more epochs (or sampling
points) are used. However, we also notice that after a certain amount of training, e.g., after 500000 epochs, the RMSE does not go down as fast as at the start of the training -- this is especially
apparent when $\lambda_{L}=1$ and $10$, for which the algorithm cannot find better NN parameters beyond 1 million epochs (as shown by the dashed lines). This is due to the vanishing gradient problem in deep learning, meaning that it is becoming increasingly difficult to find optimal NN parameters to reduce the loss. When $\lambda_{L}=10$, the vanishing gradient problem is alleviated; $\lambda_{L}=100$ helps to improve the performance further; but the performance gets worse when  $\lambda_{L}=1000$. When the number of epochs used is greater than 500000, we see that $\lambda_{L}=100$ produces the best training results for the given time domain  ($t=0.25,0.5,0.75,1.0)$. Overall, we find that $\lambda_{L}=100$ produces accurate results for our NN approximation framework.

\subsection{Performance across maturities\label{subsec:Performance-across-maturity}}


\begin{figure}[h]
\centering{}\includegraphics[scale=0.29]{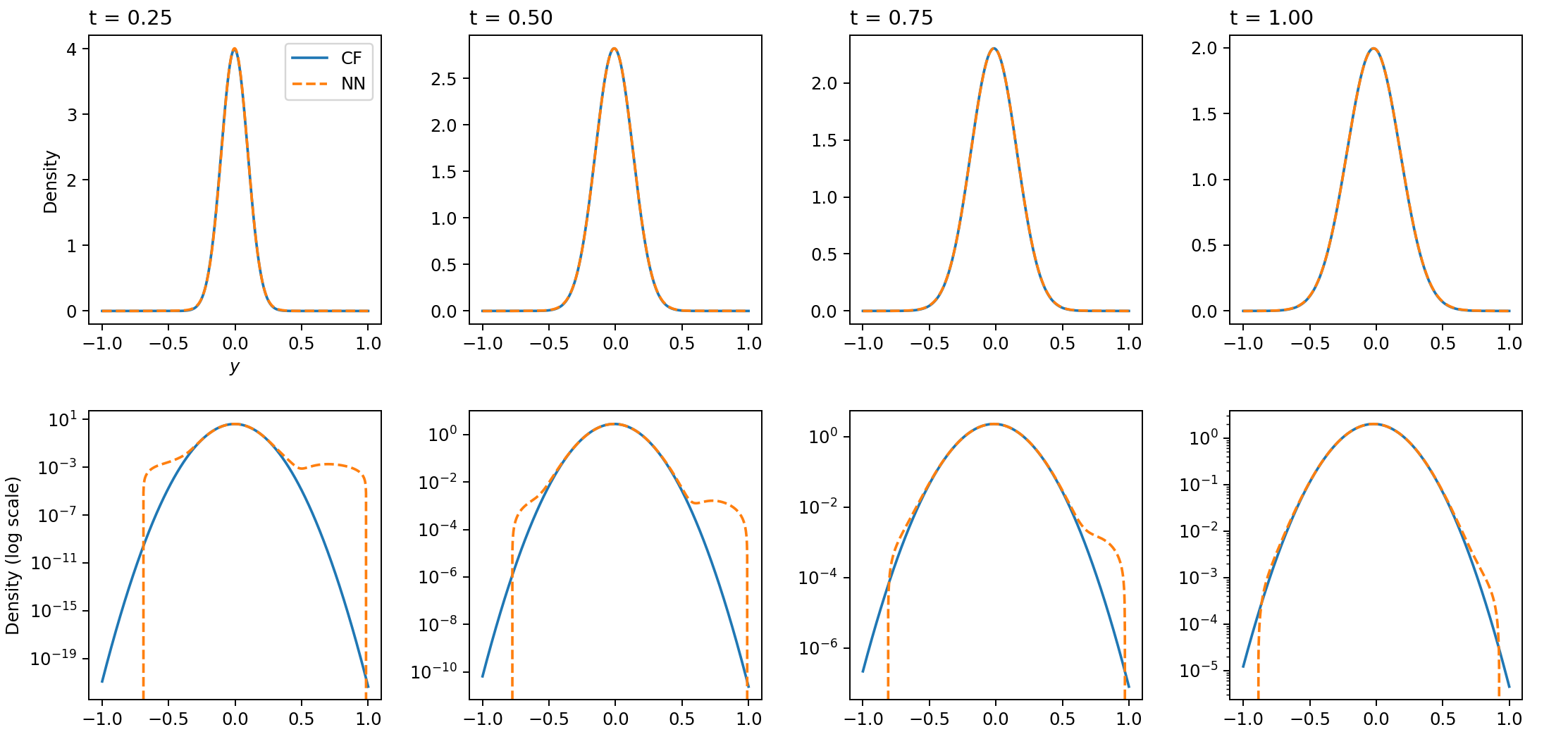}\caption{Gaussian TPDF, closed form vs. NN approximation is shown in the first row. The second row shows the same results in log scale of density. The figures shown correspond to the parameters $x=0$, $\sigma=0.2$, the time to maturity $t=0.25,0.5,0.75,1.0$, and $\lambda_{L}=100$. The domain used to train the network is $x,y\in[-2.3,2.3]$, $\sigma\in[0,0.6]$,
$t\in[0,1.2].$}
\label{Fig: BSM density graphs}
\end{figure}

Figures \ref{Fig: lambda_L Epoch against RMSE} and \ref{Fig: BSM density graphs} demonstrate that the NN approximation can closely approximate the exact TPDF in terms of RMSE. However, when considering the accuracy of the relative approximation, the log scale plots in Figure \ref{Fig: BSM density graphs} reveal that the approximation is accurate around the centre but not for the tails. Specifically, the approximation cannot yield  highly accurate results for density values smaller than $10^{-3}$. This limitation may be attributed to the error measure employed during training. As has been discussed in Section~\ref{sec:constructing_loss}, the training is designed to minimise the $L^{2}$ loss, i.e., the mean-square error. Hence, it is unsurprising that the results are more accurate in terms of RMSE than relative error, and that the approximations of the tails yield large percentage errors, which is a limitation of our approach.

 We also see from  Figure \ref{Fig: lambda_L Epoch against RMSE} that when maturity $t$ is decreased, the approximation accuracy decreases. This observation is not surprising, since when $t$ is made smaller, the TPDF concentrates around the initial state (i.e., $x=0$ in this example) and the derivatives of the CDF $C$ with respect to $y$ tend to infinity when $t$ goes to 0 (see Eqs. \eqref{eq: C to p} and
\eqref{eq:BS2}), affecting the accuracy of the numerical differentiation.
If the training range in $x$ is too wide ($[-2.3,2.3]$ in this example), then the narrow range around the zero initial state may not acquire sufficient data points, resulting in less accurate approximations compared with those obtained for larger $t$. If we did not train specifically for $t$ being close to 0, the approximation accuracy for this region could be low. Moreover, it is important to note that, since we are dealing with the backward Kolmogorov equation with a terminal condition, when translating the time to maturity, $t$, to the time variable used in the loss function, it becomes $T-t$, where $T$ is the terminal time. That means, as $t\rightarrow0$, the time variable in the training $T-t\rightarrow T$.

\begin{figure}
\begin{centering}
\includegraphics[scale=0.20]{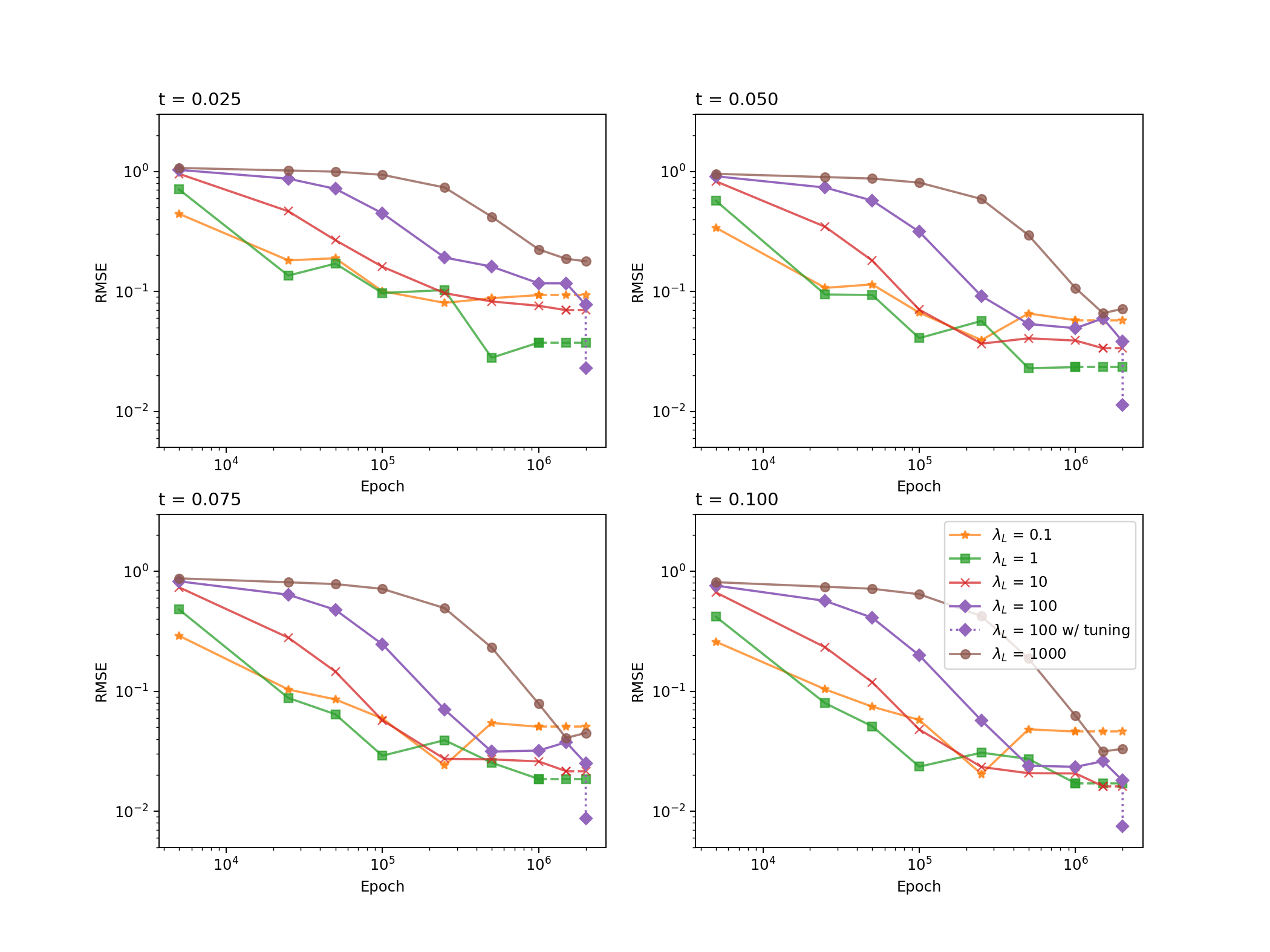}
\par\end{centering}
\caption{RMSE are calculated using the NN approximate
TPDF benchmarked against the true TPDF from Eq. \eqref{eq: BSM closed form density}. This figure plots
epoch against RMSE for different $\lambda_{L}=0.1$, $1$, $10$, $100$, $1000$
across various times $t=0.025$, $0.05$, $0.075$, $0.1$.
The epoch checkpoints are $5000$, $25000$, $50000$, $100000$, $250000$, $500000$,
$1000000$, $1500000$, $2000000$. Each checkpoint represents the NN with
the smallest training loss up to that epoch checkpoint. The dashed line in the graph means no better NN can be found and we use the best
NN in the previous checkpoint. The dotted line for $\lambda_L = 100$ represented the fine tuned model after $500$ epochs.
The domain $Q$ used to train the network is $x,y\in[-2.3,2.3]$, $\sigma\in[0,0.6]$, $t\in[0,1.2].$
In terms of validation, for each $t$, $\sigma=0.1$, $0.15$, $0.2$, $0.25$, $0.3$, $0.35$, $0.4$, $0.45$, $0.5$, $0.55$, $0.6$.
The initial log price $x_{0}=0$, and the range of $y$ for each $(t,\sigma)$
set is $[-2.3,2.3]$.}
\label{Fig: small time lambda_L Epoch against RMSE }
\end{figure}

As discussed earlier, when $\lambda_{L}=100$, the emphasis is on the differential
operator $L_{1}$ term in the loss function $L$ rather than on the terminal condition $L_{2}$ term, which increases the overall accuracy of the approximated results while simultaneously it exacerbates the inaccuracy for $t\rightarrow0$. In
Figure~\ref{Fig: small time lambda_L Epoch against RMSE }, we repeat the same tuning as in Figure \ref{Fig: lambda_L Epoch against RMSE} but this time we show the plot for various small times to maturity $t=0.025$, $0.05$, $0.075$, $0.1$. We see that for these small times to maturity, $\lambda_{L}=1$ is the clear winner. As $t$ increases, the accuracy gap narrows, and at $t=0.1$ the differences between the results obtained with $\lambda_{L}=1$, $10$, $100$ are very small. The reason why $\lambda_{L}=1$ performs best for small times to maturity is that equal emphasis is put on the terminal condition term  and the differential operator term, resulting in higher accuracy for the region close to $T$ (or time to maturity close to zero). It should be noted that when $\lambda_{L}=0.1$, although more training emphasis is placed on the terminal condition term, performance is much worse due to the lack of emphasis on the main body of the approximated solution (i.e., on the differential operator term).

\begin{figure}
\begin{centering}
\includegraphics[scale=0.29]{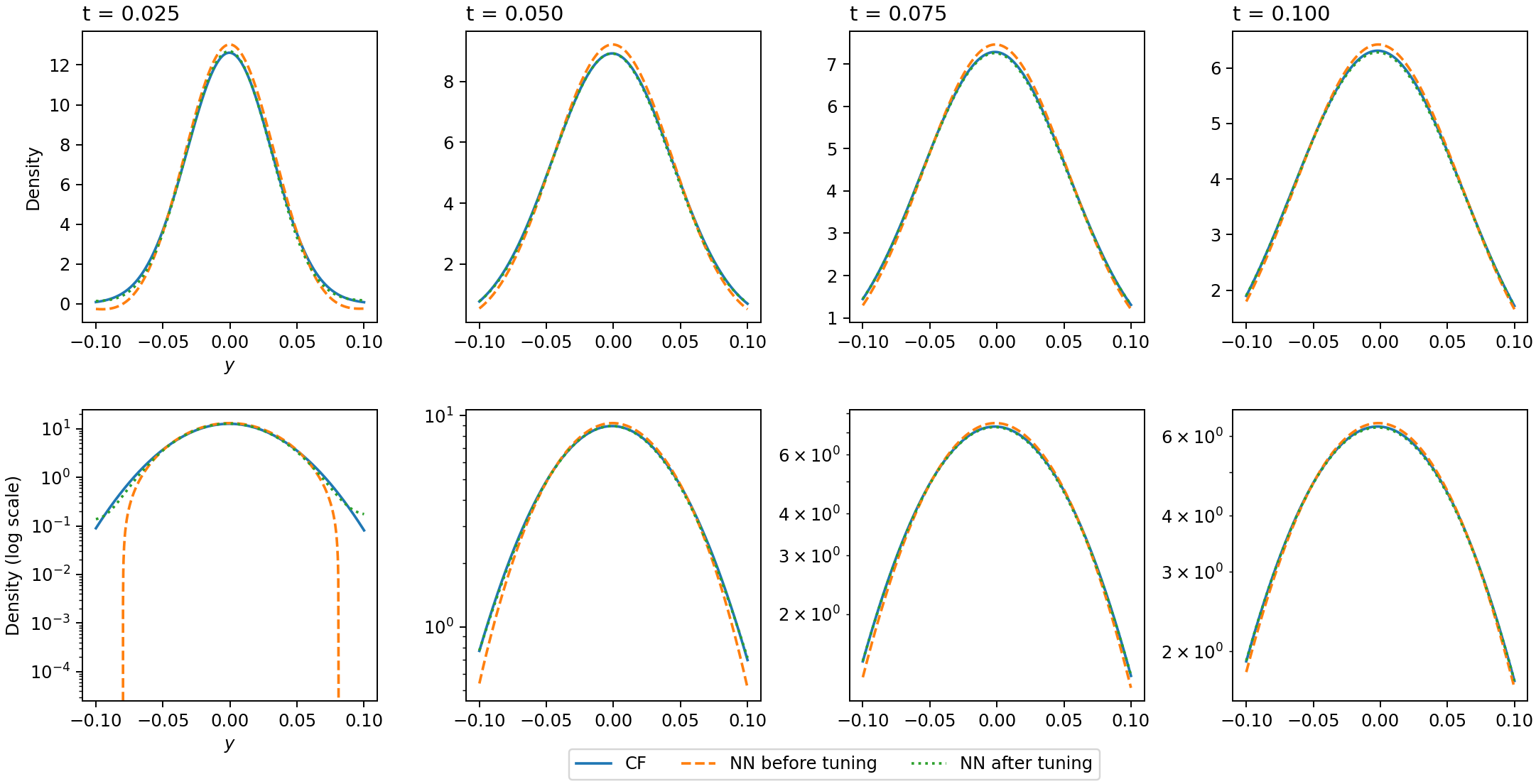}
\par\end{centering}
\caption{The first row shows the exact Gaussian TPDF and the  neural TPDF before and
after tuning for small time to maturity. The second row shows the same results in log scale of density. $x=0$, $\sigma=0.2$, the
time to maturity $T=0.025$, $0.05$, $0.075$, $0.1$. The NN results
after 500 epochs tuning (and we select the best trained NN).}
\label{Fig: Transfer learning}
\end{figure}

{\it Transfer learning} allows us to address the additional complexity arising from NN training across different maturities as discussed above. The use of transfer learning is a common strategy to train a new NN for a problem using the knowledge from an already trained NN for a related problem (see, e.g., \citealt{goodfellow2016deep}). Here, we transfer
the NN we already have from one time range to another. Specifically, if the TPDF with smaller $T$ is what we require, we can use  an already trained NN on a larger time range to train for  smaller
$T$ in order to fine tune the NN for a small time-to-maturity domain. Transfer learning is much more efficient than training a NN from scratch. For instance, here we use the NN trained on the domain as defined above (training done with $\lambda_{L}=100)$ and then change the domain $Q$
to the new domain $x,y\in[-0.75,0.75]$, $\sigma\in[0,0.6]$, $t\in[1.08,1.2].$
We only train a NN using another 2500 epochs and we select the NN with the smallest total loss. Consequently, we achieve an accurate NN approximation that is dedicated to smaller maturities $T.$ The performance is shown in Figure~\ref{Fig: small time lambda_L Epoch against RMSE }. We see that the performance of the NN obtained via transfer learning  is better than that of the NN trained with $\lambda_{L}=1$. Note that we fine tune the NN for the new domain using only a small number of
epochs since we have altered the weight in the loss function from $\lambda_{L}=100$ to $1$; training with more epochs would make the NN lose the traits of overall accuracy brought by the NN trained with $\lambda_{L}=100$ and, instead, converge to the NN trained with $\lambda_{L}=1$ as more epochs are used. Figure~\ref{Fig: Transfer learning} illustrates the TPDF before and after tuning and the improvement is visible. Similarly, when we need a result that is out of the training domain range, we can use transfer learning to obtain the required
NN efficiently. This transfer learning idea is applied to all the stochastic models considered in this paper.

\subsection{Pricing performance}\label{subsec: BSM_Pricing}

In this subsection we test the accuracy of the neural TPDF by using it in option pricing. For illustrative purposes, we begin with a  simple example. Table \ref{Tab: BSM_put} shows the pricing performance for put options for a given set of parameters under the GBM process. Overall, the pricing is quite accurate in terms of absolute error, with all absolute errors less than $5\times10^{-4}.$ However,
the relative errors for deep out-of-the-money options are large. This is because the prices of deep out-of-the-money closed form put options are very close to zero. If the approximate TPDF does not reach the required level of accuracy, it will inevitably fail to accurately price the out-of-the-money put options. In addition, for deep out-of-the-money options, when the prices are very small, the approximate TPDF could give negative prices, although the negative prices are very close to zero. This can be seen as a limitation of the NN approximated density.

\begin{table}
\caption{The put option prices calculated from GBM's closed
form TPDF (CF) and neural network approximate TPDF (NN). Parameters
used: $S_{0}=1,V_{0}=0.2,T=1.0$. We calculate option prices of 11
different strikes and the value of strike $K=0.5,0.6,\dots,1.5$.
The semi-closed form prices are calculated using closed form Black-Scholes-Merton European option pricing formula. We also include the absolute
errors (abs err.) and percentage error (\% err) of the calculations. The average time is in seconds.}
\label{Tab: BSM_put}
\scriptsize
\centering{}{}%
\begin{tabular}{>{\centering}p{1.5cm}>{\centering}p{2cm}>{\centering}p{2cm}>{\centering}p{1.5cm}>{\centering}p{2cm}}
\toprule
{Strikes} & \multicolumn{2}{c}{{Price}} & \multicolumn{2}{c}{{Error}}\tabularnewline
\cmidrule{2-5} \cmidrule{3-5} \cmidrule{4-5} \cmidrule{5-5}
 & {CF} & {NN} & {Abs $($$\times10^{-4})$} & {\%}\tabularnewline
\midrule
{0.5} & {9.4310$\times10^{-6}$} & {$-2.524\times10^{-4}$} & {$2.62$} & {2776.14\%}\tabularnewline
{0.6} & {2.6111$\times10^{-4}$} & {$-6.865\times10^{-5}$} & {$3.30$} & {126.29\%}\tabularnewline
{0.7} & {2.4811$\times10^{-3}$} & {2.1189$\times10^{-3}$} & {$3.62$} & {14.60\%}\tabularnewline
{0.8} & {0.0119} & {0.0115} & {$3.73$} & {3.14\%}\tabularnewline
{0.9} & {0.0359} & {0.0355} & {$3.91$} & {1.09\%}\tabularnewline
{1.0} & {0.0797} & {0.0792} & {$4.26$} & {0.53\%}\tabularnewline
{1.1} & {0.1429} & {0.1425} & {$4.61$} & {0.32\%}\tabularnewline
{1.2} & {0.2215} & {0.2210} & {$4.89$} & {0.22\%}\tabularnewline
{1.3} & {0.3101} & {0.3096} & {$4.97$} & {0.16\%}\tabularnewline
{1.4} & {0.4045} & {0.4040} & {$4.78$} & {0.11\%}\tabularnewline
{1.5} & {0.5019} & {0.5015} & {4.37} & {0.09\%}\tabularnewline
\midrule
{Ave. time} & {0.002} & {0.245} &  & \tabularnewline
\bottomrule
\end{tabular}
\end{table}

An important feature of the NN approximation approach is that, once the model is trained in the parametric space, it can quickly produce the TPDF given a set of parameters. To assess its performance, we compare the pricing errors of the neural TPDF against the closed form solutions over the randomly generated parameter sets across different moneyness positions and maturities. \citet{zhou2022option} defines moneyness ranges from a comprehensive data set of all US-listed equity call and put options from 1996 to 2019, and we base our definition of moneyness ranges for put options on this: Deep in-the-money (DITM) (1.20-1.40); In-the-money (ITM) (1.05-1.20); At-the-money (ATM) (0.95-1.05); Out-of-the-money (OTM) (0.80-0.95); Deep out-of-the-money (DOTM) (0.60-0.80). We acknowledge that this classification can only be applied to normal market conditions and not to extreme market conditions. We do not take into account the extreme strikes at either end, where pricing accuracy is lower, typically for any generic approximation method for a density.
Here we use the price percentage error (PrPCTE) and the price root-mean-square
error (PrRMSE) to evaluate the pricing performance with the NN approximated
TPDF. The results shown in Table \ref{Tab: Black-Scholes-Merton}
confirm our observation from the example in Table \ref{Tab: BSM_put}.
Pricing is accurate in terms of RMSE but less accurate in terms
of absolute error. We recall that, in the training of the Black-Scholes-Merton
CDF, the training loss reaches $10^{-3}$ and the RMSE of the density
reaches $2\times10^{-3}$. It is not surprising that the RMSE of the
prices also reaches a similar level of accuracy. However, this is more
of a problem with using an approximated density to price options rather
than a problem specific to the NN approach.

We have shown in Section~\ref{subsec:Performance-across-maturity} that the neural TPDF accuracy decreases as the time to maturity decreases, and the option price evaluations in Table~\ref{Tab: Black-Scholes-Merton} confirm this observation in general but the impact in terms of RMSE is still small. We emphasise that although the pricing error is of the order $10^{-4}$, the relative error can still be large since put prices are close to zero when strikes and volatility are small. This can be seen from the extreme PrPCTE values in Table~\ref{Tab: Black-Scholes-Merton}. The extreme PrPCTE is produced because the closed form solutions can give results with high accuracy, whereas the results using the NN approximated TPDF can only give a certain level of accuracy, depending on how small the loss achieved in training is. For example, if the closed form price is $10^{-16}$ but its approximation is  $10^{-4}$, the percentage error is about $10^{12}$ while the absolute error is still $10^{-4}$. This is a general limitation of using an approximate TPDF for pricing options regardless of the method used to approximate the TPDF. In terms of pricing performance across different volatilities, there is no significant difference in accuracy even when the volatility, $\sigma$, is as small as 0.05 and, thus, we do not present the results here.

Once the NN is trained, retrieving the neural TPDF is very quick and this brings us to the important feature of replacing previous calculation engines for TPDF with one developed through deep learning: accessing a trained NN to retrieve the density takes very little time, meaning that calculating a single option price using neural TPDF with QUAD is as fast as using the closed form density with QUAD. This preserves the immense speed advantage of not just the original
QUAD shown in \citet{andricopoulos2003universal} but for more complex models of the underlying which had previously slowed the QUAD calculations; e.g. in \citet{chen2014advancing}.

\begin{table}[h]
\caption{The pricing errors of the NN estimated density of the Black-Scholes-Merton
model, benchmarked against BSM's closed form pricing solutions. We
randomly generate 100 sets of parameters in the parametric space:
$V_{0}\in[0.1,0.5]$. We compare five moneyness ranges: Deep in-the-money
(DITM) (1.20-1.40); In-the-money (ITM) (1.05-1.20); At-the-money (ATM)
(0.95-1.05); Out-of-the-money (OTM) (0.80-0.95); Deep out-of-the-money
(DOTM) (0.60-0.80). We choose four maturity times $T=0.25,0.5,0.75,1.0$.
The pricing performance is in the form of price percentage error (PrPCTE),
price root-mean-square error (PrPMSE).}
\label{Tab: Black-Scholes-Merton}
\scriptsize
\centering{}%
\begin{tabular}{ccccccc}
\toprule
{Maturity} & {Error Type} & {DOTM} & {OTM} & {ATM} & {ITM} & {DITM}\tabularnewline
\midrule
\multirow{2}{*}{{$t=0.25$}} & {PrPCTE} & {5.662445$\times10^{12}$} & {2.933544} & {0.006059} & {0.003086} & {0.001537}\tabularnewline
 & {PrRMSE} & {0.000265} & {0.000204} & {0.000331} & {0.000463} & {0.000539}\tabularnewline
\midrule
\multirow{2}{*}{{$t=0.5$}} & {PrPCTE} & {1.448508$\times10^{5}$} & {0.083452} & {0.003751} & {0.002292} & {0.001373}\tabularnewline
 & {PrRMSE} & {0.000248} & {0.000218} & {0.000279} & {0.000409} & {0.000518}\tabularnewline
\midrule
\multirow{2}{*}{{$t=0.75$}} & {PrPCTE} & {3.338457$\times10^{2}$} & {0.025648} & {0.002880} & {0.001746} & {0.001089}\tabularnewline
 & {PrRMSE} & {0.000237} & {0.000225} & {0.000264} & {0.000369} & {0.000470}\tabularnewline
\midrule
\multirow{2}{*}{{$t=1.0$}} & {PrPCTE} & {14.51919} & {0.013960} & {0.002432} & {0.001437} & {0.000899}\tabularnewline
 & {PrRMSE} & {0.000232} & {0.000228} & {0.000253} & {0.000337} & {0.000429}\tabularnewline
\bottomrule
\end{tabular}
\end{table}

In the next section, we provide further illustrations of the proposed approach to approximating TPDFs using deep learning by applying it to several stochastic volatility and jump diffusion models.

\section{Neural TPDFs for stochastic volatility and jump diffusion models}
\label{sec:main_examples}

In the previous section, we introduced our approach using the simplest model - the geometric Brownian motion, a.k.a., the original Black-Scholes-Merton setup.
In this section, we apply the proposed deep learning approach to  a number of more complex, representative models of underliers.  We first consider the standard Heston model (we also successfully applied our methodology to a Heston model with time dependent parameters but the corresponding results are not included in the paper). Next, considering the SABR model, we show that non-affine models can be tackled by our approach. To further confirm the universality of our approach, we add jumps to models for underliers and demonstrate how partial integral differential equations (PIDEs) can be approximated efficiently by NNs. 

\subsection{Heston model \label{sec:The-Heston-process}}
\label{sec:illustr}


Under the forward measure, the Heston model \citep{heston1993closed}  written for the log stock price,
$X$, and variance, $V,$ of the stock price has the form
\begin{align}
dX & =-\frac{V}{2}dt+\sqrt{V}dW_{1},\label{eq:Heston stock process}\\
dV & =\kappa(\omega-V)dt+\xi\sqrt{V}dW_{2}, \notag \\
dW_{1}dW_{2} & =\rho dt, \notag
\end{align}
where $W_{1}$ and $W_{2}$ are Wiener processes with correlation $\rho$,
$\omega$ represents the long term variance, $\kappa$ is the rate
at which $V$ reverts to $\omega$, $\xi$ is the volatility of the
volatility (vol of vol). We require that $2\kappa\omega\ge\xi^{2}$
to ensure that the variance process is strictly positive at any finite
time.

The backward Kolmogorov equation for the joint CDF $C(t,x,v)=C(t,x,v;T,y,z)$
is
\begin{equation}
\frac{\partial C}{\partial t}-\frac{v}{2}\frac{\partial C}{\partial x}+(\kappa(\omega-v))\frac{\partial C}{\partial v}+\frac{1}{2}v\frac{\partial^{2}C}{\partial x}+\frac{1}{2}\xi^{2}v\frac{\partial^{2}C}{\partial v^{2}}+\rho\xi v\frac{\partial^{2}C}{\partial x\partial v}=0,t\in[0,T),\:x\in\mathbb{R},\:v>0,\label{eq:HestonKol}
\end{equation}
with terminal condition
\begin{equation}
C(T,x,v)=\mathbbm{1}(x\leq y,v\leq z)=\begin{cases}
1, & x\leq y\thinspace\text{\thinspace and}\thinspace\thinspace v\leq z,\\
0, & \text{otherwise.}
\end{cases}\label{eq:HestonKol2}
\end{equation}
Note that this PDE problem is of the form \eqref{eq:pde1}-\eqref{eq:pde2}.

We build a NN, $f$, to approximate the CDF, $C$. For the Heston model, the neural network $f$ is trained in the $(t,x,y,v,z,\kappa,\omega,\xi,\rho)$-domain $Q\subset\mathbb{R}^{9}$,
which is high-dimensional. Traditional methods aimed at solving PDEs under fixed parameters are uncompetitive in this challenging parametric PDE setting in comparison with the deep learning algorithm. We note that the high dimensionality here is of the parametric space. Recall that we need to solve high-dimensional parametric PDEs to find TPDFs for a range of values in the time-state and parametric spaces in order to effectively use the TPDFs for financial engineering tasks such as pricing options. The problem \eqref{eq:pde1}-\eqref{eq:pde2}
is higher dimensional than the geometric Brownian
motion case in Section \ref{sec:geometric-Brownian-motion}. Therefore,
it can be expected that the number of epochs and sample points for NN training are larger.
We follow the same training process as in the previous
section with natural adjustments.

Similar to the geometric Brownian motion case, we take the spot log-price $x_{0}=0$. Further, we choose $Q$ so that $x,y\in[-3.5,3.5]$, $v,z\in[0,1.0]$,
$\kappa\in[0.8,1.2]$, $\omega\in[0.1,0.5]$, $\xi\in[0,0.5]$, $\rho\in[-0.5,0.5]$,
$t\in[0,1.2]$. We compare the approximate joint TPDF with the semi-closed
form of the TPDF from \citet{lewis2016option}.

\begin{figure}
\begin{centering}
\subfloat[True and neural joint TPDFs.]{\begin{centering}
\includegraphics[scale=0.33]{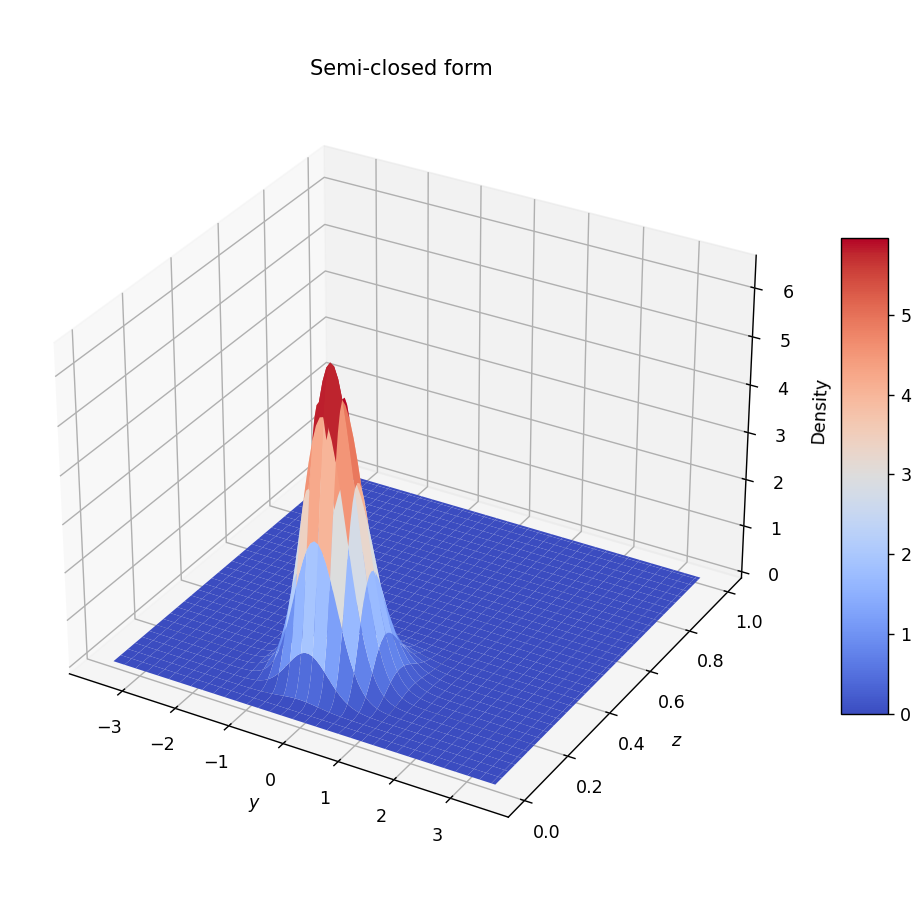}\includegraphics[scale=0.33]{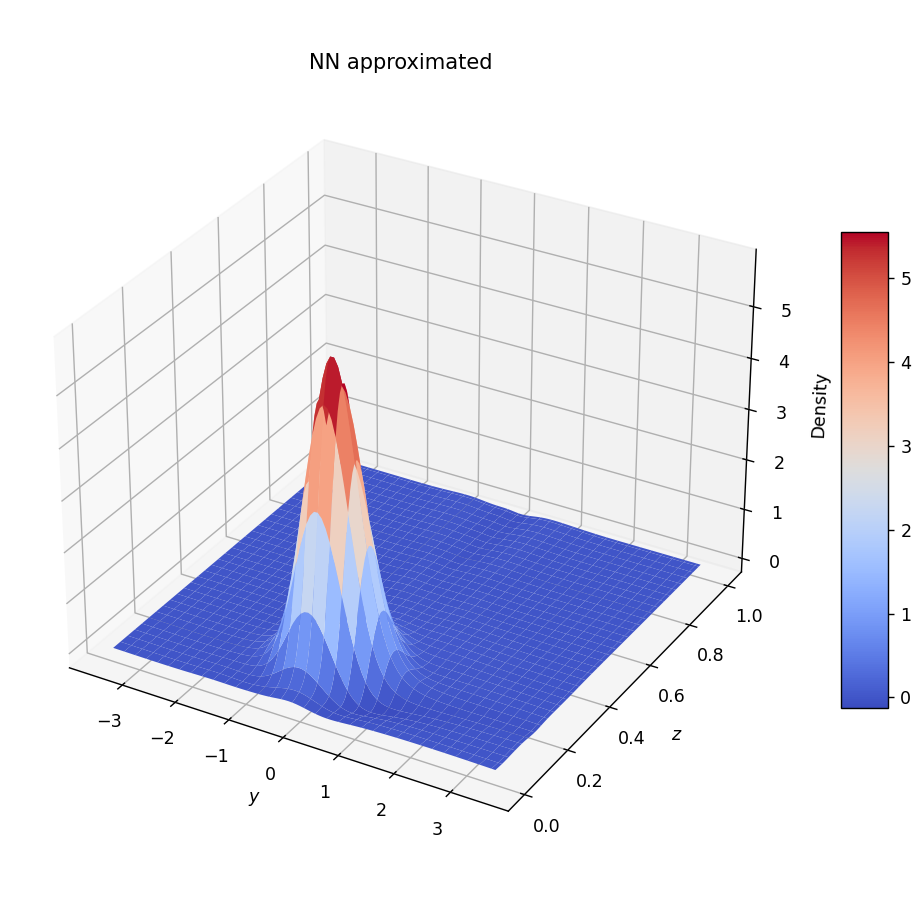}
\par\end{centering}
}
\par\end{centering}
\begin{centering}
\subfloat[The corresponding marginal densities of $X(T)$ calculated from the joint TPDFs and the same graph in log scale (right).]{\begin{centering}
\includegraphics[scale=0.35]{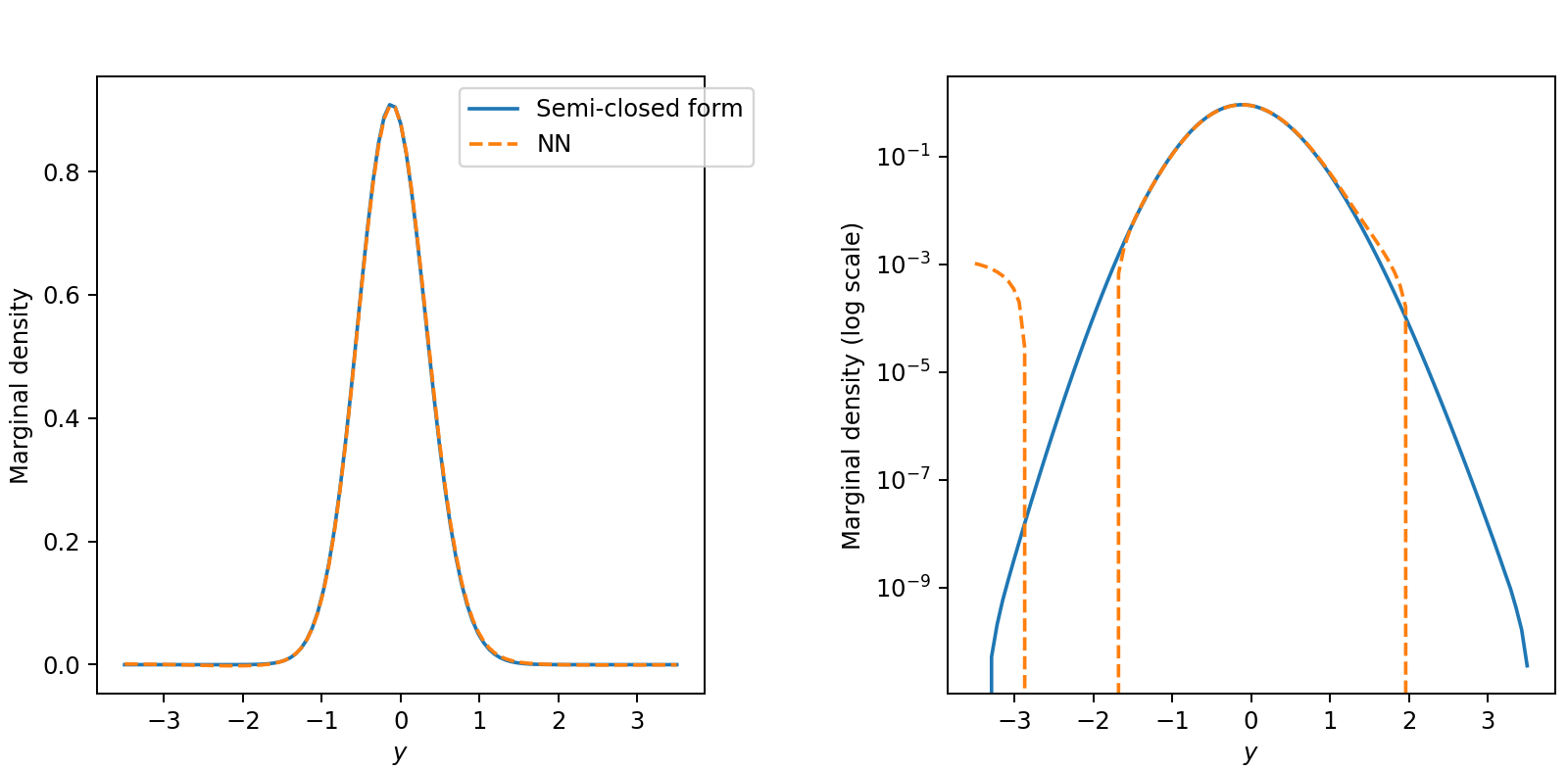}
\par\end{centering}

}
\par\end{centering}
\caption{Joint TPDF graphs according to the semi-closed form and to the NN approximation and the corresponding marginal densities.
The marginal densities are calculated according to Eq. \eqref{eq: Marginal density}.
The NN  is trained with $500000$ epochs. The parameters are $x=0$,
$v=0.2$, $\kappa=1.0$, $\omega=0.2$, $\xi=0.2$, $\rho=0.2$. Time
to maturity $T=1.0$. For the NN approximation,
$\lambda_{L}=100$. The joint TPDF is plotted using 200 points across
both $y$ and $z$ axes.}
\label{Fig: Heston Joint density vs marginal }
\end{figure}

\begin{figure}
\begin{centering}
\includegraphics[scale=0.25]{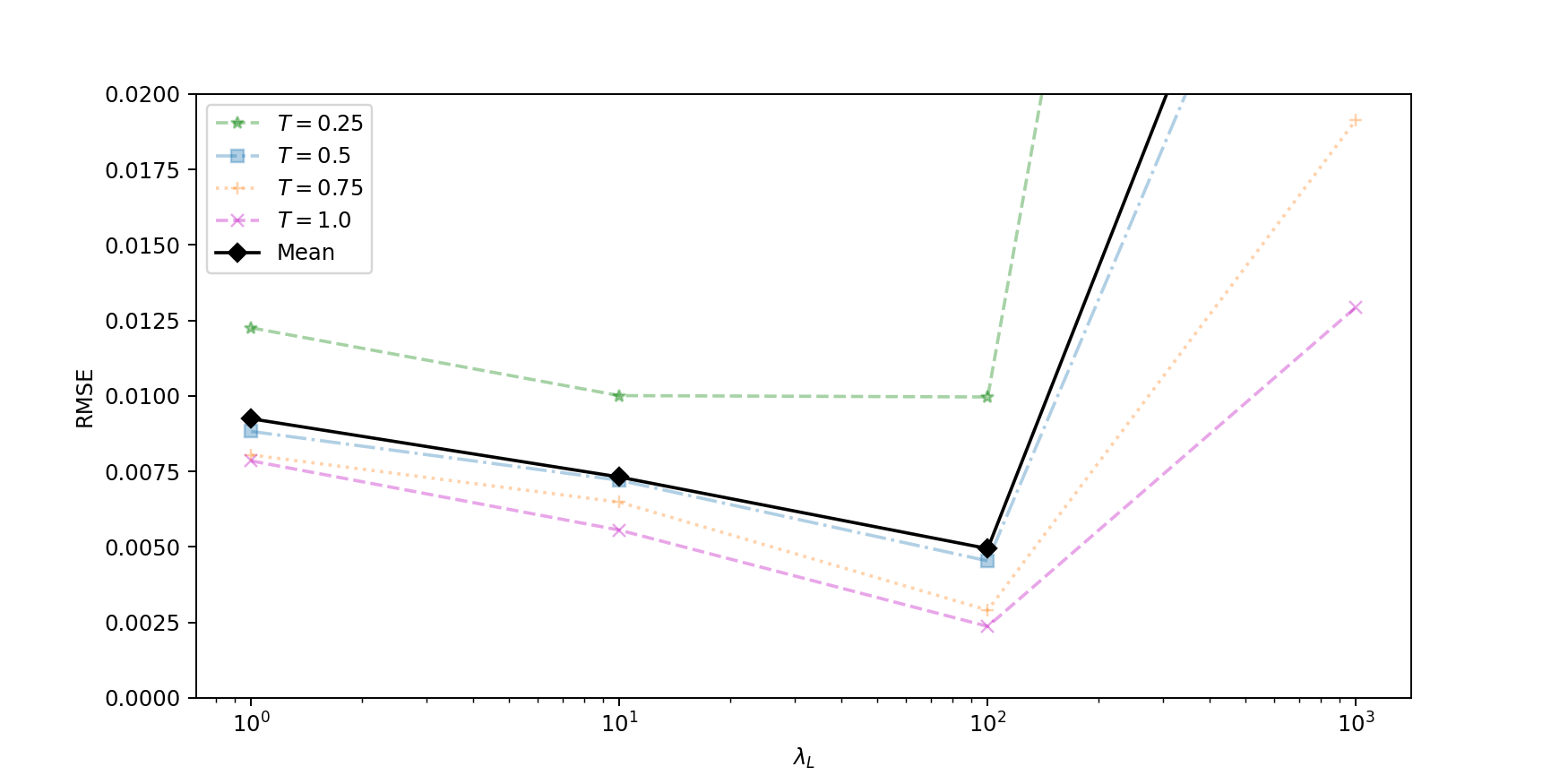}
\par\end{centering}
\caption{RMSE plot against $\lambda_{L}$ for
the Heston model with $\lambda_{L}=1,10,100,1000$. The set
up is the same as in Table~\ref{Tab: Tuning lambda_L for Heston}.
The mean result is calculated using the mean values of all four maturities
across $\lambda_{L}$.}
\label{Fig: Heston Training losses}
\end{figure}

We first tune the hyperparameter $\lambda_{L}$.  Although we ultimately calculate the joint TPDF $p(t,x,v;T,y,z)$ for the Heston model, it is convenient to perform tuning based on the accuracy of the NN approximation of the marginal density $p_{X}(t,x;T,y)$. The marginal density  is calculated by integrating the joint density $p(t,x,v;T,y,z)$ from 0 to $\infty$ with respect to $z$, i.e., \begin{equation} p_{X}(t,x;T,y)=\int_{0}^{\infty}p(t,x,v;T,y,z)dz.\label{eq: Marginal density}
\end{equation}
The integration is done using the trapezium rule. The NN training is performed by running 500000 epochs. Figure~\ref{Fig: Heston Joint density vs marginal } displays the joint TPDF and the corresponding marginal density.
We find that the approximate TPDF is close to the true TPDF. It is difficult to make a direct visual comparison using the two-dimensional joint TPDF plots; the marginal density, on the other hand,
gives a better visualisation and shows that the neural TPDF provides
a very good approximation of the true TPDF obtained from the semi-closed formula as shown in Figure~\ref{Fig: Heston Joint density vs marginal } - the two marginal density plots are indistinguishable. In addition, the marginal density itself is more closely related to European option pricing and extracting the volatility smile. Thus, here we tune $\lambda_{L}$ according to the marginal density. The tuning results are summarised in Table \ref{Tab: Tuning lambda_L for Heston}. We find that $\lambda_{L}=1$ to $100$ produce similar results although $\lambda_{L}=100$ performs slightly better while $\lambda_{L}=1000$ performs worse than the rest. Moreover, $\lambda_{L}=100$ has an advantage compared with $\lambda_{L}=1$ or 10. Indeed, from Figure~\ref{Fig: Heston Training losses}, one can see that when $\lambda_{L}=100$ the $L_{2}$ loss  still shows a stronger downward sloping trend and is less volatile whereas for $\lambda_{L}=1$ and 10 the $L_{2}$ losses  are already flattening and more volatile. This means that further training for $\lambda_{L}=100$ can improve the $L_{2}$ loss while its $L_{1}$ loss is already smaller than for the other values of $\lambda$. Thus, we again observe that the introduction of  $\lambda_{L}>1$ helps to mitigate the vanishing gradient problem commonly encountered in machine learning training. This tuning result agrees with the one obtained for the GBM model in Section~\ref{sec:tuning}. Note that we tested 500000 epochs here, and the performance of $\lambda_{L}=100$ is already better than the others. It can be assumed that with more training, the difference in accuracy will become larger, as has been shown in the GBM model case.  Thus, we recommend the setting $\lambda_L = 100$ for training all NNs.

\begin{table}
\caption{Tuning $\lambda_{L}$. Minimum loss table for $\lambda_{L}=1,10,100,1000$
and the corresponding RMSE of the NN marginal density compared to the semi-closed form marginal density across various
maturity times $T=0.25$, $0.5$, $0.75$, $1.0$. We denote $L_{\min}$ the minimum total loss
value per $500000$ epochs. The domain $Q$ used in the NN training  is $x,y\in[-3.5,3.5]$, $v,z\in[0,1.0]$, $\kappa\in[0.8,1.2]$,
$\omega\in[0.1,0.5]$, $\xi\in[0,0.5]$, $\rho\in[-0.5,0.5]$, $t\in[0,1.2]$.
We use $51$ points across the price direction $y$ and across the variance direction $z$. In terms of validation, for each $t$, we choose five
initial variance values $v_{0}=0.1$, $0.2$, $0.3$, $0.4$, $0.5$. The initial log
price $x_{0}=0$, and we fix $\kappa=1,$ $\omega=0.2$, $\xi=0.2$, $\rho=0.2$.}
\scriptsize
\centering{}{}%
\begin{tabular}{ccccccccc}
\toprule
{$\lambda_{L}$} & {$L_{\min}$} & {$L_{1}$} & {$L_{2}$} & {$T=0.25$} & {$T=0.5$} & {$T=0.75$} & {$T=1.0$} & {Mean}\tabularnewline
\midrule
{1} & {$1.22\times10^{-3}$} & {$2.14\times10^{-4}$} & {$1.23\times10^{-3}$} & {0.012} & {0.008} & {0.008} & {0.007} & {0.009}\tabularnewline

{10} & {$1.94\times10^{-3}$} & {$5.21\times10^{-5}$} & {$1.49\times10^{-3}$} & {0.010} & {0.007} & {0.006} & {0.005} & {0.007}\tabularnewline

{100} & {$3.13\times10^{-3}$} & {$7.10\times10^{-6}$} & {$2.37\times10^{-3}$} & {0.009} & {0.004} & {0.002} & {0.002} & {0.004}\tabularnewline

{1000} & {$1.03\times10^{-2}$} & {$2.25\times10^{-6}$} & {$8.47\times10^{-3}$} & {0.078} & {0.033} & {0.019} & {0.012} & {0.035}\tabularnewline
\bottomrule
\end{tabular}{\label{Tab: Tuning lambda_L for Heston}}{\footnotesize\par}
\end{table}

Following the discussion above, we choose $\lambda_{L}=100$ and, in order to gain higher accuracy,
we train 2 million epochs and 5000 points every five epochs, which is equivalent to using 2 billion sample points to train this NN. Using a single NVIDIA Tesla P100 GPU, each epoch takes about 0.038 seconds, which is slightly more than for the one dimensional Black-Scholes-Merton example in the previous section. Completing the whole 2 million epochs of training takes about 21 hours. This time can be greatly reduced by using more powerful GPUs. Far more important, we emphasise that this is a training time which is not part of the computation times for later use of the neural TPDF, e.g. for option pricing calculations - just as the time to write computer code for options calculations is not part of the later option pricing calculation times. NN training is done separately (``offline''). Once the NN is trained, it is used as a highly convenient engine (``online'') to obtain the approximate TPDF quickly. Indeed, next, we illustrate by showing calculations of option prices using the neural TPDF that are faster than when the semi-closed form of the TPDF is used.

We compare the pricing performance incorporated into QUAD. As can be seen in Table~\ref{Tab: Heston put prices and volatility smile}, option pricing using the neural joint TPDF produces sufficiently accurate results when compared with the semi-closed formulas. At this point, we highlight a considerable advantage of the NN over the semi-closed form of TPDF. Based on QUAD calculation with 51 mesh points per direction, QUAD with the neural TPDF is 7 times faster than with the semi-closed form TPDF. If we increase the number of mesh points per direction to 101, that is 10201 points in the QUAD mesh, the speed advantage is about 11 times. The speed difference is larger with more points because the computing time in accessing the NN  is independent of the number of points used (see Appendix~\ref{appdx: calculation_time}). This is a good property brought by the NN method. In practice, QUAD with 51 mesh points per direction can produce very accurate results. The neural TPDF is very fast to evaluate and it is a perfect match with QUAD.

\begin{figure}
\begin{centering}
\includegraphics[scale=0.25]{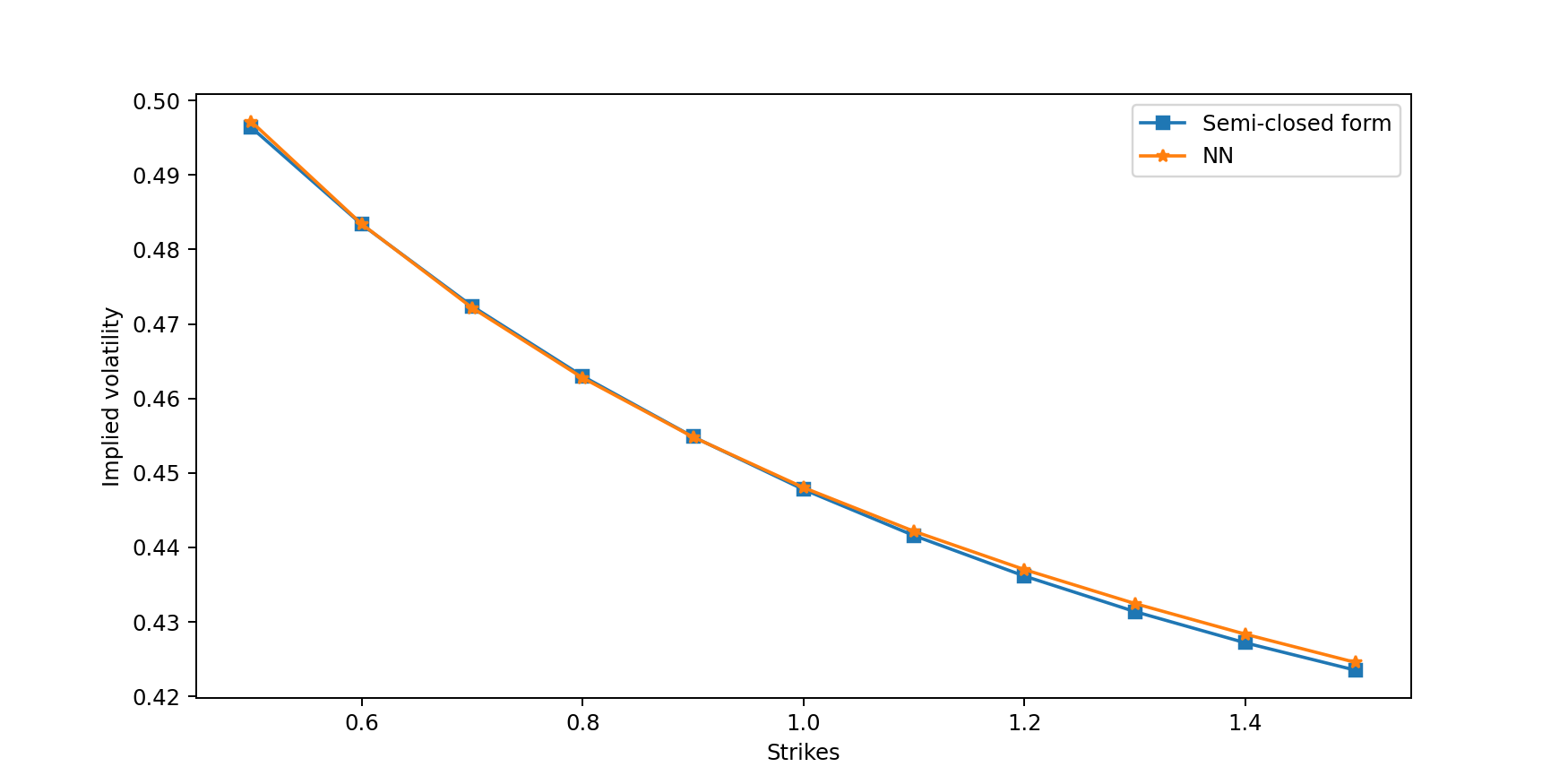}
\par\end{centering}
\caption{Heston's volatility smile obtained through put prices under the semi-closed
form TPDF and neural TPDF. Parameters used: $S_{0}=1,V_{0}=0.16,T=1.0,\kappa=1.0,\omega=0.3,\xi=0.35,\rho=-0.5$.
We calculate option prices with 11 different strikes $K=0.5,0.6,\dots,1.5$.}

\label{Fig: Heston smile}
\end{figure}

In this illustration, we calculate put prices and then obtain the volatility smile by inverting the  Black-Scholes-Merton formula. We use an 11-point calculation and interpolate the smile. Figure~\ref{Fig: Heston smile} shows the implied volatility smiles using the semi-closed form TPDF and the neural TPDF. The results are also presented in Table~\ref{Tab: Heston put prices and volatility smile}. The neural TPDF can reproduce a smile close to the semi-closed form pricing formula for the Heston model. The smile is accurate for both deep in-the-money options and out-of-the-money options.

\begin{table}
\caption{The put option prices and option implied volatility (IV) calculated
from semi-closed form TPDF (CF) and neural TPDF
(NN). Parameters used: $S_{0}=1$, $V_{0}=0.16$, $T=1.0$, $\kappa=1.0$, $\omega=0.3$, $\xi=0.35$, $\rho=-0.5$.
We calculate option prices with 11 different strikes $K=0.5,0.6,\dots,1.5$. The closed form prices are calculated
using the closed form European option pricing formula from \citet{heston1993closed}.
We also include the absolute errors (abs) and percentage error
(\%) of the calculations.}
\scriptsize
\centering{}{}%
\begin{tabular}{>{\centering}p{1.25cm}>{\centering}p{1cm}>{\centering}p{1cm}>{\centering}p{1.5cm}>{\centering}p{1cm}>{\centering}p{1cm}>{\centering}p{1cm}>{\centering}p{1.5cm}>{\centering}p{1cm}}
\toprule
{Strikes} & \multicolumn{2}{c}{{Price}} & \multicolumn{2}{c}{{Error}} & \multicolumn{2}{c}{{Implied Volatility}} & \multicolumn{2}{c}{{Error}}\tabularnewline
\cmidrule{2-9} \cmidrule{3-9} \cmidrule{4-9} \cmidrule{5-9} \cmidrule{6-9} \cmidrule{7-9} \cmidrule{8-9} \cmidrule{9-9}
 & {CF} & {NN} & {Abs $(\times10^{-4})$} & {\%} & {CF } & {NN} & {Abs $(\times10^{-4})$} & {\%}\tabularnewline
\midrule
{0.5} & {0.0127} & {0.0128} & {$3.40$} & {0.64\%} & {0.4964} & {0.4971} & {$7.86$} & {0.16\%}\tabularnewline

{0.6} & {0.0274} & {0.0274} & {$0.10$} & {0.04\%} & {0.4834} & {0.4834} & {$0.57$} & {0.01\%}\tabularnewline

{0.7} & {0.0506} & {0.0505} & {$0.54$} & {0.11\%} & {0.4724} & {0.4722} & {$2.20$} & {0.05\%}\tabularnewline

{0.8} & {0.0831} & {0.0830} & {$0.69$} & {0.08\%} & {0.4630} & {0.4628} & {$2.24$} & {0.05\%}\tabularnewline

{0.9} & {0.1254} & {0.1254} & {$0.13$} & {0.01\%} & {0.4549} & {0.4548} & {$0.38$} & {0.01\%}\tabularnewline

{1.0} & {0.1772} & {0.1772} & {$1.00$} & {0.06\%} & {0.4478} & {0.4480} & {$2.56$} & {0.06\%}\tabularnewline

{1.1} & {0.2377} & {0.2379} & {$2.31$} & {0.10\%} & {0.4416} & {0.4422} & {$5.80$} & {0.13\%}\tabularnewline

{1.2} & {0.3059} & {0.3063} & {$3.40$} & {0.11\%} & {0.4361} & {0.4370} & {$8.70$} & {0.20\%}\tabularnewline

{1.3} & {0.3808} & {0.3812} & {$3.99$} & {0.10\%} & {0.4314} & {0.4325} & {$10.8$} & {0.25\%}\tabularnewline

{1.4} & {0.4613} & {0.4617} & {$3.92$} & {0.09\%} & {0.4272} & {0.4283} & {$11.6$} & {0.27\%}\tabularnewline

{1.5} & {0.5462} & {0.5465} & {$3.16$} & {0.06\%} & {0.4235} & {0.4245} & {$10.5$} & {0.25\%}\tabularnewline
\bottomrule
\end{tabular}\label{Tab: Heston put prices and volatility smile}
\end{table}

Similar to Section~\ref{subsec: BSM_Pricing}, to further investigate the accuracy in the parametric space, we compare the pricing errors of the NN approximated density against the semi-closed form solutions over the random parameter sets across different moneyness positions and maturities.

\begin{table}[h]
\caption{The pricing errors of the NN estimated density of the Heston model, benchmarked
against Heston's semi-closed form pricing solutions. We randomly generate
100 sets of parameters in the parametric space: $V_{0}\in[0.05,0.4],\kappa\in[0.8,1.2],\omega\in[0.1,0.3],\xi\in[0.05,0.4],\rho\in[-0.5,0.5]$.
We compare five ranges of moneyness: Deep in-the-money (DITM) (1.20-1.40);
In-the-money (ITM) (1.05-1.20); At-the-money (ATM) (0.95-1.05); Out-of-the-money
(OTM) (0.80-0.95); Deep out-of-the-money (DOTM) (0.60-0.80). We choose
four maturities $T=0.25,0.5,0.75,1.0$. Pricing performance
is in the form of price percentage error (PrPCTE), price root-mean-square
error (PrPMSE), implied volatility percentage error (IVPCTE), implied
volatility root-mean-square error (IVRMSE).}
\label{Tab: Heston_pricing_error}
\scriptsize
\begin{centering}
\begin{tabular}{ccccccc}
\toprule
Maturity & Error Type &    DITM &       ITM &       ATM &       OTM &      DOTM \\
\midrule
$t=0.25$ & PrPCTE &  0.586593 &  0.042004 &  0.020751 &  0.016033 &  0.013519 \\
      & PrRMSE &  0.000860 &  0.001432 &  0.002250 &  0.003219 &  0.006042 \\
      & IVPCTE &  0.044854 &  0.017662 &  0.021136 &  0.034856 &  0.149867 \\
      & IVRMSE &  0.011698 &  0.004474 &  0.005225 &  0.008336 &  0.040449 \\

\midrule
$t=0.5$ & PrPCTE &  0.052484 &  0.021202 &  0.015676 &  0.013270 &  0.011460 \\
      & PrRMSE &  0.000965 &  0.001522 &  0.002280 &  0.003137 &  0.004891 \\
      & IVPCTE &  0.014657 &  0.012225 &  0.015972 &  0.022808 &  0.055374 \\
      & IVRMSE &  0.005128 &  0.004220 &  0.005310 &  0.007488 &  0.023155 \\

\midrule
$t=0.75$ & PrPCTE &  0.032845 &  0.019015 &  0.015003 &  0.012673 &  0.010561 \\
      & PrRMSE &  0.001143 &  0.001790 &  0.002541 &  0.003409 &  0.004622 \\
      & IVPCTE &  0.011712 &  0.012189 &  0.015216 &  0.019762 &  0.030584 \\
      & IVRMSE &  0.004717 &  0.004844 &  0.005946 &  0.007948 &  0.012789 \\

\midrule
$t=1.0$ & PrPCTE &  0.024775 &  0.016752 &  0.013907 &  0.012213 &  0.010638 \\
      & PrRMSE &  0.001182 &  0.001875 &  0.002775 &  0.003654 &  0.005394 \\
      & IVPCTE &  0.009928 &  0.011274 &  0.014246 &  0.017971 &  0.027622 \\
      & IVRMSE &  0.004528 &  0.005025 &  0.006504 &  0.008436 &  0.015080 \\
\bottomrule
\end{tabular}
\par\end{centering}
\end{table}

Table \ref{Tab: Heston_pricing_error} shows the results of pricing errors
for the Heston model. The pricing performance with the NN approximated
density is good overall, showing that the NN approximated density is sufficiently accurate to produce reasonable prices and implied volatilities. Similar to the results shown for the Black-Scholes-Merton model, a number of features can be observed from these results. First, the pricing performance is better at larger $t$, because the CDF has smoother turns, making it easier for the NN to approximate. Second, the approximation is only accurate to a certain extent. For example, if the training is accurate up to
$10^{-3}$, then this will translate to a similar level of accuracy in the density approximation. Therefore, when the approximated density is subsequently used to price options, the accuracy may be lower
due to the loss of accuracy incurred in either the density calculation step of the CDF or the QUAD method. Thus, it is seen that when pricing a deep out-of-the-money option with a small maturity, although the RMSE of the price is small, the percentage errors of the prices can still be quite large because the prices of deep out-of-the-money put options are usually very small, making it difficult to obtain prices with low relative (percentage) errors. This is a problem with using approximate density to price options rather than a problem specific to the NN approach.

\subsection{SABR model}\label{sec:SABR}

The Black-Scholes-Merton model and the standard Heston model are affine. For these, closed forms (or semi-closed forms) of European option prices and the corresponding TPDFs are known. Non-affine models, on the other hand, have neither closed form European prices nor TPDFs. The popular SABR model
\citep{hagan2002managing} is non-affine when the parameter $\beta<1$   and it does not have
a closed form or semi-closed form TPDF. \citet{hagan_arbitragefree_2014}
subsequently derived an approximate effective forward equation for SABR's marginal density over price using an asymptotic expansion and the marginal density can then be calculated numerically using finite difference techniques. The joint TPDF for the SABR model, however, is still difficult to evaluate, even with numerical techniques. In this section, we show that our method can tackle the SABR model and can approximate the joint TPDF successfully.

The SABR model for forward price $F$ and volatility $\sigma$ is written as \citep{hagan2002managing}:
\begin{align}dF & =\sigma F^{\beta}dW_{1}, \,\, F(t)=x,\label{eq: SABR forward price}\\
d\sigma & =\nu\sigma dW_{2},  \,\, \sigma(t)=\sigma_0, \label{eq:SABR volatility}\\
dW_{1}dW_{2} & =\rho dt. \notag
\end{align}

Here we consider the more challenging case of this model when $0<\beta<1$.
It is known (\cite{rebonato2009sabr,lewis2016option}) that, for $0<\beta<1$, $F=0$ is an attainable barrier for the process $F(t)$. Since for $1/2 \leq \beta<1$ the solution $(F(t)$, $\sigma(t))$ of (\ref{eq: SABR forward price})-(\ref{eq:SABR volatility}) is unique, $F=0$ is an absorbing boundary.   For $0<\beta<1/2$, the absorbing boundary condition is prescribed at $F=0$ to satisfy the no arbitrage condition.

For training purposes, it is easier to train if we transform the volatility from $\sigma$. Let $\omega=\ln\sigma$, then Eq. \eqref{eq:SABR volatility}
becomes
\begin{equation}\label{eq:omega}
d\omega=-\frac{\nu^{2}}{2}dt+\nu dW_{2}, \,\, \omega(t)=\ln\sigma_0=v.
\end{equation}
Consequently, Eq. (\ref{eq: SABR forward price}) takes the form
\begin{equation}\label{eq:Fch}
dF  =\exp(\omega(t)) F^{\beta}dW_{1}, \,\, F(t)=x.
\end{equation}

To be specific, consider the pricing of a European option with payoff $f(x)$ and maturity $T$. Its price at time $t$, with spot price $x$ and spot log-volatility $v$ can be expressed as
\begin{eqnarray}
u(t,x,v) &=&Ef(F_{t,x,v}(T))  \label{eq:SABRnew1} \\
&=&E\left[ f(F_{t,x,v}(T))I\{\tau _{t,x,v}\geq T\}+f(0)I\{\tau _{t,x,v}<T\}%
\right]  \notag \\
&=&E\left[ f(F_{t,x,v}(T))I\{\tau _{t,x,v}\geq T\}\right] +f(0)E\left[
I\{\tau _{t,x,v}<T\}\right] ,  \notag
\end{eqnarray}%
where $\tau _{t,x,v}$ is the Markov time when $F_{t,x,v}(\tau _{t,x,v})$ from (\ref{eq:omega})-(\ref{eq:Fch})
hits the value zero.

To compute the first expectation in (\ref{eq:SABRnew1}), we need to find the
transition probability function $C(t,x,v)=C(t,x,v;T,y,z)$:
\begin{eqnarray*}
C(t,x,v;T,y,z): &=&\text{Prob}(F_{t,x,v}(T)\leq y,\omega _{t,x,v}(T)\leq
y,\tau _{t,x,v}\geq T) \\
&=&\int_{0}^{y}\int_{0}^{z}p(t,x,v;T,y^{\prime },z^{\prime })dy^{\prime
}dz^{\prime },
\end{eqnarray*}%
where $p(t,x,v;T,y^{\prime },z^{\prime })$ is the corresponding density. The
function $C(t,x,v)$ satisfies the backward Kolmogorov equation
\begin{equation} \label{eq: SABR differential operator}
-\frac{\partial C}{\partial t}=\frac{1}{2}\exp(2v)x^{2\beta }\frac{\partial
^{2}C}{\partial x^{2}}
-\frac{\nu^{2}}{2}\frac{\partial C}{\partial v}
+\frac{\nu^2}{2} \frac{\partial ^{2}C}{\partial
v^{2}}+\rho \nu \exp(v) x^{\beta }\frac{\partial ^{2}C}{\partial x\partial v}%
,\,\,x>0,v>0,
\end{equation}%
with the terminal condition
\begin{equation}
C(T,x,v)=%
\begin{cases}
1, & x\leq y\text{ and }v\leq z, \\
0, & \text{otherwise,}%
\end{cases}
\label{eq: SABR terminal condition}
\end{equation}%
and the Dirichlet boundary condition
\begin{equation}
C(t,0,v)=0.  \label{eq: SABR boundary condition}
\end{equation}

To compute the second expectation in (\ref{eq:SABRnew1}), we need to find
the probability
\begin{equation*}
\mathcal{P}(t,x,v)=E\left[ I\{\tau _{t,x,v}<T\}\right] =P(\{\tau
_{t,x,v}<T\}),
\end{equation*}%
which satisfies the same backward Kolmogorov equation
\begin{equation} \label{eq:P1}
-\frac{\partial \mathcal{P}}{\partial t}=\frac{1}{2} \exp(2v) x^{2\beta }\frac{%
\partial ^{2}\mathcal{P}}{\partial x^{2}}
-\frac{\nu^{2}}{2}\frac{\partial \mathcal{P}}{\partial v}
+\frac{1}{2}\nu \frac{\partial
^{2}\mathcal{P}}{\partial v^{2}}+\rho \nu \exp(v) x^{\beta }\frac{\partial ^{2}%
\mathcal{P}}{\partial x\partial v},\,\,x>0,v>0,
\end{equation}%
with the terminal condition
\begin{equation} \label{eq:P2}
\mathcal{P}(T,x,v)=0
\end{equation}%
and the Dirichlet boundary condition
\begin{equation} \label{eq:P3}
\mathcal{P}(t,0,v)=1.
\end{equation}

The solutions to each of the Dirichlet problems, above, need to be approximated by their own appropriately trained NN. The problem (\ref{eq:P1})-(\ref{eq:P3}) is easier to solve as it has fewer parameters. In order not to make the presentation overly intricate, we will price call options for which $f(0)=0$, so that only the problem (\ref{eq: SABR differential operator})-(\ref{eq: SABR boundary condition}) needs to be solved.

In contrast with the models of underlyings considered earlier, here we have to prescribe the boundary condition for the function $C(t,x,v)$. Consequently, we need to add an extra term corresponding to the boundary condition in the loss function $L(f)$ (here $f$ is a NN as usual):
\begin{align}
L_{1}(f) & =\left\Vert \frac{\partial}{\partial t}f(t,\boldsymbol{x};\boldsymbol{\mathbf{\theta}})+\mathcal{L}f(t,\boldsymbol{x};\boldsymbol{\mathbf{\theta}})\right\Vert _{[0,T]\times G,\nu_{1}}^{2},\\
L_{2}(f) & =\left\Vert f(T,\boldsymbol{x};\boldsymbol{\mathbf{\theta}})-C(T,x,v)\right\Vert _{G,\nu_{2}}^{2},\\
L_{3}(f) & =\left\Vert f(t,\boldsymbol{x};\boldsymbol{\mathbf{\theta}})-C(t,0,v;T,y,z)\right\Vert _{[0,T]\times\partial G,\nu_{3}}^{2},\\
L(f) & =\lambda_{L}L_{1}(f)+L_{2}(f)+L_{3}(f),\label{eq: SABR loss function}
\end{align}
where $L_{1}$ loss is for the differential operator term Eq. \eqref{eq: SABR differential operator}
and $L_{2}$ loss is for the terminal condition term Eq. \eqref{eq: SABR terminal condition}, and $L_{3}$ loss is for the Dirichlet boundary condition Eq. \eqref{eq: SABR boundary condition} with $\nu_{3}(y)$ being the probability density for the sampling distribution on the time-price
domain $\mathcal{Y}=[0,T]\times\partial G.$ As before, $\lambda_{L}$ is the hyperparameter
for the loss function. We assume that both the terminal condition and the Dirichlet boundary condition are of equal importance and, thus, have the same weight in the loss function.

For the SABR model, the NN $f$ is trained in the  $(t,x,v,y,z,\beta,\rho)$-domain
$Q\subset\mathbb{R}^{7}$. We simplify the training by removing the parameter $\nu$ since we can choose $\nu=1$ in (\ref{eq: SABR differential operator}) and recover the solution for arbitrary $\nu>0$ through the change of variables:
\[
C(t,x,v,y,z,\beta,\rho,\nu)=C\bigg(t\nu^{2},x,v-\ln(\nu),y,z-\ln(\nu),\beta,\rho,1\bigg).
\]
We choose the domain $Q$ so that $t\in[0,1.2]$, $x,y\in$$[0.0,5.0]$, $v\in[-5.0,0.0]$,
$z\in[-7.0,3.0]$, $\beta\in[0.5,1.0]$, $\text{\ensuremath{\rho=[-1.0,0.0]}}$. The range for the terminal volatility value, $z$, is higher in the SABR case since the marginal density of the volatility process has a
fat tail, observed in the lognormal distribution.  As with the Heston model, to attain the required accuracy, we choose $\lambda_{1}=100$ and $\lambda_{2}=\lambda_{3}=1$, and we train 2.5 million epochs and generate 5000 independent points every five epochs, which is equivalent to using 2.5 billion sample points to train this NN. We select the NN $f$ according to the epoch  with the lowest loss. Using a single NVIDIA Tesla P100 GPU, each epoch takes about 0.043 seconds, slightly longer than for the Heston model due to the increased complexity. It takes about 30 hours to complete 2.5 million epochs of training.

Since there are no closed form or semi-closed form solutions for the SABR model, as the benchmark, we use option prices calculated numerically by the simplest random walk algorithm for solving Dirichlet problems from  \citep{MT03} (see also \citep{MTbook21}) combined with the  Monte Carlo technique;  we take the very small time step $0.0005$ and  $10^{8}$ Monte Carlo runs. We also compare NN results for the TPDF and option prices with the ones obtained by the finite difference method (implemented in QuantLib\footnote{\url{https://www.quantlib.org/}, see FdSabrVanillaEngine class.} with 200 points across the time dimension, 800 points across the price dimension and 200 points across the volatility dimension), by the approximate marginal density for price calculated using the finite difference method  (\citealp{hagan_arbitragefree_2014}), and by the approximate implied volatility formula from \citeauthor{hagan2002managing} \citeyearpar{hagan2002managing,hagan_arbitragefree_2014}.

\begin{figure}
\begin{centering}
\includegraphics[scale=0.67]{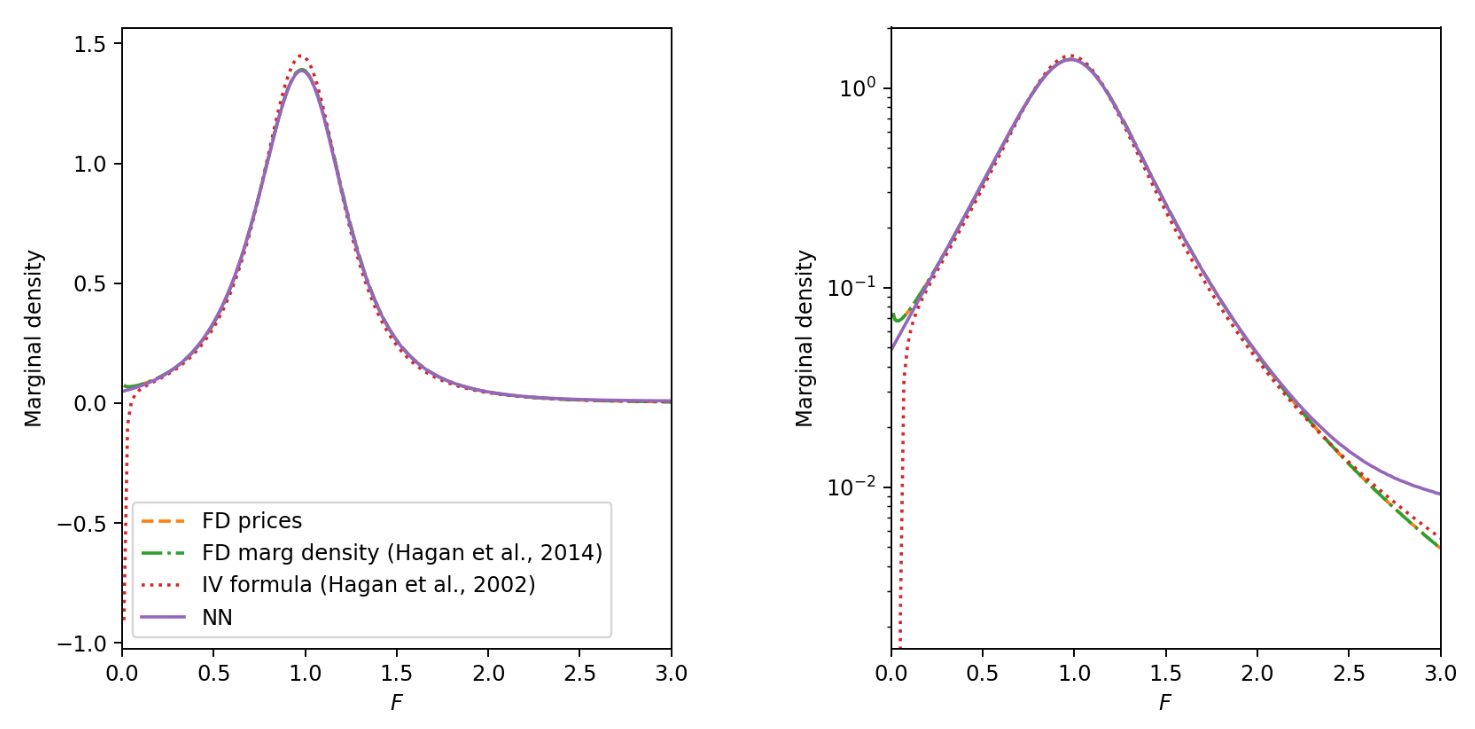}
\par\end{centering}
\caption{Comparison of marginal densities for price and for volatility. The
parameters used for the SABR model are the initial price $F_{0}=1$,
the initial volatility $\sigma_{0}=0.35,$ $\beta=0.6$, $\rho=-0.1$,
$t=1.0$. We use 201 points interpolation. The NN marginal densities are calculated by numerical integration from the joint neural TPDF.}
\label{Fig: SABR marginal density}
\end{figure}

We see in Figure~\ref{Fig: SABR marginal density} that the marginal density for the stock price produced by the NN is very close to the results obtained using the finite difference methods. \citet{hagan_arbitragefree_2014} report that the implied volatility formula can produce negative
densities as price $F\rightarrow 0$, which violates the no arbitrage condition, and the same result is also observed here. The NN approximation has no such deficiency.

\begin{figure}
\begin{centering}
\includegraphics[scale=0.7]{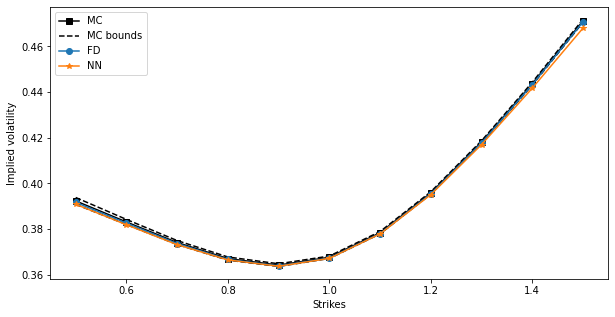}
\par\end{centering}
\caption{SABR volatility smile obtained through put prices using the QUAD method
with TPDF computed using finite difference methods and with the neural
TPDF; Monte Carlo simulations; and approximate implied volatility formula. The parameters used: $F_{0}=1,\sigma_{0}=0.35,T=1.0,\beta=0.6,\rho=-0.1$.
We calculate option prices for 11 strikes  $K=0.5,0.6,\dots,1.5$.}
\label{Fig: SABR volatility smiles}
\end{figure}

Figure \ref{Fig: SABR volatility smiles} shows that the approximate SABR implied volatility formula, although very convenient to use, does not produce accurate results, compared with results using the other methods. The implied volatility smile obtained through the neural TPDF is accurate and it is very close to the ones generated by the Monte Carlo method and the finite difference method -- see Table~\ref{Tab: SABR call option prices and implied vol} and Figure~\ref{Fig: SABR volatility smiles}.

\begin{table}
\begin{centering}
\caption{The call option prices and option implied volatility calculated from the three numerical approaches: Monte Carlo method (MC), finite difference method (FD) and QUAD with the neural TPDF (NN). The benchmark values are the call prices calculated using the Monte Carlo method as explained in the text. The Monte Carlo (statistical) error for the benchmark is also given. The finite difference method uses 200, 800, 200 mesh points in the time direction, price direction, volatility direction,
respectively. The QUAD method uses 251 points in both price and volatility directions. The parameters used: $F_{0}=1,\sigma_{0}=0.35,T=1.0,\beta=0.6,\rho=-0.1$. We calculate option prices for 11 strikes $K=0.5,0.6,\dots,1.5$. We  include the absolute errors of the calculations. 
}
\label{Tab: SABR call option prices and implied vol}
\par\end{centering}
\scriptsize
\centering{}%
\begin{tabular}{c>{\centering}p{1.75cm}>{\centering}p{2.25cm}>{\centering}p{2.25cm}c>{\centering}p{2.25cm}>{\centering}p{2.25cm}}
\toprule
{Strikes} & {MC price} & {FD price} & {NN price} & {MC IV} & {FD IV} & {NN IV}\tabularnewline
 & {MC est. err $(\times10^{-4})$} & {Abs err $(\times10^{-4})$ / \% err} & {Abs err $(\times10^{-4})$ / \% err} &  & {Abs err $(\times10^{-3})$ / \% err} & {Abs err $(\times10^{-3})$ / \% err}\tabularnewline
\midrule
{0.5} & {0.5189} & {0.5187} & {0.5186} & {0.392} & {0.392} & {0.391}\tabularnewline
 & {$2.66$} & {$1.14$ / 0.02\%} & {$2.58$ / 0.05\%} &  & {$0.64$ / 1.64\%} & {$1.46$ / 0.37\%}\tabularnewline
\addlinespace
{0.6} & {0.4293} & {0.4292} & {0.4291} & {0.383} & {0.390} & {0.382}\tabularnewline
 & {$2.58$} & {$1.20$ / 0.03\%} & {$2.81$ / 0.07\%} &  & {$0.52$ / 1.36\%} & {$1.22$ / 0.32\%}\tabularnewline
\addlinespace
{0.7} & {0.3448} & {0.3447} & {0.3446} & {0.374} & {0.373} & {0.373}\tabularnewline
 & {$2.49$} & {$1.25$ / 0.04\%} & {$2.67$ / 0.08\%} &  & {$0.43$ / 0.12\%} & {$0.92$ / 0.25\%}\tabularnewline
\addlinespace
{0.8} & {0.2677} & {0.2675} & {0.2675} & {0.367} & {0.366} & {0.366}\tabularnewline
 & {2.37} & {$1.31$ / 0.05\%} & {$2.05$ / 0.08\%} &  & {$0.38$ / 0.10\%} & {$0.60$ / 0.16\%}\tabularnewline
\addlinespace
{0.9} & \centering{}{0.2008} & \centering{}{0.2006} & \centering{}{0.2006} & {0.364} & {0.364} & {0.364}\tabularnewline
 & \centering{}{$2.24$} & \centering{}{$1.27$ / 0.06\%} & \centering{}{$1.91$ / 0.10\%} &  & {$0.33$ / 0.09\%} & {$0.50$ / 0.14\%}\tabularnewline
\addlinespace
{1.0} & {0.1467} & \centering{}{0.1466} & {0.1465} & {0.368} & \centering{}{0.367} & \centering{}{0.367}\tabularnewline
 & {$2.10$} & {$1.32$ / 0.09\%} & {$1.98$ / 0.13\%} &  & \centering{}{$0.33$ / 0.09\%} & \centering{}{$0.50$ / 0.13\%}\tabularnewline
\addlinespace
{1.1} & {0.1062} & {0.1061} & {0.1060} & {0.378} & {0.378} & {0.378}\tabularnewline
 & {$1.95$} & {$1.35$ / 0.13\%} & {$1.95$ / 0.18\%} &  & {$0.35$ / 0.09\%} & {$0.51$ / 0.14\%}\tabularnewline
\addlinespace
{1.2} & {0.0777} & {0.0775} & {0.0774} & {0.396} & {0.395} & {0.395}\tabularnewline
 & {$1.81$} & {$1.31$ / 0.17\%} & {$2.61$ / 0.34\%} &  & {$0.38$ / 0.09\%} & {$0.74$ / 0.19\%}\tabularnewline
\addlinespace
{1.3} & {0.0580} & {0.0579} & {0.0576} & {0.418} & {0.417} & {0.417}\tabularnewline
 & {$1.69$} & {$1.16$ / 0.20\%} & {$3.52$ / 1.17\%} &  & {$0.38$ / 0.09\%} & {$1.14$ / 0.27\%}\tabularnewline
\addlinespace
{1.4} & {0.0444} & {0.0443} & {0.045} & {0.444} & {0.443} & {0.439}\tabularnewline
 & {$1.58$} & {$1.08$ / 0.24\%} & {$5.21$ / 1.17\%} &  & {$0.41$ / 0.09\%} & {$1.97$ / 0.44\%}\tabularnewline
\addlinespace
{1.5} & {0.0349} & {0.0348} & {0.0342} & {0.471} & {0.471} & {0.468}\tabularnewline
 & {$1.49$} & {$1.04$ / 0.30\%} & {$7.25$ / 2.08\%} &  & {$0.46$ / 0.10\%} & {$3.20$ / 0.68\%}\tabularnewline
\bottomrule
\end{tabular}
\end{table}

Table \ref{Tab: SABR_pricing_error} shows the pricing errors for the SABR model. Similar to the Heston model, the SABR performance shows that the NN approximated density can be used to obtain option prices up to a certain level of accuracy. Although the performance in terms of absolute pricing accuracy (PrRMSEs) is consistent across different option moneyness positions and maturities, the relative pricing accuracy (PrPCTE) is lower for OTM options, especially at shorter maturities, because the values of OTM options at shorter maturities are minuscule and prices obtained using the NN approximated density have difficulty in accurately capturing those minuscule prices.

\begin{table}[h]
\caption{The pricing errors of the NN estimated density of the SABR model, benchmarked
against Monte Carlo's pricing solutions (with an average approximation error $2\times10^{-4}$). We randomly generate
100 sets of parameters in the parametric space: $\sigma_0\in[0.1,0.5], \beta\in[0.5,1.0],\rho\in[-0.8,0.0]$.
We compare five ranges of moneyness: Deep out-of-the-money (DOTM) (1.20-1.40);
Out-of-the-money (OTM) (1.05-1.20); At-the-money (ATM) (0.95-1.05); In-the-money
(ITM) (0.80-0.95); Deep in-the-money (DITM) (0.60-0.80). We choose
four maturities $T=0.25,0.5,0.75,1.0$. Pricing performance
is in the form of price percentage error (PrPCTE), price root-mean-square
error (PrPMSE), implied volatility percentage error (IVPCTE), implied
volatility root-mean-square error (IVRMSE). The QUAD method uses 101 points in both price and volatility directions. A few Monte Carlo results fail to calculate implied volatility as is usual for the SABR model, so we exclude those results from the calculations. }
\label{Tab: SABR_pricing_error}
\scriptsize
\begin{centering}
\begin{tabular}{ccccccc}
\toprule
Maturity & Error Type &      DITM &       ITM &       ATM &       OTM &       DOTM \\
\midrule
$t=0.25$ & PrPCTE &  0.015778 &  0.035042 &  0.115320 &  1.796568 &  69.396973 \\
      & PrRMSE &  0.005581 &  0.006020 &  0.006196 &  0.005959 &   0.005344 \\
      & IVPCTE &  0.280015 &  0.152160 &  0.106639 &  0.191038 &   0.390133 \\
      & IVRMSE &  0.097729 &  0.046520 &  0.031690 &  0.048050 &   0.108325 \\
\midrule
$t=0.5$ & PrPCTE &  0.012689 &  0.025158 &  0.063435 &  0.254321 &   4.199479 \\
      & PrRMSE &  0.004747 &  0.004961 &  0.005062 &  0.004892 &   0.004330 \\
      & IVPCTE &  0.134565 &  0.068937 &  0.060878 &  0.081447 &   0.163783 \\
      & IVRMSE &  0.050319 &  0.021951 &  0.018137 &  0.022068 &   0.044683 \\
\midrule
$t=0.75$ & PrPCTE &  0.009436 &  0.017624 &  0.040875 &  0.119415 &   0.577437 \\
      & PrRMSE &  0.003792 &  0.003790 &  0.003839 &  0.003706 &   0.003268 \\
      & IVPCTE &  0.055559 &  0.040761 &  0.039432 &  0.048818 &   0.093607 \\
      & IVRMSE &  0.019857 &  0.012927 &  0.011206 &  0.013033 &   0.027393 \\
\midrule
$t=1.0$ & PrPCTE &  0.010708 &  0.017620 &  0.034019 &  0.076755 &   0.229785 \\
      & PrRMSE &  0.005316 &  0.005027 &  0.004942 &  0.004816 &   0.004577 \\
      & IVPCTE &  0.043792 &  0.033750 &  0.033411 &  0.038157 &   0.059842 \\
      & IVRMSE &  0.019643 &  0.013648 &  0.012439 &  0.013042 &   0.020889 \\
\bottomrule
\end{tabular}
\par\end{centering}
\end{table}

\subsection{Adding jumps}

When it comes to computing TPDFs, jump-diffusion processes pose a unique challenge:
the Kolmogorov backward equations for jump-diffusion processes involve the integral term  due to the presence of jumps in the models and so the PDEs occurring in the case of the diffusion processes considered in the previous sections become partial integral differential equations (PIDEs). While solving PIDEs may be awkward by other numerical techniques such as finite difference methods, the deep learning approach presented in this paper can solve them efficiently. The use of deep learning to solve (especially parametric) PIDE problems has rarely been explored in the literature. Here we illustrate the deep learning approach to finding approximate TPDFs arising from jump-diffusion models by considering the two well-known jump-diffusion models: Kou's double exponential jump-diffusion model \citep{kou2002jump} and the stochastic volatility jump-diffusion model \citep{bates1996jumps}.

\subsubsection{Kou's double exponential jump diffusion.}


Kou's model for the asset price $S$  under a forward measure can be written in the following form \citep{kou2002jump}:
\begin{equation}
dS=S(t-)[-\lambda kdt+\sigma dW+JdN], \label{eq:kou}
\end{equation}
where $W(t)$ is a standard Wiener process, $N(t)$ is a Poisson process with intensity $\lambda >0$, and the
jumps $J_{1},J_{2},\ldots $ are such that
\begin{equation*}
J_{i}+1=:\tilde{J}_{i}=\exp (\xi _{i})
\end{equation*}%
with $\xi _{i}$ being i.i.d. double-exponential random variables with density
\begin{equation}
\rho_{\xi}(x)=p\cdot\eta_{1}e^{-\eta_{1}x}\mathbbm{1}_{\{x\geq0\}}+q\cdot\eta_{2}e^{\eta_{2}x}\mathbbm{1}_{\{x<0\}}.
\label{eq: rho for kou's}
\end{equation}
Here the parameters $p,q\geq0,$ $p+q=1$, $\eta_{1}>1$, $\eta_{2}>0$, and $\mathbbm{1}_{\{A\}}$ is the indicator function of a set $A$. The jumps with random sizes $J_{1},J_{2},\ldots $ occur at jump times $\tau_1, \tau_2, \ldots$ at which the Poisson process $N(t)$ increases by $1$.
The constant
\[
k= E(J)=\frac{p\eta _{1}}{\eta _{1}+1}+%
\frac{q\eta _{2}}{\eta _{2}-1}-1.
\]
It is assumed that the Wiener process $W(t),$ the Poisson process $N(t),$ and the jumps $%
J_{i}$ are independent.

Let $X(t)=\ln S(t)$ and re-write the SDE (\ref{eq:kou}) accordingly (we do not present this SDE here). The CDF $C(t,x)=C(t,x;T,y)$  for the $X(T)$ satisfies the backward Kolmogorov equation \citep{Apple09}:
\begin{equation}
\frac{\partial C}{\partial t}+\bigg(-\lambda k-\frac{\sigma^{2}}{2} \bigg)\frac{\partial C}{\partial x}+\frac{\sigma^{2}}{2}\frac{\text{\ensuremath{\partial^{2}C}}}{\partial x^{2}}+\lambda\int_{\mathbb{R}}C(t,x+z)\rho_{\xi}(z)dz-\lambda C(t,x)=0 \label{eq: Jump CDF}
\end{equation}
with terminal condition
\begin{equation}
C(T,x)=\mathbbm{1}(x\leq y)=\begin{cases}
1, & x\leq y,\\
0, & x>y,
\end{cases}\:\:\:x\in\mathbb{R}.
\end{equation}
Note that (\ref{eq: Jump CDF}) is a PIDE.

We re-write the integral term of  (\ref{eq: Jump CDF}) as
\begin{equation}
\lambda\int_{\mathbb{R}}C(t,x+z)\rho_{\xi}(z)dz=
\lambda \Bigg[p\int_{0}^{\infty}C \bigg(t,x+\frac{z}{\eta_{1}} \bigg)e^{-z}dz+q\int_{0}^{\infty}C \bigg(t,x-\frac{z}{\eta_{2}} \bigg)e^{-z}dz \Bigg],
\label{eq: Kou's Gauss-Laguerre}
\end{equation}
where the integrals on the right-hand side can be approximated using  a Gauss-Laguerre quadrature in the implementation of the NN training. However, the range of abscissas in a Gauss-Laguerre quadrature rule is too wide for NN training purposes. For instance, the seven-point Gauss-Laguerre rule needs to have abscissa up to 19.40 and, as a result, $z/\eta_{1}$
should have a much wider range than is required  for the price range of interest if e.g. $\eta_{1}=2$. In comparison, the seven-point Gauss-Hermite rule has an abscissa range of $\pm2.65$, which is well within the price range in the NN training. Therefore, we further transform the integral term  \eqref{eq: Kou's Gauss-Laguerre}:
\begin{equation}
\lambda\int_{\mathbb{R}}C(t,x+z)\rho_{\xi}(z)dz=
\lambda \Bigg[2p\int_{0}^{\infty}zC\bigg(t,x+\frac{z^{2}}{\eta_{1}}\bigg)e^{-z^{2}}dz+2q\int_{0}^{\infty}zC \bigg(X-\frac{z^{2}}{\eta_{2}} \bigg)e^{-z^{2}}dz \Bigg].\label{eq: Kou's Gauss-Hermite}
\end{equation}
We note that the above integrals cannot be approximated by standard Gaussian-Hermite quadratures since the lower bound is zero rather than infinity. Instead, we use Gaussian quadrature weights and abscissas suggested in \citep{steen1969gaussian} to approximate $\int_{0}^{\infty}f(z)e^{-z^{2}}dz$.
We find that using seven quadrature points suffices in terms of accuracy
and efficiency.\footnote{The abscissas of the 7 quadrature points are 0.0637164846067008, 0.318192018888619,
0.724198989258373, 1.23803559921509, 1.83852822027095, 2.53148815132768,
3.37345643012458; and their corresponding weights are 0.160609965149261,
0.306319808158099, 0.275527141784905, 0.120630193130784, 0.0218922863438067,
0.00123644672831056, 0.000110841575911059.}

The domain $Q$ used for training this model is $x,y\in[-5.0,5.0]$,
$\sigma\in[0,0.5]$, $\lambda\in[0.0,2.0]$, $p\in[0.0,1.0]$, $\eta_{1}\in[1.1,20.0]$, $\eta_{2}\in[0.1,20.0]$, $t\in[0,1.2]$. We choose $\lambda_{L}=100$
and, in order to get higher accuracy, we train the NN using 2 million epochs with 5000
points every five epochs. We choose the NN with the smallest total loss.
The training time per epoch using an NVIDIA P100 GPU is about 0.047 seconds
and it takes about 26 hours to complete the training of 2 million epochs.

\begin{figure}
\begin{centering}
\includegraphics[scale=0.29]{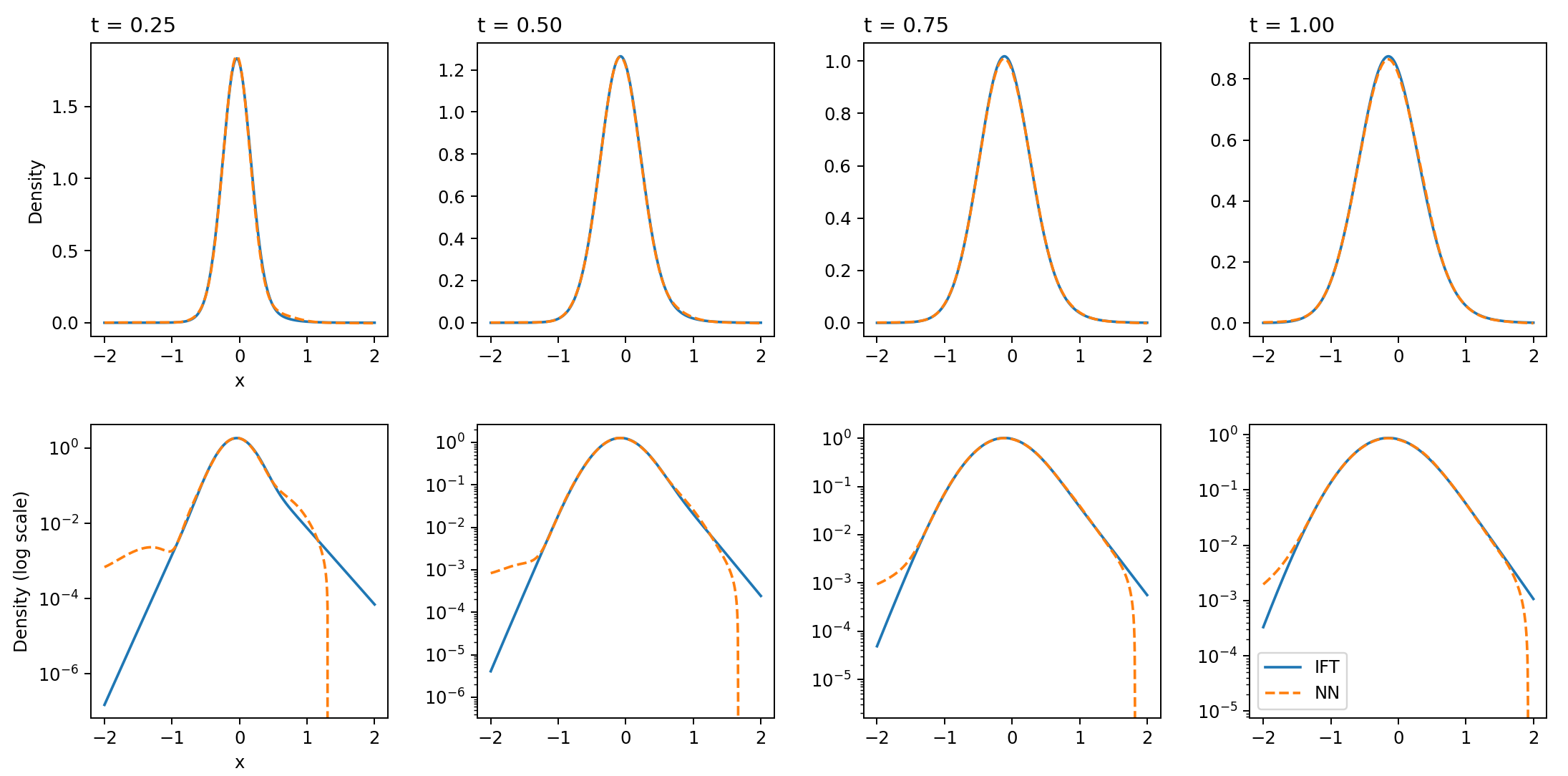}
\par\end{centering}
\caption{The first row shows the true TPDF of Kou's double jump exponential model calculated by inverse Fourier transform of the characteristic function (IFT) vs. the neural TPDF (NN). The second row shows the same graph in log scale. The initial
price is $x=0$ and the other parameters used are similar to the ones from \citep{kou2002jump} (but under a forward measure): $\sigma=0.4$, $\lambda=1.0$, $p=0.4$, $q=0.6$, $\eta_{1}=10$,
$\eta_{2}=5$. The time to maturity $t=0.25$, $0.5$, $0.75$, $1.0$. The hyperparameter $\lambda_{L}=100$.
The domain used to train the NN is $x,y\in[-5.0,5.0]$, $\sigma\in[0,0.5]$,
$t\in[0,1.2]$, $\lambda\in[0.0,2.0]$, $p\in[0.0,1.0]$, $\eta_{1}\in[1.1,20.0]$, $\eta_{2}\in[0.1,20.0]$.
We plot the graphs with 100 points.}
\label{Fig:DEJD TPDF}

\end{figure}

Since Kou's model does not have a closed form TPDF, we use the
inverse Fourier transform of its characteristic function to derive
the true TPDF in the log price space,
before using it to compare with the neural TPDF.
Figure~\ref{Fig:DEJD TPDF} shows the comparisons. As can be
seen from this figure, the neural TPDF provides a close approximation to the true density. Similar to the discussion in Section~\ref{subsec: BSM_Pricing}, the neural TPDF for Kou's double exponential jump diffusion is accurate in terms of RMSE. However, when it comes to relative error, it performs well around the centre but its accuracy decreases towards the tails where the density values are low.

\begin{figure}
\begin{centering}
\includegraphics[scale=0.35]{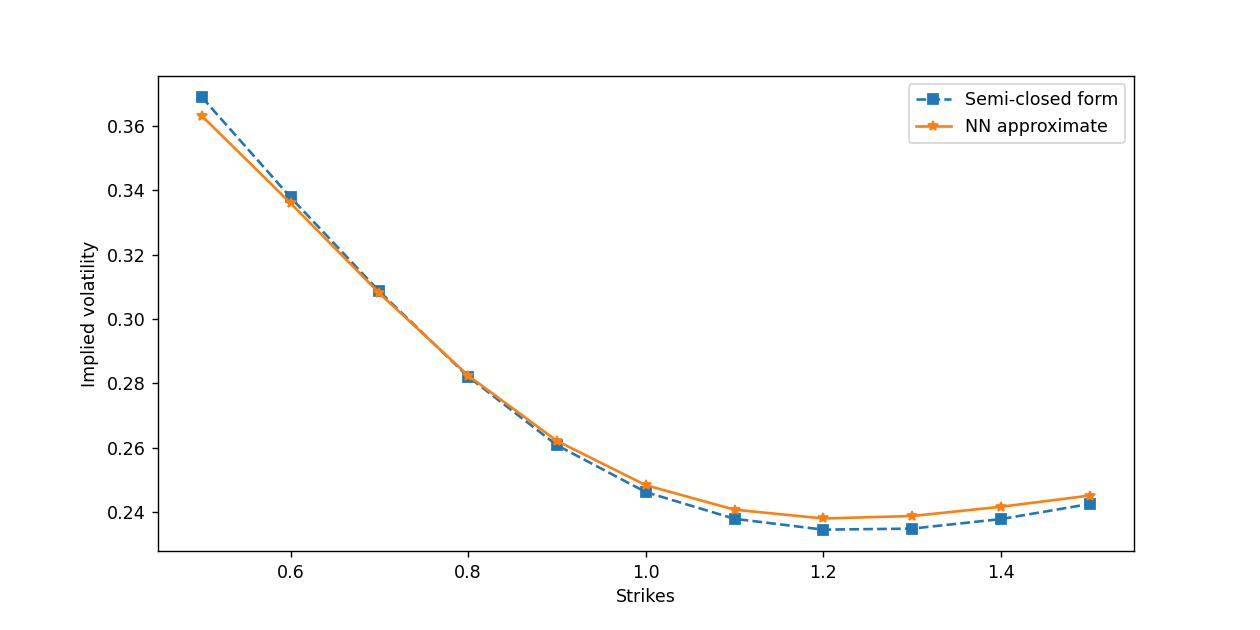}
\par\end{centering}
\caption{Kou's double exponential jump diffusion volatility smile obtained through put prices using the semi-closed form European option pricing formula \citep{kou2002jump} and the QUAD method with the neural  TPDF. The parameters used: $S_{0}=1$, $\sigma_{0}=0.16$, $T=1.0$, $\lambda=1.0$, $p=0.4$, $\eta_{1}=10$, $\eta_{2}=5$.
We calculate option prices for 11 strikes $K=0.5,0.6,\dots,1.5$.}

\label{Fig: Kou's DEJD implied vol smile}
\end{figure}

We also measure the performance of the trained NN using option prices and the implied volatility smile against the semi-closed form option prices. Table~\ref{Tab: Kou put prices and smile} gives the results. We observe that option prices obtained through the neural TPDF are accurate according to the benchmarking against the semi-closed form solutions; the implied volatility error is not larger than 1.7\% (see also the implied volatility smile in Figure~\ref{Fig: Kou's DEJD implied vol smile}).
The highlight here is the immense speed advantage of the combined QUAD plus the neural TPDF over the semi-closed form solution: the semi-closed form solutions require a number of expensive numerical procedures and take over 100 seconds to evaluate 11 options, whereas our deep learning parametric approach delivers very high speed option valuations, costing only about 0.15 second to calculate 11 options.

\begin{table}
\caption{The put option prices and option implied volatility calculated from
semi-closed form option prices (see \citet{kou2002jump}) and the neural TPDF with QUAD in the case of Kou's double exponential jump diffusion
model. The parameters used: $S_{0}=1$, $\sigma_{0}=0.16$, $T=1.0$, $\lambda=1.0$, $p=0.4$, $\eta_{1}=10$, $\eta_{2}=5$.
We calculate option prices for 11 strikes $K=0.5$, $0.6$, $\dots$, $1.5$.
We include the absolute errors of the calculations. We record the total calculation time
for calculation 11 options using the semi-closed form formula and
the QUAD method with neural TPDF. The average
time is in seconds.}
\label{Tab: Kou put prices and smile}
\scriptsize
\centering{}{}%
\begin{tabular}{>{\centering}p{1.5cm}>{\centering}p{1cm}>{\centering}p{1cm}>{\centering}p{1.5cm}>{\centering}p{1cm}>{\centering}p{1cm}>{\centering}p{1cm}>{\centering}p{1.5cm}>{\centering}p{1cm}}
\toprule
{Strikes} & \multicolumn{2}{c}{{Price}} & \multicolumn{2}{c}{{Error}} & \multicolumn{2}{c}{{Implied volatility}} & \multicolumn{2}{c}{{Error}}\tabularnewline
\cmidrule{2-9} \cmidrule{3-9} \cmidrule{4-9} \cmidrule{5-9} \cmidrule{6-9} \cmidrule{7-9} \cmidrule{8-9} \cmidrule{9-9}
 & {S-CF} & {NN} & {Abs $(\times10^{-4})$} & {\%} & {S-CF} & {NN} & {Abs $(\times10^{-3})$} & {\%}\tabularnewline
\midrule
{0.5} & {0.0030} & {0.0028} & {$2.68$} & {8.91\%} & {0.369} & {0.363} & {$5.80$} & {1.57\%}\tabularnewline

{0.6} & {0.0074} & {0.0073} & {$1.74$} & {2.37\%} & {0.338} & {0.336} & {$1.81$} & {0.54\%}\tabularnewline

{0.7} & {0.0157} & {0.0158} & {$0.95$} & {0.61\%} & {0.308} & {0.308} & {$0.56$} & {0.18\%}\tabularnewline

{0.8} & {0.0306} & {0.0309} & {$0.87$} & {0.28\%} & {0.282} & {0.282} & {$0.34$} & {0.12\%}\tabularnewline

{0.9} & {0.0565} & {0.0571} & {$4.62$} & {0.82\%} & {0.261} & {0.262} & {$1.33$} & {0.51\%}\tabularnewline

{1.0} & {0.0980} & {0.0991} & {$8.52$} & {0.87\%} & {0.246} & {0.248} & {$2.15$} & {0.87\%}\tabularnewline

{1.1} & {0.1573} & {0.1586} & {$10.88$} & {0.69\%} & {0.238} & {0.241} & {$2.84$} & {1.19\%}\tabularnewline

{1.2} & {0.2320} & {0.2334} & {$11.15$} & {0.48\%} & {0.235} & {0.238} & {$3.46$} & {1.47\%}\tabularnewline

{1.3} & {0.3177} & {0.3189} & {$9.55$} & {0.30\%} & {0.235} & {0.239} & {$3.91$} & {1.66\%}\tabularnewline

{1.4} & {0.4100} & {0.4109} & {$6.69$} & {0.16\%} & {0.238} & {0.242} & {$3.82$} & {1.61\%}\tabularnewline

{1.5} & {0.5058} & {0.5064} & {$3.18$} & {0.06\%} & {0.243} & {0.245} & {$2.61$} & {1.08\%}\tabularnewline
\midrule
{Avg time} & {9.43} & {0.014} &  &  &  &  &  & \tabularnewline
\bottomrule
\end{tabular}{\tiny\par}
\end{table}

We further investigate pricing accuracy in the parametric space. To this end, we train Kou's model in the parametric space and compare the pricing
errors of the NN approximated density against the semi-closed form solutions
over the random parameter sets across different moneyness positions
and maturities. We use the same moneyness definition as in Section~\ref{sec:The-Heston-process}.

From Table \ref{Tab: DEJD_pricing_error}, we see that the pricing performance is consistent in terms of RMSE (reaching $10^{-3})$. However, if we switch to the relative error, we observe that the performance drops notably for deep out-of-the-money options and small maturities. This is caused by the prices for deep out-of-the-money options under
Kou's model being very small (about $10^{-3}$ or less), and hence option prices calculated using QUAD with approximated density will inevitably produce high percentage errors, depending on the approximation accuracy of the density.

\begin{table}
\caption{The pricing errors of the NN approximated density of Kou's double exponential jump diffusion model, compared with Kou's semi-closed form pricing solutions. We randomly generate 100 parameter sets in the parametric space: $V_{0}\in[0.05,0.4]$,$\lambda\in[0.5,1.5],p\in[0.0,1.0],\eta_{1}=[5,13],\eta_{2}\in[5,13]$.
We compare five moneyness ranges: Deep in-the-money (DITM) (1.20-1.40);
In-the-money (ITM) (1.05-1.20); At-the-money (ATM) (0.95-1.05); Out-of-the-money
(OTM) (0.80-0.95); Deep out-of-the-money (DOTM) (0.60-0.80). We choose
four maturityies $T=0.25,0.5,0.75,1.0$. Pricing performance
is in the form of price percentage error (PrPCTE), price root-mean-square
error (PrPMSE), implied volatility percentage error (IVPCTE), implied
volatility root-mean-square error (IVRMSE).}
\label{Tab: DEJD_pricing_error}
\scriptsize
\centering{}%
\begin{tabular}{ccccccc}
\toprule
Maturity & Error Type &      DOTM &       OTM &       ATM &       ITM &      DITM \\
\midrule
$t=0.25$ & PrPCTE &  2.594857 &  1.320442 &  0.097307 &  0.025130 &  0.010503 \\
      & PrRMSE &  0.004886 &  0.004260 &  0.006870 &  0.005054 &  0.004784 \\
      & IVPCTE &  0.194727 &  0.130941 &  0.086532 &  0.100043 &  0.276032 \\
      & IVRMSE &  0.054412 &  0.019972 &  0.016152 &  0.015340 &  0.060144 \\
\midrule
$t=0.5$ & PrPCTE &  1.067554 &  0.594614 &  0.045973 &  0.018796 &  0.009077 \\
      & PrRMSE &  0.009018 &  0.004287 &  0.004066 &  0.004337 &  0.004006 \\
      & IVPCTE &  0.098684 &  0.086295 &  0.044884 &  0.046043 &  0.154606 \\
      & IVRMSE &  0.041641 &  0.016527 &  0.009546 &  0.011344 &  0.048207 \\
\midrule
$t=0.75$ & PrPCTE &  0.565688 &  0.756655 &  0.037224 &  0.015442 &  0.008830 \\
      & PrRMSE &  0.001821 &  0.004245 &  0.004281 &  0.003629 &  0.003836 \\
      & IVPCTE &  0.054287 &  0.077844 &  0.037561 &  0.032271 &  0.076492 \\
      & IVRMSE &  0.019996 &  0.016563 &  0.010004 &  0.008988 &  0.027049 \\
\midrule
$t=1.0$ & PrPCTE &  0.558983 &  1.444879 &  0.032576 &  0.014928 &  0.007891 \\
      & PrRMSE &  0.002411 &  0.004463 &  0.004840 &  0.003781 &  0.003585 \\
      & IVPCTE &  0.059034 &  0.067676 &  0.032462 &  0.027322 &  0.050652 \\
      & IVRMSE &  0.022499 &  0.015685 &  0.011293 &  0.008994 &  0.020515 \\
\bottomrule
\end{tabular}

\end{table}

\subsubsection{Stochastic volatility jump diffusion.}

In this section we consider a stochastic volatility model with jumps. Jumps can be added to the underlying alone or to both the underlying and the volatility. \citet{gatheral2011volatility} points out that stochastic volatility with jumps or SVJ (i.e., jumps in the underlying only), suggested by \citet{bates1996jumps}, performs much better empirically than stochastic volatility with simultaneous jumps in both, the price and the volatility (SVJJ). The main reason, as the author notes, is that the SVJJ model has more parameters, making it harder to fit to observed option prices. Here we focus on Bates' SVJ model, for which the TPDF is available in a semi-closed form \citealp{bates1996jumps} and hence allowing us to perform insightful benchmarking of our approach to illustrate its accuracy and speed.

Bates' SVJ model written for the log stock price, $X$, and the variance, $V,$ of the stock price under a forward measure takes the form
\begin{align}
dX & =(-\lambda k-\frac{V}{2})dt+\sqrt{V}dW_{1}+JdN, \label{eq:SVJ}\\
dV & =\kappa(\omega-V)dt+\xi\sqrt{V}dW_{2}, \notag \\
dW_{1}dW_{2} & =\rho dt, \notag
\end{align}
where $W_1(t)$ and $W_2(t)$ are correlated Wiener processes, $N(t)$ is a Poisson process with intensity $\lambda >0$, and the independent jumps $J_{1},J_{2},\ldots $ are so that
\begin{equation}
\eta:=\ln(1+J)\sim\mathcal{N}(\mu_{J},\sigma_{J}^{2}).
\end{equation}
Here, $\mu_{J}$ is the mean logarithmic jump amplitude and $\sigma_{J}$ is the jump volatility, and the density of $\eta$ is \begin{equation}
\rho_{\eta}(x)=\frac{\exp\left[-(x-\mu_{J})^{2}/2\sigma_{J}^{2}\right]}{\sqrt{2\pi\sigma_{J}^{2}}}.\label{eq: jump density term}
\end{equation}
The constant $k= E(J)=\exp(\mu_{J}+\sigma^{2}/2)-1.$ The rest of the notation in (\ref{eq:SVJ}) is the same as in the Heston model (\ref{eq:Heston stock process}).

The backward Kolmogorov equation for the joint CDF $C(t,x,v)=C(t,x,v;T,y,z)$
is
\begin{align}
\frac{\partial C}{\partial t}-(\lambda k+\frac{v}{2})\frac{\partial C}{\partial x}+(\kappa(\omega-v))\frac{\partial C}{\partial v}+\frac{1}{2}v\frac{\partial^{2}C}{\partial x}+\frac{1}{2}\xi^{2}v\frac{\partial^{2}C}{\partial v^{2}}+\rho\xi v\frac{\partial^{2}C}{\partial x\partial v}\nonumber \\
+\lambda\int_{\mathbb{R}} C(t,x+u)\rho_{\eta}(u)du-\lambda C & =0,t\in[0,T),\:x\in\mathbb{R},\:v>0,\label{eq:SVJ backward equation}
\end{align}
with terminal condition
\begin{equation}
C(T,x,v)=\mathbbm{1}(x\leq y,v\leq z)=\begin{cases}
1, & x\leq y\thinspace\text{\thinspace and}\thinspace\thinspace v\leq z,\\
0, & \text{otherwise.}
\end{cases}
\end{equation}

We approximate the integral term in Eq. \eqref{eq:SVJ backward equation} by Gauss-Hermite quadrature with 7 points.
The domain $Q$ used here for NN training is $x,y\in[-3.5,3.5]$, $v,z\in[0,0.6]$, $\kappa\in[0.8,1.2]$, $\omega\in[0.1,0.3]$, $\xi\in[0,0.4]$, $\rho\in[-0.5,0.5]$, $\lambda\in[0.0,2.0]$, $\mu_{J}\in[-0.5,0.5],$ $\sigma_{J}\in[0.0,0.5]$, $t\in[0,1.2]$.
We choose $\lambda_{L}=100$, and in order to get higher accuracy, we train 2 million epochs and sample 5000 points every five epochs. As usual, we choose the NN with the smallest total loss. The training time per epoch using
an NVIDIA P100 GPU is about 0.051 seconds and it takes about 28.5 hours to complete the NN training of 2 million epochs.

\begin{figure}
\begin{centering}
\includegraphics[scale=0.35]{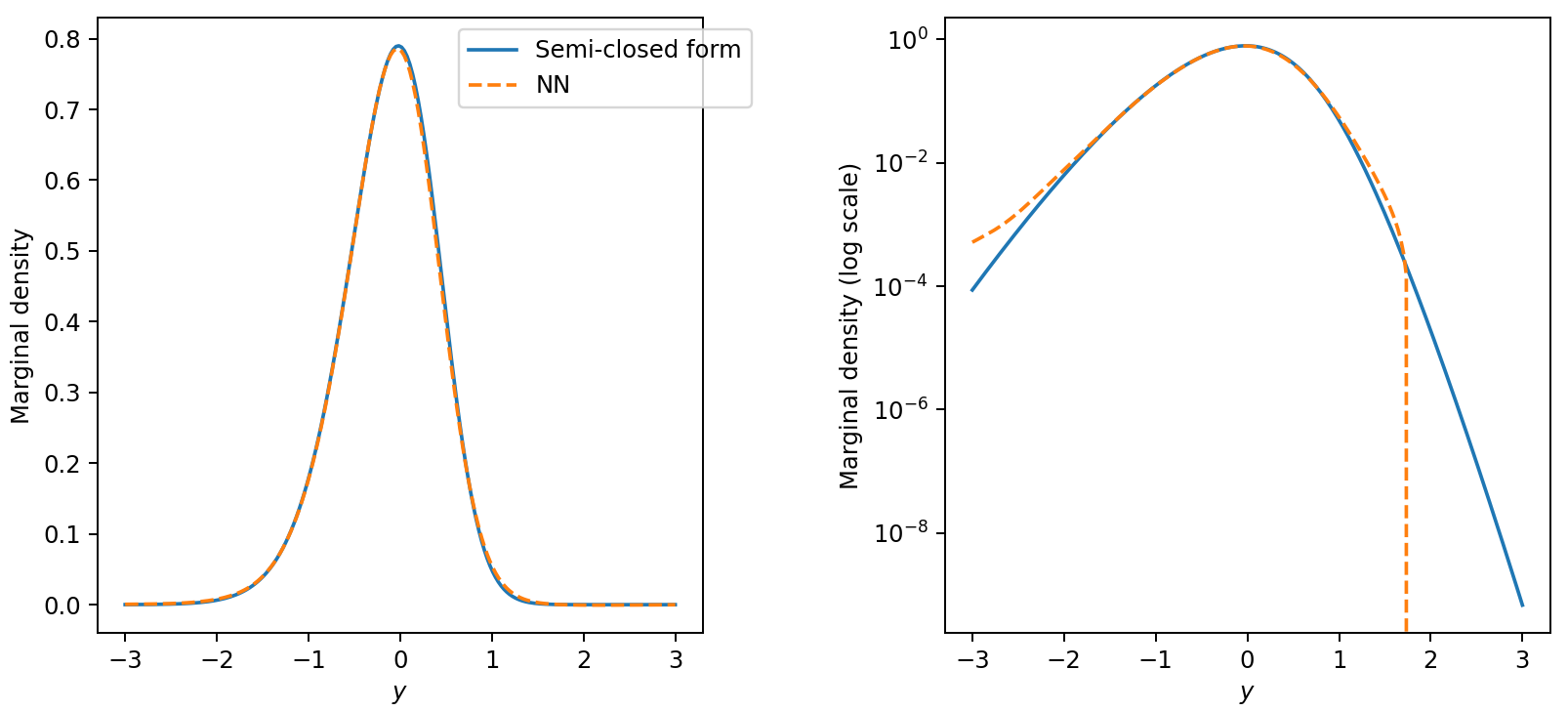}
\par\end{centering}
\caption{SVJ model. Comparison of the marginal density obtained through the neural
joint TPDF with the semi-closed form marginal density. The log scale graph is shown on the right. The parameters
used: initial stock price $S_{0}=1$, $V_{0}=0.16$, $\omega=0.3$,
$\xi=0.35$, $\kappa=1.0$, $\lambda=1.0$, $\mu_{J}=-0.1$, $\sigma_{J}=0.2$,
$\rho=-0.5$. The semi-closed form marginal TPDF used in this example
follows \citet{bates1996jumps}.}
\label{Fig: Heston with jump marginal density}
\end{figure}

For Bates' SVJ model, the marginal density for $X$ is known in semi-closed form (see \citealp{bates1996jumps}). In Figure~\ref{Fig: Heston with jump marginal density}, we plot the marginal density obtained through the neural joint TPDF against the semi-closed form marginal density. We see that the marginal density calculated from neural joint TPDF gives a very accurate result benchmarked against the semi-closed form marginal density (the RMSE is $1.72\times10^{-3}$).

We further analyse the NN performance using option pricing and implied volatility smile. Once again, in terms of speed, we see that the two-dimensional QUAD method with neural TPDF is remarkably faster than the one-dimensional QUAD method with semi-closed form marginal density even when using fewer quadrature points. In terms of accuracy, the neural TPDF produces accurate option prices and implied volatility, with percentage implied volatility error smaller than 0.5\%.

\begin{table}
\caption{The put option prices and option implied volatility calculated using the one-dimensional QUAD method with semi-closed form marginal density of the SVJ model and the two-dimensional QUAD with neural joint TPDF. The number of quadrature points used in the one dimensional QUAD is 1001 and the number of quadrature points used in each dimension in the two dimensional QUAD is 101 (10,201 points in total). The parameters used: $S_{0}=1$, $V_{0}=0.16$,
$\omega=0.3$, $\xi=0.35$, $\kappa=1.0$, $\lambda=1.0$, $\mu_{J}=-0.1$, $\sigma_{J}=0.2$, $\rho=-0.5$. We calculate option prices for 11 strikes $K=0.5,0.6,\dots,1.5$.
The semi-closed form marginal density is from \citet{bates1996jumps}. We include the absolute errors of the approximations. We record the total calculation time for pricing the 11 options using the semi-closed form formula and the QUAD method with neural TPDF. The average time is in seconds.}

\centering{}{\label{Tab: Heston with jump put prices and implied vol}}%
\scriptsize
\begin{tabular}{>{\centering}p{1.5cm}>{\centering}p{1cm}>{\centering}p{1cm}>{\centering}p{1.5cm}>{\centering}p{1cm}>{\centering}p{1cm}>{\centering}p{1cm}>{\centering}p{1.5cm}>{\centering}p{1cm}}
\toprule
{Strikes} & \multicolumn{2}{c}{{Price}} & \multicolumn{2}{c}{{Error}} & \multicolumn{2}{c}{{Implied volatility}} & \multicolumn{2}{c}{{Error}}\tabularnewline
\cmidrule{2-9} \cmidrule{3-9} \cmidrule{4-9} \cmidrule{5-9} \cmidrule{6-9} \cmidrule{7-9} \cmidrule{8-9} \cmidrule{9-9}
 & {S-CF} & {NN} & {Abs $(\times10^{-4})$} & {\%} & {S-CF} & {NN} & {Abs $(\times10^{-3})$} & {\%}\tabularnewline
\midrule
{0.5} & {0.0176} & {0.0180} & {$3.45$} & {0.02\%} & {0.541} & {0.543} & {$2.87$} & {0.53\%}\tabularnewline

{0.6} & {0.0355} & {0.0359} & {$3.76$} & {0.01\%} & {0.528} & {0.530} & {$2.01$} & {0.38\%}\tabularnewline

{0.7} & {0.0619} & {0.0623} & {$4.00$} & {0.01\%} & {0.518} & {0.519} & {$1.57$} & {0.30\%}\tabularnewline

{0.8} & {0.0974} & {0.0978} & {$4.32$} & {0.00\%} & {0.509} & {0.510} & {$1.38$} & {0.27\%}\tabularnewline

{0.9} & {0.1420} & {0.1425} & {$4.70$} & {0.00\%} & {0.501} & {0.503} & {$1.31$} & {0.26\%}\tabularnewline

{1.0} & {0.1952} & {0.1957} & {$4.94$} & {0.00\%} & {0.494} & {0.496} & {$1.28$} & {0.26\%}\tabularnewline

{1.1} & {0.2563} & {0.2568} & {$4.69$} & {0.00\%} & {0.488} & {0.490} & {$1.18$} & {0.24\%}\tabularnewline

{1.2} & {0.3244} & {0.3248} & {$3.60$} & {0.00\%} & {0.483} & {0.484} & {0.91} & {0.19\%}\tabularnewline

{1.3} & {0.3985} & {0.3986} & {$1.39$} & {0.00\%} & {0.478} & {0.479} & {0.37} & {0.08\%}\tabularnewline

{1.4} & {0.4777} & {0.4775} & {$2.08$} & {0.00\%} & {0.474} & {0.474} & {$0.58$} & {0.12\%}\tabularnewline

{1.5} & {0.5611} & {0.5604} & {$6.82$} & {0.00\%} & {0.470} & {0.469} & {$0.21$} & {0.44\%}\tabularnewline
\midrule
{Avg time} & {9.70} & {4.26} &  &  &  &  &  & \tabularnewline
\bottomrule
\end{tabular}{\tiny\par}
\end{table}

\begin{figure}
\begin{centering}
\includegraphics[scale=0.35]{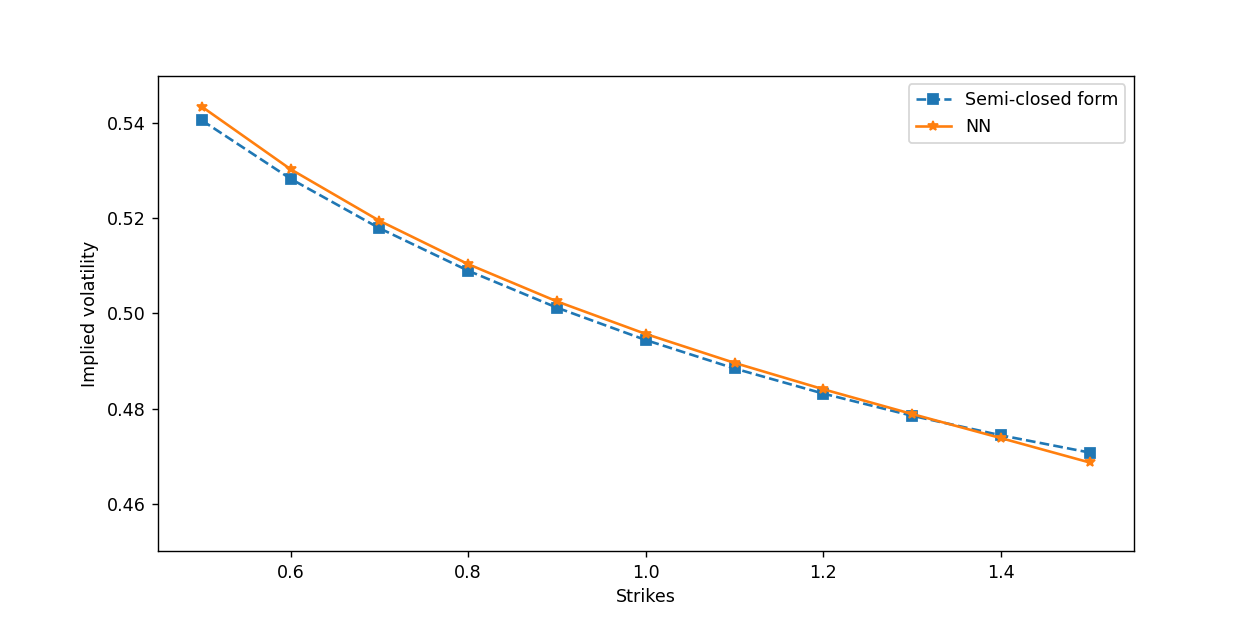}
\par\end{centering}
\caption{SVJ model. The implied volatility smile is obtained
through put prices under the QUAD method with semi-closed form
joint TPDF and with the neural TPDF. The parameters
used: $S_{0}=1$, $V_{0}=0.16$, $\omega=0.3$, $\xi=0.35$, $\kappa=1.0$,
$\lambda=1.0$, $\mu_{J}=-0.1$, $\sigma_{J}=0.2$, $\rho=-0.5$.
We calculate option prices for 11 strikes $K=0.5,0.6,\dots,1.5$.}

\label{Fig: SVJ vol smile}
\end{figure}

We train the stochastic volatility with jump model in the parametric space and compare the pricing errors of the NN approximated density against the semi-closed form solutions over the random parameter sets across different moneyness positions and maturities. We use the same moneyness definition as in Section~\ref{sec:The-Heston-process}. From Table \ref{Tab: SVJ_pricing_error}, we see that the pricing errors for the SVJ model are similar to the case of the Heston model. In fact, the performance of our approach may even be slightly better in the case of the SVJ model. This could be because with the introduction of the jump component, the density for a given maturity is more spread out, i.e., the turns are smoother for CDF, making it even easier for NN to approximate the density.

\begin{table}
\caption{The pricing errors of the NN approximated density of stochastic volatility jump diffusion model, compared with SVJ's semi-closed form pricing solutions. We randomly generate 100 parameter sets in the parametric space: $V_{0}\in[0.05,0.4]$, $\kappa\in[0.8,1.2]$, $\omega\in[0.1,0.3]$, $\xi\in[0.05,0.4]$, $\rho\in[-0.5,0.5]$, $\lambda\in[0.5,1.5]$, $\mu_{J}=[-0.5,0.5]$, our$\sigma_{J}\in[0.0,0.5]$. We compare five moneyness ranges: Deep in-the-money (DITM) (1.20-1.40); In-the-money (ITM) (1.05-1.20); At-the-money (ATM) (0.95-1.05); Out-of-the-money
(OTM) (0.80-0.95); Deep out-of-the-money (DOTM) (0.60-0.80). We choose four maturity times $T=0.25,0.5,0.75,1.0$. The pricing performance is in the form of price percentage error (PrPCTE), price root-mean-square error (PrPMSE), implied volatility percentage error (IVPCTE), implied volatility root-mean-square error (IVRMSE).}
\label{Tab: SVJ_pricing_error}
\scriptsize
\centering{}%
\begin{tabular}{ccccccc}
\toprule
Maturity & Error Type &     DOTM &       OTM &       ATM &       ITM &      DITM \\
\midrule
$t=0.25$ & PrPCTE &  0.074059 &  0.027604 &  0.016259 &  0.010797 &  0.006669 \\
      & PrRMSE &  0.001114 &  0.001704 &  0.002021 &  0.002313 &  0.002711 \\
      & IVPCTE &  0.018660 &  0.015493 &  0.016342 &  0.019114 &  0.029144 \\
      & IVRMSE &  0.005974 &  0.004728 &  0.004697 &  0.005591 &  0.010288 \\
\midrule
$t=0.5$ & PrPCTE &  0.021764 &  0.012465 &  0.009069 &  0.007021 &  0.005370 \\
      & PrRMSE &  0.000929 &  0.001383 &  0.001691 &  0.002003 &  0.002555 \\
      & IVPCTE &  0.008300 &  0.008525 &  0.009187 &  0.010263 &  0.013916 \\
      & IVRMSE &  0.003563 &  0.003611 &  0.003968 &  0.004631 &  0.006642 \\
\midrule
$t=0.75$ & PrPCTE &  0.012259 &  0.008152 &  0.006691 &  0.005799 &  0.004926 \\
      & PrRMSE &  0.000881 &  0.001319 &  0.001653 &  0.001987 &  0.002671 \\
      & IVPCTE &  0.005651 &  0.006112 &  0.006841 &  0.007879 &  0.010497 \\
      & IVRMSE &  0.003051 &  0.003412 &  0.003934 &  0.004588 &  0.006501 \\
\midrule
$t=1.0$ & PrPCTE &  0.009093 &  0.006730 &  0.005915 &  0.005329 &  0.004700 \\
      & PrRMSE &  0.000934 &  0.001405 &  0.001779 &  0.002162 &  0.002959 \\
      & IVPCTE &  0.004717 &  0.005295 &  0.006099 &  0.006986 &  0.009058 \\
      & IVRMSE &  0.003095 &  0.003659 &  0.004293 &  0.005030 &  0.007010 \\
\bottomrule
\end{tabular}
\end{table}

\begin{remark}
It is known (see, e.g., \citep{baron94}) that, for problems satisfying certain assumptions, increasing the size of the NN and the number of training points causes the NN approximation to tend asymptotically towards the exact solution of the problem. However, in contrast with conventional numerical methods (e.g., finite difference or Monte Carlo), there is currently no theory to guarantee the accuracy of a NN approximation. In this paper, we aim to illustrate the feasibility of the proposed computational approach and, consequently, we mostly consider models for underlyings where closed form or semi-closed form TPDFs or option prices are known. When such knowledge is not available, the following benchmarking is suggested.
We train a neural TPDF for a given model of underlying and compute option prices using e.g. QUAD for a small set of parameters. Then for the same set of parameters we compute option prices accurately using conventional numerical methods where there is a well-established theory on controlling the error of approximations. If the option prices produced by the neural TPDF are close to accurate results obtained by the conventional methods, then we declare that the neural TPDF is sufficiently accurate to use as a neural TPDF generator. Note that the generator is available for all points in the parametric space on which the NN is trained while the expensive sanity check of accuracy is done using traditional numerical methods for a small set of parameters just once and before the NN is considered finally trained (``offline''). Such a sanity check was illustrated on the SABR model in Section~\ref{sec:SABR}.
We also note that within our approach, a neural TPDF itself and the results produced based on it do not have a statistical error because we fix the NN parameters that the training suggests are best and use these afterwards as a neural TPDF generator post-training (``online''); in other words the error of a neural TPDF is deterministic. This is in contrast to how deep learning can be used to speed up Monte Carlo simulations in option pricing (see, e.g., \citep{Piers}).
\end{remark}

\begin{remark}
Here, we illustrate the use of neural TPDFs in conjunction with QUAD to evaluate option prices. Since QUAD is quadrature based, it is only effective for relatively low-dimensional situations (i.e., a small number of underlyings - demonstrated under geometric Brownian motion for barrier options with five underlyings  by \citet{andricopoulos2007extending} who note that it is overtaken by Monte Carlo at around fifteen underlyings).
To use neural TPDFs for option prices in higher dimensional situations, one can exploit space grid methods  \citep{smolyak,spacegrid}.
\end{remark}

\section{Conclusion}

This paper is the first in the literature to use a deep learning approach in approximating transition probability density (TPDF) for any model of the underlying in a parametric fashion. Our approach not only provides the required TPDF approximations but also its ``single solve'' capability across all parameters means that effort is expended only once in a single pre-computation, delivering a parametric neural TPDF generator that does not require recalibration.

Instead of aiming at directly solving the forward Kolmogorov (Fokker-Planck) equation for the TPDF, which would require handling the initial condition (a ``nasty'' Dirac delta function), we use an ingenious way of first solving the backward Kolmogorov equation for the cumulative probability function which involves the terminal condition, a more ``tamable'' step function, before calculating the TPDF by differentiating the cumulative probability. We train each artificial neural network (NN) by optimising a purposefully constructed cost function, thereby solving the parametric  partial differential equation problem for the cumulative distribution function of any given underlying price process.
For this, we can use as many  data points as needed in order to train the NN such that it sufficiently accurately approximates the TPDF. Once the NN is trained, we have an ultra-fast neural TPDF generator, tailored to the underlying model and portable to other calculation setups and computers.

We illustrate the approach first on the simplest Black-Scholes-Merton setup under geometric Brownian motion, noting that TPDF approximations for other one-dimensional models of underlyings can likewise be obtained. Next, we use \citeauthor{heston1993closed}'s model \citeyearpar{heston1993closed} as an example of stochastic volatility models which can be handled by the deep learning approach. We also show that this approach can deal with other scenarios including non-affine processes such as the SABR model of the underlying (benchmarked against Monte Carlo simulation for option prices). Finally, we demonstrate how neural TPDFs can be obtained for jump diffusion models (Kou's double exponential jump-diffusion \citep{kou2002jump} and stochastic volatility jump diffusion \citep{bates1996jumps}) by solving in a parametric fashion the backward Kolmogorov equations taking the form of partial integro-differential equations.

The applications of learning the TPDF using deep NNs go beyond those illustrated, such as simulating likelihood estimators and default probability calculations in credit risk management (see, e.g., \citep{filipovic2013density}). Options are ubiquitous in finance, from the large range of directly available options on exchanges and over-the-counter, to credit risk and Real Options. During the five decades since the founding framework of Black, Scholes and Merton, practitioners and academics have developed ways of dealing with option features (such as early exercise) in combination with models of the underlying but often either the models are too simple or the solutions too hard. Although parallel computing, using GPUs, speeds up conventional approaches (indeed, investment banks may use hundreds or even thousands of these to accelerate parallel calculations) nevertheless, massive calculations still need to be repeated in frequent recalibrations. The deep learning approach we have introduced, resting on finding parametric neural TPDFs, avoids this.

Our paper also makes contributions to the currently fast developing area of deep learning algorithms used for solving differential equations, which includes demonstrating how to computationally efficiently find solutions to the Fokker-Planck equations, both for diffusion and jump-diffusion models in a parametric fashion; extensive testing to demonstrate how to choose cost functions for training NNs with higher accuracy; comparing the performance of different NN architectures and the suitability of different GPUs for the task.

There is room for improvement in our method, as demonstrated by the analysis of pricing errors. A limitation of the approach presented here is its accuracy in terms of absolute errors, which means it may struggle with precise approximation of values very close to 0.0, resulting in high relative errors (percentage errors). Future research should address this issue. Additionally, it should be noted that the use of the DGM NN \citep{sirignano2018dgm} in our approach does not necessarily imply that it is the optimal network architecture for solving the underlying problems (see Appendix \ref{appdx: MLP} for a brief comparison between DGM and MLP networks for our approach). There is potential for further investigation into the search for or development of a better and more effective network architecture. Moreover, an interesting avenue for future research involves exploring model calibration modified through using market data.

The introduction of targeted deep learning for TPDF, as initiated in this paper, holds promise for the development of comprehensive and efficient programs capable of addressing a wide range of models and applications.

\bibliographystyle{informs2014} 
\bibliography{QUAD-IV} 

\begin{thebibliography}{69}
\providecommand{\natexlab}[1]{#1}
\providecommand{\url}[1]{\texttt{#1}}
\providecommand{\urlprefix}{URL }

\bibitem[{A{\"\i}t-Sahalia(2002)}]{ait2002maximum}
A{\"\i}t-Sahalia Y (2002) Maximum likelihood estimation of discretely sampled
  diffusions: a closed-form approximation approach. \emph{Econometrica}
  70:223--262.

\bibitem[{A{\"\i}t-Sahalia(2008)}]{ait2008closed}
A{\"\i}t-Sahalia Y (2008) Closed-form likelihood expansions for multivariate
  diffusions. \emph{Annals of Statistics} 36:906--937.

\bibitem[{A{\"\i}t-Sahalia \protect\BIBand{} Kimmel(2007)}]{ait2007maximum}
A{\"\i}t-Sahalia Y, Kimmel R (2007) Maximum likelihood estimation of stochastic
  volatility models. \emph{Journal of Financial Economics} 83(2):413--452.

\bibitem[{A{\"\i}t-Sahalia et~al.(2006)A{\"\i}t-Sahalia, Yu
  et~al.}]{ai2006saddlepoint}
A{\"\i}t-Sahalia Y, Yu J, et~al. (2006) Saddlepoint approximations for
  continuous-time {M}arkov processes. \emph{Journal of Econometrics}
  134(2):507--551.

\bibitem[{Al-Aradi et~al.(2018)Al-Aradi, Correia, Naiff, Jardim,
  \protect\BIBand{} Saporito}]{al2018solving}
Al-Aradi A, Correia A, Naiff D, Jardim G, Saporito Y (2018) Solving nonlinear
  and high-dimensional partial differential equations via deep learning.
  \emph{arXiv:1811.08782} .

\bibitem[{Andricopoulos et~al.(2003)Andricopoulos, Widdicks, Duck,
  \protect\BIBand{} Newton}]{andricopoulos2003universal}
Andricopoulos AD, Widdicks M, Duck PW, Newton DP (2003) Universal option
  valuation using quadrature methods. \emph{Journal of Financial Economics}
  67:447--471, (See also Corrigendum, Journal of Financial Economics 73, 603
  (2004)).

\bibitem[{Andricopoulos et~al.(2007)Andricopoulos, Widdicks, Newton,
  \protect\BIBand{} Duck}]{andricopoulos2007extending}
Andricopoulos AD, Widdicks M, Newton DP, Duck PW (2007) Extending quadrature
  methods to value multi-asset and complex path dependent options.
  \emph{Journal of Financial Economics} 83:471--499.

\bibitem[{Antoulas et~al.(2015)Antoulas, Ionutiu, Martins, ter Maten,
  Mohaghegh, Pulch, Rommes, Saadvandi, \protect\BIBand{}
  Striebel}]{Antoulas2015}
Antoulas AC, Ionutiu R, Martins N, ter Maten EJW, Mohaghegh K, Pulch R, Rommes
  J, Saadvandi M, Striebel M (2015) Model order reduction: Methods, concepts
  and properties. G{\"u}nther M, ed., \emph{Coupled Multiscale Simulation and
  Optimization in Nanoelectronics}, 159--265 (Berlin: Springer).

\bibitem[{Applebaum(2009)}]{Apple09}
Applebaum D (2009) \emph{L\'evy processes and stochastic calculus} (Cambridge:
  Cambridge University Press).

\bibitem[{Baron(1994)}]{baron94}
Baron AR (1994) Approximation and estimation bounds for artificial neural
  networks. \emph{Machine Learning} 14(1):115--133.

\bibitem[{Bates(1996)}]{bates1996jumps}
Bates DS (1996) Jumps and stochastic volatility: Exchange rate processes
  implicit in {D}eutsche {M}ark options. \emph{Review of Financial Studies}
  9(1):69--107.

\bibitem[{Beskos \protect\BIBand{} Roberts(2005)}]{beskos2005exact}
Beskos A, Roberts GO (2005) Exact simulation of diffusions. \emph{Annals of
  Applied Probability} 15(4):2422--2444.

\bibitem[{Black \protect\BIBand{} Scholes(1973)}]{black1973pricing}
Black F, Scholes M (1973) The pricing of options and corporate liabilities.
  \emph{Journal of Political Economy} 81:637--654.

\bibitem[{Chen et~al.(2014)Chen, H{\"a}rk{\"o}nen, \protect\BIBand{}
  Newton}]{chen2014advancing}
Chen D, H{\"a}rk{\"o}nen HJ, Newton DP (2014) Advancing the universality of
  quadrature methods to any underlying process for option pricing.
  \emph{Journal of Financial Economics} 114:600--612.

\bibitem[{Chen \protect\BIBand{} Huang(2013)}]{chen2013localization}
Chen N, Huang Z (2013) Localization and exact simulation of brownian
  motion-driven stochastic differential equations. \emph{Mathematics of
  Operations Research} 38(3):591--616.

\bibitem[{Cox(1996)}]{cox1996constant}
Cox JC (1996) The constant elasticity of variance option pricing model.
  \emph{Journal of Portfolio Management} 23:15--17.

\bibitem[{Dissanayake \protect\BIBand{}
  Phan-Thien(1994)}]{dissanayake1994neural}
Dissanayake M, Phan-Thien N (1994) Neural-network-based approximations for
  solving partial differential equations. \emph{Communications in Numerical
  Methods in Engineering} 10:195--201.

\bibitem[{Figlewski(2009)}]{figlewski2009estimating}
Figlewski S (2009) \emph{Estimating the implied risk-neutral density for the US
  market portfolio, volatility and time series econometrics: essays in honor of
  {Robert} {F}. {Engle}} (Oxford: Oxford University Press).

\bibitem[{Figlewski(2018)}]{figlewski2018risk}
Figlewski S (2018) Risk-neutral densities: A review. \emph{Annual Review of
  Financial Economics} 10:329--359.

\bibitem[{Filipovi{\'c} et~al.(2013)Filipovi{\'c}, Mayerhofer,
  \protect\BIBand{} Schneider}]{filipovic2013density}
Filipovi{\'c} D, Mayerhofer E, Schneider P (2013) Density approximations for
  multivariate affine jump-diffusion processes. \emph{Journal of Econometrics}
  176:93--111.

\bibitem[{Floc'h \protect\BIBand{} Kennedy(2014)}]{floc2014finite}
Floc'h L, Kennedy GJ (2014) Finite difference techniques for arbitrage free
  {SABR}. \emph{Available at SSRN 2402001} .

\bibitem[{Freidlin(1985)}]{FRE85}
Freidlin MI (1985) \emph{Functional integration and partial differential
  equations} (Princeton: Princeton Univ. Press).

\bibitem[{Garcke \protect\BIBand{} Griebel(2013)}]{spacegrid}
Garcke J, Griebel M (2013) \emph{Sparse Grids and Applications} (Springer).

\bibitem[{Gardiner(2004)}]{gardiner2004}
Gardiner CW (2004) \emph{Handbook of stochastic methods for physics, chemistry
  and the natural sciences} (Berlin: Springer).

\bibitem[{Gatheral(2011)}]{gatheral2011volatility}
Gatheral J (2011) \emph{The volatility surface: a practitioner's guide}, volume
  357 (John Wiley \& Sons).

\bibitem[{Geist et~al.(2020)Geist, Petersen, Raslan, Schneider,
  \protect\BIBand{} Kutyniok}]{Kutyniok2020}
Geist M, Petersen P, Raslan M, Schneider R, Kutyniok G (2020) Numerical
  solution of the parametric diffusion equation by deep neural networks.
  \emph{arXiv:2004.12131} .

\bibitem[{Gichman \protect\BIBand{} Skorochod(1972)}]{GIS68}
Gichman II, Skorochod AV (1972) \emph{Stochastic differential equations}
  (Berlin: Springer).

\bibitem[{Giesecke \protect\BIBand{} Schwenkler(2019)}]{giesecke2019simulated}
Giesecke K, Schwenkler G (2019) Simulated likelihood estimators for discretely
  observed jump--diffusions. \emph{Journal of Econometrics} 213(2):297--320.

\bibitem[{Giesecke \protect\BIBand{} Smelov(2013)}]{giesecke2013exact}
Giesecke K, Smelov D (2013) Exact sampling of jump diffusions. \emph{Operations
  Research} 61(4):894--907.

\bibitem[{Glasserman(2003)}]{GLA03}
Glasserman P (2003) \emph{Monte {C}arlo methods in financial engineering}
  (Berlin: Springer).

\bibitem[{Goodfellow et~al.(2016)Goodfellow, Bengio, \protect\BIBand{}
  Courville}]{goodfellow2016deep}
Goodfellow I, Bengio Y, Courville A (2016) \emph{Deep Learning} (MIT press).

\bibitem[{Guay \protect\BIBand{} Schwenkler(2021)}]{guay2021efficient}
Guay F, Schwenkler G (2021) Efficient estimation and filtering for multivariate
  jump--diffusions. \emph{Journal of Econometrics} 223(1):251--275.

\bibitem[{Hagan et~al.(2014)Hagan, Kumar, Lesniewski, \protect\BIBand{}
  Woodward}]{hagan_arbitragefree_2014}
Hagan PS, Kumar D, Lesniewski A, Woodward D (2014) Arbitrage-free {SABR}.
  \emph{Wilmott Magazine} 2014:60--75, ISSN 1541-8286.

\bibitem[{Hagan et~al.(2002)Hagan, Kumar, Lesniewski, \protect\BIBand{}
  Woodward}]{hagan2002managing}
Hagan PS, Kumar D, Lesniewski AS, Woodward DE (2002) Managing smile risk.
  \emph{The Best of Wilmott} 1:249--296.

\bibitem[{Henry-Labord{\`e}re(2008)}]{henry2008analysis}
Henry-Labord{\`e}re P (2008) \emph{Analysis, geometry, and modeling in finance:
  Advanced methods in option pricing} (CRC Press).

\bibitem[{Henry-Labordere et~al.(2017)Henry-Labordere, Tan, \protect\BIBand{}
  Touzi}]{henry2017unbiased}
Henry-Labordere P, Tan X, Touzi N (2017) Unbiased simulation of stochastic
  differential equations. \emph{Annals of Applied Probability}
  27(6):3305--3341.

\bibitem[{Heston(1993)}]{heston1993closed}
Heston SL (1993) A closed-form solution for options with stochastic volatility
  with applications to bond and currency options. \emph{Review of Financial
  Studies} 6:327--343.

\bibitem[{Heston(1997)}]{heston1997simple}
Heston SL (1997) A simple new formula for options with stochastic volatility.
  \emph{Available at SSRN 86074} .

\bibitem[{Hinds \protect\BIBand{} Tretyakov(2022)}]{Piers}
Hinds PD, Tretyakov MV (2022) Neural variance reduction for stochastic
  differential equations. \emph{arXiv:2209.12885} .

\bibitem[{Hochreiter \protect\BIBand{} Schmidhuber(1997)}]{hochreiter1997long}
Hochreiter S, Schmidhuber J (1997) Long short-term memory. \emph{Neural
  Computation} 9:1735--1780.

\bibitem[{Hutchinson et~al.(1994)Hutchinson, Lo, \protect\BIBand{}
  Poggio}]{hutchinson1994nonparametric}
Hutchinson JM, Lo AW, Poggio T (1994) A nonparametric approach to pricing and
  hedging derivative securities via learning networks. \emph{Journal of
  Finance} 49:851--889.

\bibitem[{Khoo et~al.(2017)Khoo, Lu, \protect\BIBand{} Ying}]{Khoo17}
Khoo Y, Lu J, Ying L (2017) Solving parametric {PDE} problems with artificial
  neural networks. \emph{arXiv:1707.03351} .

\bibitem[{Kingma \protect\BIBand{} Ba(2014)}]{kingma2014adam}
Kingma DP, Ba J (2014) Adam: A method for stochastic optimization.
  \emph{arXiv:1412.6980} .

\bibitem[{Kou(2002)}]{kou2002jump}
Kou SG (2002) A jump-diffusion model for option pricing. \emph{Management
  Science} 48(8):1086--1101.

\bibitem[{Kutyniok et~al.(2019)Kutyniok, Petersen, Raslan, \protect\BIBand{}
  Schneider}]{Kutyniok2019}
Kutyniok G, Petersen P, Raslan M, Schneider R (2019) A theoretical analysis of
  deep neural networks and parametric {PDE}s. \emph{arXiv:1904.00377} .

\bibitem[{Ladyzhenskaya et~al.(1968)Ladyzhenskaya, Solonnikov,
  \protect\BIBand{} Ural'tseva}]{Ladyzh}
Ladyzhenskaya O, Solonnikov V, Ural'tseva NN (1968) \emph{Linear and
  quasi-linear equations of parabolic type}, volume~23 of \emph{Trans. Math.
  Monog.} (Providence, RI: American Mathematical Society).

\bibitem[{Lagaris et~al.(1998)Lagaris, Likas, \protect\BIBand{}
  Fotiadis}]{lagaris1998artificial}
Lagaris IE, Likas A, Fotiadis DI (1998) Artificial neural networks for solving
  ordinary and partial differential equations. \emph{IEEE Transactions on
  Neural Networks} 9:987--1000.

\bibitem[{Lagaris et~al.(2000)Lagaris, Likas, \protect\BIBand{}
  Papageorgiou}]{lagaris2000neural}
Lagaris IE, Likas AC, Papageorgiou DG (2000) Neural-network methods for
  boundary value problems with irregular boundaries. \emph{IEEE Transactions on
  Neural Networks} 11:1041--1049.

\bibitem[{Lee \protect\BIBand{} Kang(1990)}]{lee1990neural}
Lee H, Kang IS (1990) Neural algorithm for solving differential equations.
  \emph{Journal of Computational Physics} 91:110--131.

\bibitem[{Lewis(2016)}]{lewis2016option}
Lewis AL (2016) Option valuation under stochastic volatility \uppercase{II}.
  \emph{Finance Press} .

\bibitem[{Lord et~al.(2008)Lord, Fang, Bervoets, \protect\BIBand{}
  Oosterlee}]{lord2008fast}
Lord R, Fang F, Bervoets F, Oosterlee CW (2008) A fast and accurate
  \uppercase{FFT}-based method for pricing early-exercise options under
  {L{\'e}vy} processes. \emph{SIAM Journal on Scientific Computing}
  30:1678--1705.

\bibitem[{Malliaris \protect\BIBand{}
  Salchenberger(1993)}]{malliaris1993beating}
Malliaris M, Salchenberger L (1993) Beating the best: A neural network
  challenges the {Black}-{Scholes} formula. \emph{Proceedings of 9th IEEE
  Conference on Artificial Intelligence for Applications}, 445--449 (IEEE).

\bibitem[{Merton(1973)}]{merton1973theory}
Merton RC (1973) Theory of rational option pricing. \emph{Bell Journal of
  Economics and Management Science} 4:141--183.

\bibitem[{Milstein \protect\BIBand{} Tretyakov(2003)}]{MT03}
Milstein G, Tretyakov M (2003) The simplest random walks for the {D}irichlet
  problem. \emph{Theory Probab. Appl.} 47:53--68.

\bibitem[{Milstein \protect\BIBand{} Tretyakov(2021)}]{MTbook21}
Milstein GN, Tretyakov MV (2021) \emph{Stochastic numerics for mathematical
  physics.} (Switzerland: Springer), second edition.

\bibitem[{O'Sullivan(2005)}]{o2005path}
O'Sullivan C (2005) Path dependant option pricing under {L{\'e}vy} processes.
  \emph{EFA 2005 Moscow Meetings Paper}.

\bibitem[{Rackauckas et~al.(2020)Rackauckas, Ma, Martensen, Warner, Zubov,
  Supekar, Skinner, \protect\BIBand{} Ramadhan}]{rackauckas2020universal}
Rackauckas C, Ma Y, Martensen J, Warner C, Zubov K, Supekar R, Skinner D,
  Ramadhan A (2020) Universal differential equations for scientific machine
  learning. \emph{arXiv:2001.04385} .

\bibitem[{Rebonato et~al.(2009)Rebonato, McKay, \protect\BIBand{}
  White}]{rebonato2009sabr}
Rebonato R, McKay K, White R (2009) \emph{The SABR/LIBOR Market Model: Pricing,
  calibration and hedging for complex interest-rate derivatives} (John Wiley \&
  Sons).

\bibitem[{Schmidhuber(2015)}]{schmidhuber2015deep}
Schmidhuber J (2015) Deep learning in neural networks: An overview.
  \emph{Neural Networks} 61:85--117.

\bibitem[{Sirignano \protect\BIBand{} Spiliopoulos(2018)}]{sirignano2018dgm}
Sirignano J, Spiliopoulos K (2018) \uppercase{DGM}: A deep learning algorithm
  for solving partial differential equations. \emph{Journal of Computational
  Physics} 375:1339--1364.

\bibitem[{Smolyak(1963)}]{smolyak}
Smolyak S (1963) Quadrature and interpolation formulas for tensor products of
  certain classes of functions. \emph{Soviet Mathematics Doklady} 4:240--243.

\bibitem[{Srivastava et~al.(2015)Srivastava, Greff, \protect\BIBand{}
  Schmidhuber}]{srivastava2015training}
Srivastava RK, Greff K, Schmidhuber J (2015) Training very deep networks.
  \emph{arXiv:1507.06228} .

\bibitem[{Steen et~al.(1969)Steen, Byrne, \protect\BIBand{}
  Gelbard}]{steen1969gaussian}
Steen N, Byrne G, Gelbard E (1969) Gaussian quadratures for the integrals $
  \int_0^\infty\exp(-x^2)f(x)dx $ and $ \int_0^b\exp(-x^2)f(x)dx $.
  \emph{Mathematics of Computation} 661--671.

\bibitem[{Su et~al.(2017)Su, Chen, \protect\BIBand{} Newton}]{su2017option}
Su H, Chen D, Newton DP (2017) Option pricing via \uppercase{QUAD}: from
  {Black}-{Scholes}-{Merton} to {Heston} with jumps. \emph{Journal of
  Derivatives} 24:9--27.

\bibitem[{Su \protect\BIBand{} Newton(2020)}]{su2020option}
Su H, Newton DP (2020) Widening the range of underlyings for derivatives
  pricing with \uppercase{QUAD} by using finite difference to calculate
  transition densities - demonstrated for the no-arbitrage {SABR} model.
  \emph{Journal of Derivatives} 28.

\bibitem[{van Milligen et~al.(1995)van Milligen, Tribaldos, \protect\BIBand{}
  Jim{\'e}nez}]{van1995neural}
van Milligen BP, Tribaldos V, Jim{\'e}nez J (1995) Neural network differential
  equation and plasma equilibrium solver. \emph{Physical Review Letters}
  75:3594.

\bibitem[{Yadav et~al.(2015)Yadav, Yadav, \protect\BIBand{}
  Kumar}]{yadav2015introduction}
Yadav N, Yadav A, Kumar M (2015) \emph{An introduction to neural network
  methods for differential equations} (Springer).

\bibitem[{Yu(2007)}]{yu2007closed}
Yu J (2007) Closed-form likelihood approximation and estimation of
  jump-diffusions with an application to the realignment risk of the chinese
  yuan. \emph{Journal of Econometrics} 141(2):1245--1280.

\bibitem[{Zhou(2022)}]{zhou2022option}
Zhou Y (2022) Option trading volume by moneyness, firm fundamentals, and
  expected stock returns. \emph{Journal of Financial Markets} 58:100648.

\end{thebibliography}

\begin{appendices}

\section{Training and Calculation Times}

Training NNs to replace other numerical methods can be expensive but, once trained, the NNs can be used, e.g. in option pricing, with no further major computational effort to evaluate TPDFs, as illustrated in our paper. The speed of training depends on the GPU and the deep learning implementation library used. Multiple GPUs can, of course, accelerate the training and it is also possible to use server training for asynchronous training. Theoretically speaking, using such an asynchronous training strategy, $n$ GPUs can increase the speed by $n$ times. For our work, we did not have access to computers with multiple high-end GPUs but, nevertheless, our paper shows that training can still be achieved using a single GPU, either via the Google Colab Pro server or simply using a gaming computer (since gaming computers come with relatively powerful GPUs). More powerful computers could considerably accelerate the training but this is unimportant since a trained network can be used without further computational effort and becomes an ultra-fast generator of TPDFs, insertable into various numerical techniques (we illustrate using QUAD for option pricing) and, also, is portable to less powerful computers.

In Appendix \ref{appdx:GPU}, we compare training speeds using 10 GPUs. Of these, 5 GPUs are from the gaming GPU category and the other 5 are from the data centre and high performance computing category. We show in Appendix~\ref{appdx: calculation_time} that the online accessing time is very quick and independent of the number of points on which TPDFs need to be evaluated.

\subsection{Training time using various GPUs}\label{appdx:GPU}

In this test, we select 5 NVIDIA gaming GPUs (RTX 3090, RTX 3080, RTX 2080 Ti, RTX 2070, RTX 1080 Ti) and 5 NVIDIA data centre GPUs (A100 SXM4, A40, V100, P100, T4). We refer readers to the NVIDIA official web page for their detailed specifications. The GPU memory size limits the amount of data which can be used in NN training as well as the complexity of NN architecture. For all the applications in this paper, GPU memory size was not a limiting factor; but to tackle more complex problems (e.g., with more parameters to train or training for higher dimensional models of underliers) GPUs with more memory might be required. At the same time, memory size is not a deciding factor for training speed which is determined by the number of GPU cores and the GPU architecture. It can be expected that more cores can accelerate parallel computation while newer generation GPUs deliver faster computations. In particular, RTX 3090, RTX 3080, A100 SXM4 and A40 are equipped with the latest NVIDIA Ampere GPU architecture (at the time of the writing) and all the other GPUs have previous generation GPU architectures. The older generation of the TensorFlow library (TensorFlow 2.3 or lower) cannot be used on Ampere architecture GPUs. However, we find that for GPUs with previous generation architectures, TensorFlow 2.2 and TensorFlow 2.3 work best, while newer or older versions perform worse. Thus, we use TensorFlow 2.3 on GPUs with previous generation architectures and TensorFlow 2.5 on GPUs with Ampere GPU architecture. The speed of training depends solely on the GPUs where thousands of NVIDIA's Compute Unified Device Architecture (Tensor or CUDA) cores perform highly efficient parallel computing. It might be expected that newer GPUs should perform much better than older ones and that higher end models are better, e.g., RTX 3090 better than RTX 3080 and RTX 3080 better than RTX 2080 Ti. This turns out not to be entirely the case, as can be seen from the test results below.

\begin{figure}
\begin{centering}
\includegraphics[scale=0.3]{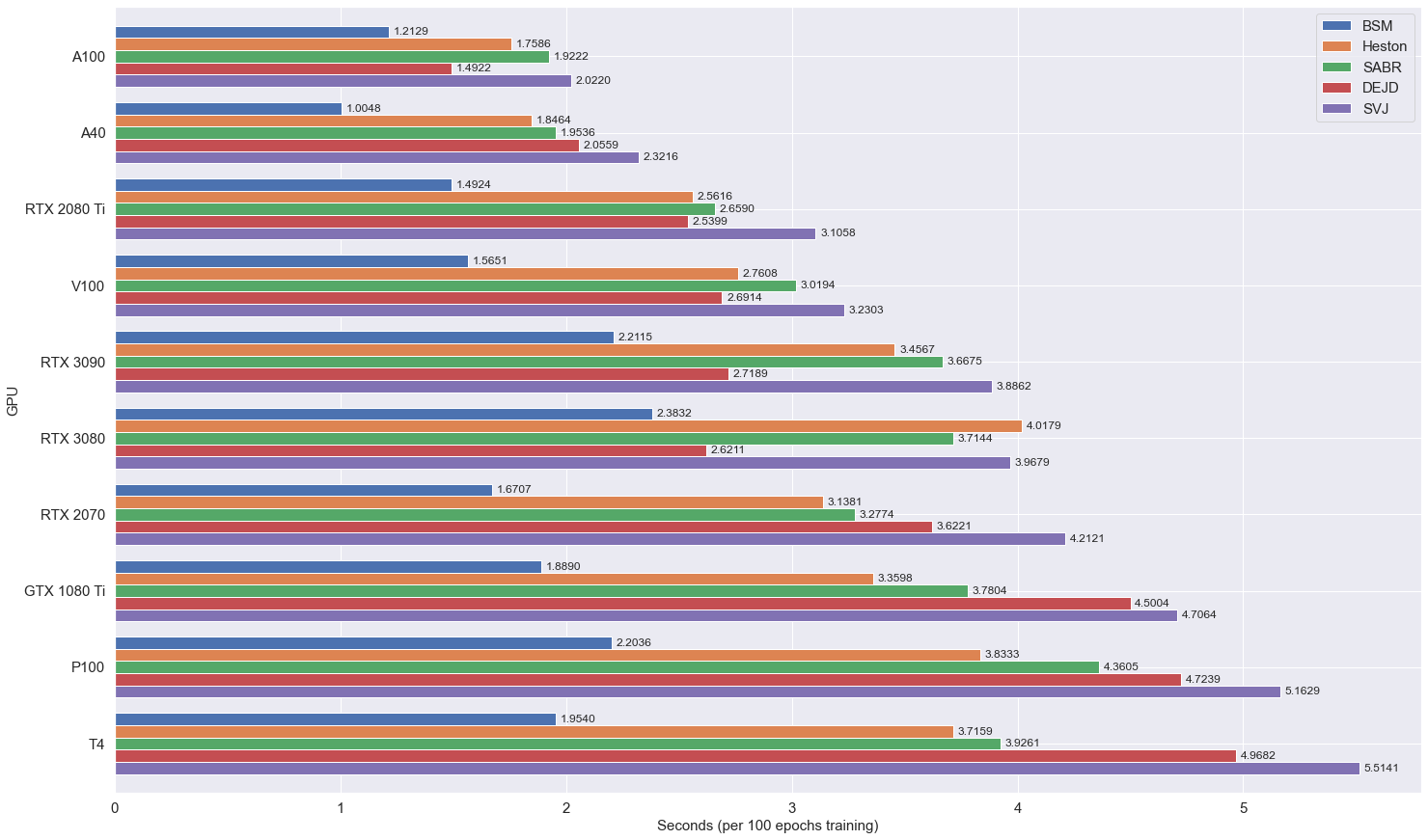}
\par\end{centering}
\caption{Training time of the 5 models of underliers considered in this paper per 100 epochs using
various GPUs. Ranked by the speed of the stochastic volatility jump
diffusion model (SVJ).}
\label{Fig: GPUs_speed}
\end{figure}

Figure \ref{Fig: GPUs_speed} shows the training times using various
GPUs. One can see that the most powerful data centre GPU A100 (at
the time of the writing) tops this comparison, followed by
another data centre GPU A40. Among gaming GPUs, the RTX
2080 Ti performs the best, even better than its successors RTX 3080
and 3090. This could be because TensorFlow 2.2 and 2.3 are more optimised
for training the NN architecture used in this paper.
If we use the same TensorFlow 2.5 on RTX 2080 Ti as on the Ampere GPUs,
the computing time increases to 3.001, 4.9469, 4.8508, 3.7418, 5.6915 seconds on the five different models of underliers, respectively  -- a 50\% to 100\% increase compared with using TensorFlow 2.3, making it worse than RTX 3080 and 3090. However, since RTX 3080 and 3090 cannot use TensorFlow 2.3 or below, the older GPUs such as RTX 2080 Ti are still better than the newer ones. Additionally, we find that the previous generation top gaming
GPU RTX 2080 Ti performs a little bit better than the same generation
data centre server GPU V100. This demonstrates that the more accessible and cheaper gaming GPUs are already sufficient for the purposes of this paper.

To conclude, we have shown that the NNs exploited in this paper can be
trained using a single gaming GPU. As computing power advances,
the training speed will certainly be improved. It is worth emphasising that the results in this paper come from training using Google Colab Pro servers (the two types of GPUs available in Google Colab Pro servers are P100 and T4). These two older GPUs can still be used to train NNs, although that may take much more time than using the
latest GPUs. However, training is done offline (i.e., before the NN is used in computational finance), and in finance applications the offline training speed is not important in comparison with the online speed of accessing TPDFs.
Next, we will show how fast online accessing speed is.

\subsection{Calculation Time}
\label{appdx: calculation_time}

\begin{figure}
\begin{centering}
\includegraphics[scale=0.3]{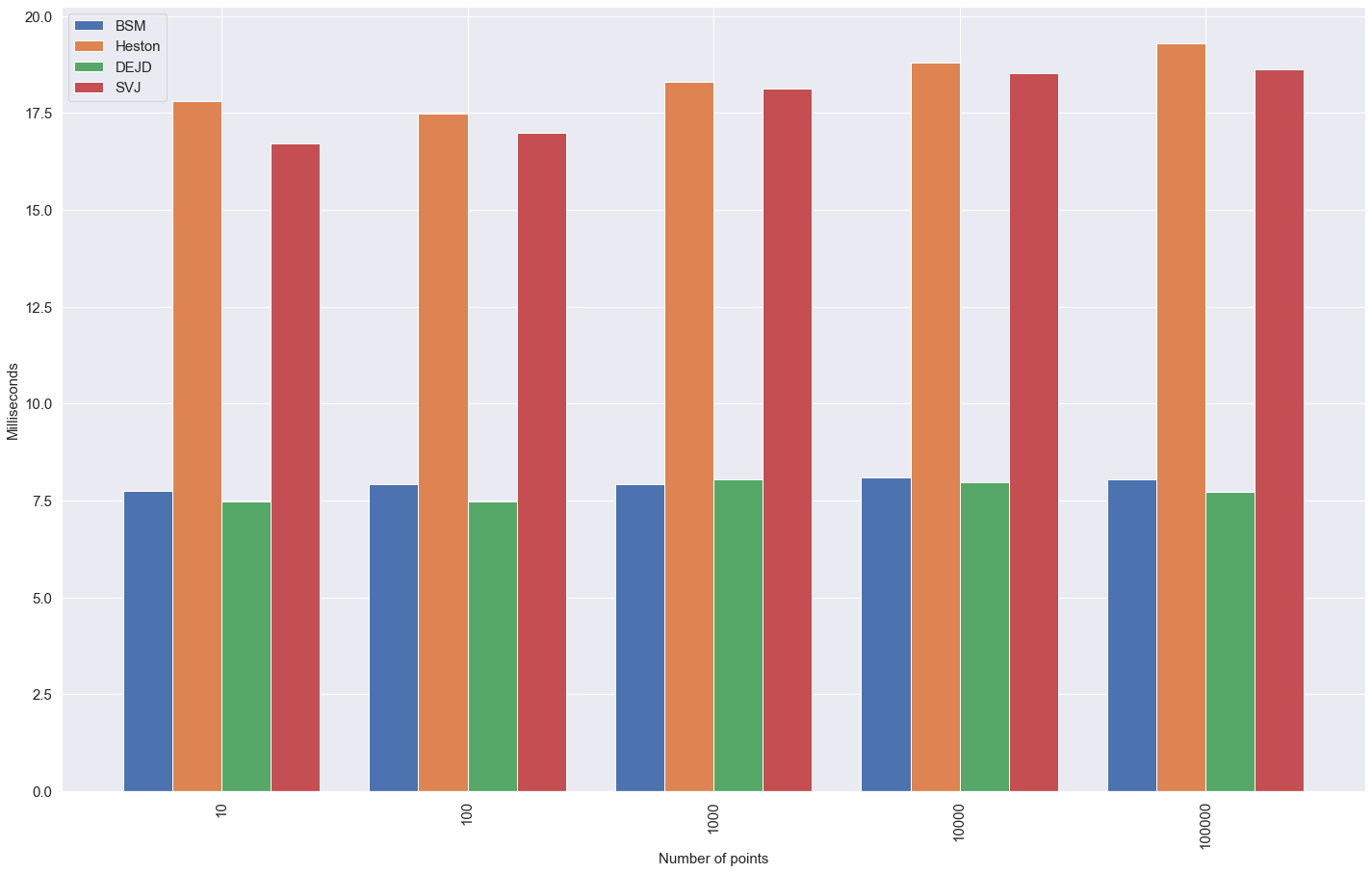}
\par\end{centering}
\caption{Computing time of TPDFs for various models of underliers on a single
GPU RTX 2070. The number of points the NN calculates at the same
time are 10, 100, 1000, 10000, 100000.}

\label{Fig: Computing_speed}
\end{figure}

Once a NN is trained and ready to use, the application calculation (`online') time, delivering densities, is very fast. Similar to the training, the calculations are also done in parallel with thousands of CUDA cores, meaning that the calculation speed does not depend on the number of points calculated. In Figure \ref{Fig: Computing_speed}, we calculate the NN approximated densities using a single GPU RTX 2070. The NN computes the densities in a matter of a few milliseconds; the calculation time remains roughly the same regardless of the number of points calculated. Here, we note that the maximum number of points a GPU can handle depends on its
memory size. The fast online usage time is another useful property
of the NN approach shown here. The difference in computing time between one dimensional and two dimensional models of underliers is mainly due to the differentiation of the CDF to obtain the TPDF.

\section{The performance of using Multilayer Perceptrons} \label{appdx: MLP}

In this appendix, we briefly discuss the performance of an alternative
NN setup. In Sections~\ref{sec:geometric-Brownian-motion}-\ref{sec:main_examples}, we utilised the DGM NN, which was introduced alongside the DGM algorithm by \citet{sirignano2018dgm}. In this appendix we explore whether simpler network architectures can also be effective for training in the context of approximating TPDFs. Specifically, we examine one of the simplest NN structures, known as Multilayer Perceptrons (MLPs). MLPs are widely used in deep learning and are a type of feedforward NN (see, e.g., \cite{goodfellow2016deep}).

\begin{figure}
\begin{centering}
\includegraphics[scale=0.8]{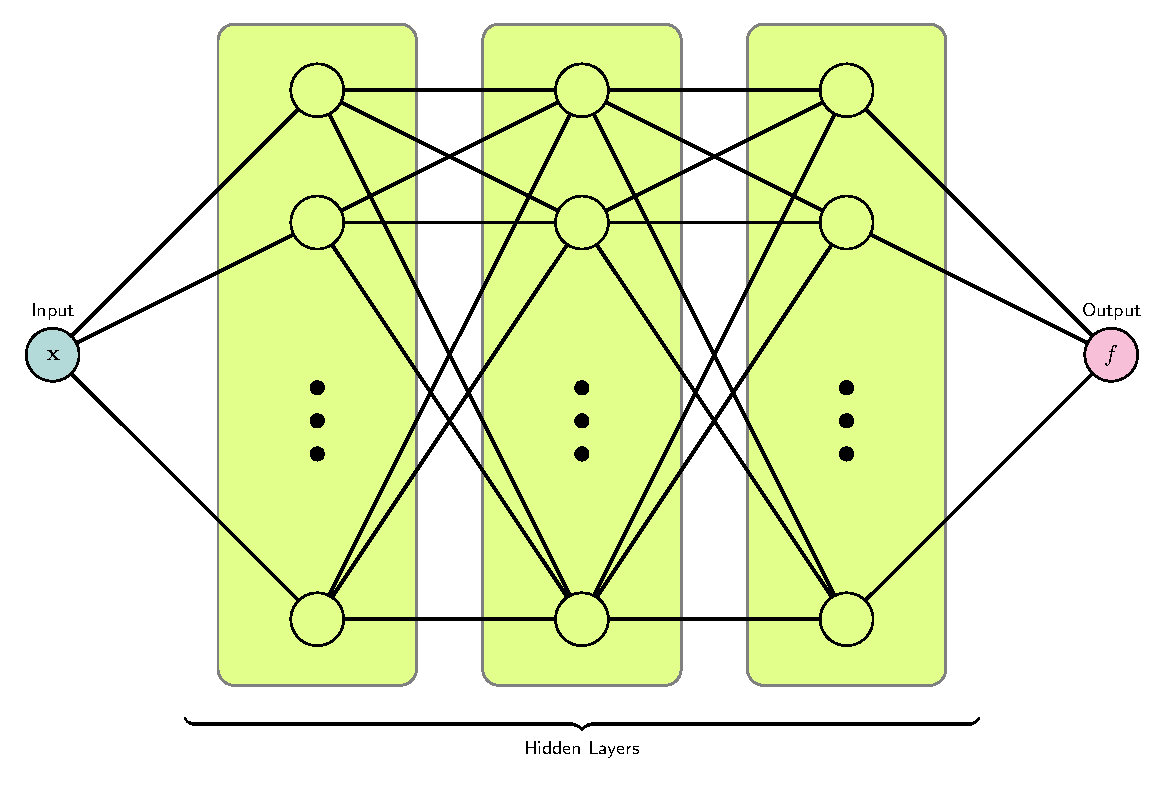}
\par\end{centering}
\caption{Illustration of the Multilayer Perceptrons NN architecture.}

\label{fig: MLP}
\end{figure}

MLPs consist of multiple layers of nodes, where each node in a layer is connected to every node in the subsequent layer. Figure \ref{fig: MLP} provides a visualisation of the MLP neural network architecture. The initial layer is referred to as the input layer, while the final layer is known as the output layer. Any layers positioned between these two are referred to as hidden layers. The number of nodes in the input and output layers is determined by the specific problem, while the number of nodes in the hidden layers is a hyperparameter that must be selected.

Each node within a layer is connected to every node in the subsequent layer through weighted connections. During training, these weights are adjusted to enable the neural network to learn how to perform the desired task, such as classification or regression. MLPs derive their name from ``perceptrons" because each node calculates a weighted sum of its inputs and applies an activation
function to the result. The activation function introduces nonlinearity to the network, allowing it to model complex relationships between inputs and outputs.

Using the same notations as in Section~\ref{sec:DGMNN}, the MLP neural network
can be expressed as follows:
\begin{align*}
\mathbf{S}^{1} & =\vartheta(\mathbf{W}^{0}\mathbf{x}+\mathbf{b^{0}})\\
\mathbf{\mathbf{S}}^{\ell+1} & =\vartheta(\mathbf{W}^{\ell}\mathbf{S}^{\ell}+\mathbf{b}^{\ell}),\thinspace\thinspace\ell=1,\dots,L-1,\\
f(\boldsymbol{\mathbf{x}};\boldsymbol{\theta}) & =\mathbf{W}\mathbf{S}^{L}+\mathbf{b},
\end{align*}
where $\mathbf{x}$ represents the input layer, $\vartheta(\cdot)$
denotes the activation function, $L$ signifies the number of hidden
layers, and $\mathbf{W},\mathbf{b}$ are the network parameters, i.e.,
weights and biases. The input $\mathbf{x}$ and the output $f$ are
described in Section~\ref{sec:DGMNN}. From the above expressions, we can observe
that the MLP structure is simpler compared to the DGM network.

While making direct comparisons between two different network structures
is challenging, we opted to use a MLP network with 3 hidden layers
and 300 nodes per layer. This choice aimed to minimise the computational
time difference with the DGM NN used in Sections~\ref{sec:geometric-Brownian-motion}-\ref{sec:main_examples}\footnote{During our testing, we found that GPUs such as the NVIDIA RTX 2070
and NVIDIA Tesla T4 yielded identical computational times for both
networks. However, on more powerful server GPUs like the NVIDIA RTX
A6000, the MLP network demonstrated a computational time approximately
30\% faster than that of the DGM NN.}.

Tables~\ref{Tab: MLP_DEJD_pricing_error} and \ref{Tab: MLP_SVJ_pricing_error} present the pricing errors of two models: Kou's double exponential
jump diffusion and stochastic volatility jump diffusion (SVJ). The
results indicate that the MLP network performs slightly better on
Kou's model, while the DGM NN excels on SVJ. This intriguing discovery
challenges the assumption that a more complex neural network structure
always yields superior performance. One possible explanation for this
phenomenon is that the DGM NN was specifically designed to tackle
high-dimensional problems, thereby achieving better results on SVJ
(a 12-dimensional parametric problem) than Kou's model (an
8-dimensional parametric problem).

\begin{table}
\caption{Under the MLP network, the pricing errors of the NN approximated density of Kou's double exponential jump diffusion model, compared with Kou's semi-closed form pricing solutions. The other table information remains consistent with Table~\ref{Tab: DEJD_pricing_error}.}
\label{Tab: MLP_DEJD_pricing_error}
\scriptsize
\centering{}%
\begin{tabular}{ccccccc}
\toprule
Maturity & Error Type &     DOTM &       OTM &       ATM &       ITM &      DITM \\
\midrule
$t=0.25$ & PrPCTE &  0.565540 &  0.227200 &  0.047374 &  0.020492 &  0.011626 \\
      & PrRMSE &  0.000309 &  0.001625 &  0.004176 &  0.003275 &  0.004250 \\
      & IVPCTE &  0.053758 &  0.036993 &  0.046664 &  0.078632 &  0.195837 \\
      & IVRMSE &  0.026923 &  0.019267 &  0.021644 &  0.024183 &  0.076983 \\
\midrule
$t=0.5$ & PrPCTE &  0.183634 &  0.054303 &  0.032206 &  0.013901 &  0.008919 \\
      & PrRMSE &  0.000357 &  0.004457 &  0.002425 &  0.002385 &  0.003170 \\
      & IVPCTE &  0.024350 &  0.019254 &  0.030406 &  0.037600 &  0.072384 \\
      & IVRMSE &  0.010135 &  0.020723 &  0.008940 &  0.010598 &  0.023723 \\
\midrule
$t=0.75$ & PrPCTE &  0.152235 &  0.088177 &  0.022879 &  0.011666 &  0.007300 \\
      & PrRMSE &  0.000435 &  0.001322 &  0.002684 &  0.002234 &  0.002597 \\
      & IVPCTE &  0.018415 &  0.018745 &  0.023755 &  0.026259 &  0.042757 \\
      & IVRMSE &  0.006708 &  0.007255 &  0.007833 &  0.007467 &  0.013185 \\
\midrule
$t=1.0$ & PrPCTE &  0.128098 &  0.028252 &  0.023524 &  0.011519 &  0.006743 \\
      & PrRMSE &  0.001195 &  0.001718 &  0.003825 &  0.003100 &  0.002656 \\
      & IVPCTE &  0.020543 &  0.012923 &  0.025150 &  0.023839 &  0.030379 \\
      & IVRMSE &  0.008059 &  0.005255 &  0.009625 &  0.009151 &  0.009602 \\
\bottomrule
\end{tabular}

\end{table}

\begin{table}[h]
\caption{Under the MLP network, the pricing errors of the NN approximated density of stochastic volatility jump diffusion model, compared with SVJ's semi-closed form pricing solutions. The other table information remains consistent with Table~\ref{Tab: SVJ_pricing_error}.}
\label{Tab: MLP_SVJ_pricing_error}
\scriptsize
\centering{}%
\begin{tabular}{ccccccc}
\toprule
Maturity & Error Type &     DOTM &       OTM &       ATM &       ITM &      DITM \\
\midrule
$t=0.25$ & PrPCTE &  0.355213 &  0.053095 &  0.020794 &  0.012279 &  0.008007 \\
      & PrRMSE &  0.002246 &  0.002587 &  0.002647 &  0.002708 &  0.003373 \\
      & IVPCTE &  0.056589 &  0.027055 &  0.020781 &  0.021206 &  0.045405 \\
      & IVRMSE &  0.017434 &  0.007637 &  0.006148 &  0.006497 &  0.020049 \\
\midrule
$t=0.5$ & PrPCTE &  0.066232 &  0.018693 &  0.010047 &  0.007895 &  0.006452 \\
      & PrRMSE &  0.001825 &  0.001859 &  0.001935 &  0.002292 &  0.003204 \\
      & IVPCTE &  0.020812 &  0.012015 &  0.010197 &  0.011691 &  0.017303 \\
      & IVRMSE &  0.008963 &  0.005024 &  0.004530 &  0.005312 &  0.008535 \\
\midrule
$t=0.75$ & PrPCTE &  0.029266 &  0.010795 &  0.007516 &  0.006662 &  0.006045 \\
      & PrRMSE &  0.001542 &  0.001628 &  0.001888 &  0.002377 &  0.003339 \\
      & IVPCTE &  0.011648 &  0.007788 &  0.007708 &  0.009246 &  0.013340 \\
      & IVRMSE &  0.006291 &  0.004281 &  0.004482 &  0.005495 &  0.008182 \\
\midrule
$t=1.0$ & PrPCTE &  0.017908 &  0.008097 &  0.006694 &  0.006325 &  0.006134 \\
      & PrRMSE &  0.001418 &  0.001638 &  0.002033 &  0.002582 &  0.003563 \\
      & IVPCTE &  0.008275 &  0.006303 &  0.006951 &  0.008483 &  0.012112 \\
      & IVRMSE &  0.005223 &  0.004297 &  0.004902 &  0.006005 &  0.008426 \\
\bottomrule
\end{tabular}
\end{table}

However, determining the best NN for solving parametric PDEs extends beyond the scope of this paper.

\end{appendices}

\end{document}